\documentstyle[epsfig,12pt]{article}
\title{ 
\mbox{}\hfill \normalsize LPNHE 98--XX\\
 \Large Reconstruction and Particle Identification for  a DIRC System }
\author{M.  Benayoun$^1$, L. Del Buono, Ph. Leruste \\[1.0cm]
\normalsize $^1$ LPNHE des Universit\'es Paris VI et VII, IN2P3--CNRS,
Paris, France }
 
\date{April 1st, 1996}
\date{\today}
\begin{document}
\maketitle
\def\hhht{\rule[ 0.mm]{0.mm}{6.mm}}
\def\hhha{\rule[-3.mm]{0.mm}{7.mm}}
\def\hhhb{\rule[-3.mm]{0.mm}{9.mm}}
\def\hhhu{\rule[-3.mm]{0.mm}{12.mm}}
\def\hhhv{\rule[-3.mm]{0.mm}{10.mm}}
\vspace{1cm}
 
{\small 
{\bf Abstract}~: We study the reconstruction and particle identification (PID) problem for
Ring Imaging devices providing a good knowledge of the direction of the Cerenkov photons,
as the DIRC system, on which we specialize. We advocate first the use of the stereographic
projection as a tool allowing a suitable representation of the photon data, as it allows
to represent the Cerenkov cone always as a circle. We set up an algorithm able to perform
reliably a fit of circle arcs of small angular opening, by minimising a true 
$\chi^2$ expression. The system we develop for PID
relies on this algorithm and on a procedure able to remove background photons with
a high efficiency. We thus show that, even when the background is large, it is possible
to perform an efficient PID by means of a fit algorithm which finally provides all the circle parameters~;
these are connected with the charged track direction and its Cerenkov angle.
It is shown that background effects can be dealt without spoiling significantly 
the reconstruction probability distributions.
}  
 
\vspace{1.0cm}
{\bf Keywords~:}  data analysis methods, event shape analysis, particle identification.

{\bf Free Keywords~:}  Rich detector reconstruction, DIRC  reconstruction.
 
\newpage

\vspace{1.cm}

\section{Introduction}

\indent \indent Cerenkov Ring Imaging devices settle specific problems of pattern recognition.
Generally, devices are designed in such a way that one has to look for photons
projected onto a circle \cite{YPSILANTIS, SEGUINOT}, the radius of which being simply connected 
to the Cerenkov angle of the charged
track radiating these photons. In RICH detectors (as for DELPHI), the main problem is due to
background photons which come together with the real signal photons and render difficult
pattern recognition, basically because the number of independent constraints which can be used is 
small. Traditionally, current methods in pattern recognition follow more or less
the ideas from Baillon \cite{BAILLON}, assume the knowledge of the charged track information 
and perform a maximum likelihood fit to the number of background and/or signal photons, under the
various mass assumptions.

It is not of current knowledge to attempt a full fit of the ring parameters, mainly because
background photons are expected to spoil the fit quality, and anyway prevent to define a 
reasonable $\chi^2$ to be minimised and then a fit probability having the properties expected for
a probability distribution. This is mainly due to the difficulties encountered when trying
to select a set of photons associated with a track and possibly still affected by
contamination from background photons.

Motivated by the DIRC device \cite{DIRC0,DIRC}, which will be used by the BaBar Collaboration \cite{BABAR} 
at the SLAC PEP--II B--factory, we reexamine this problem with the goal of associating to a given charged
track, a set of photons with a low level of contamination, having in mind to be able to check in each
photon sample the effect of a residual background contamination. This provides several problems to be
solved in order to have a procedure able to run under realistic conditions.

In a DIRC device, one can define with a controlled accuracy the Cerenkov photon directions.
This allows to reconsider the problem of data representation, with the aim of having 
a well defined geometrical figure to look for. We advocate here that the stereographic
projection provides such a suitable tool. Whatever are the errors, it allows to look always
for an image of the Cerenkov ring which is surely a circle arc, without any special 
hardware request on the DIRC device considered. This is true even if the 
charged track direction is poorly known, possibly even unknown. In real life however, because
of background, one cannot completely avoid some knowledge -- even approximate and/or subject
to systematic errors -- of informations about the charged track direction. In this representation,
the radius of this circle is connected in a simple way with the Cerenkov angle, and the
circle center with the charged track direction. The stereographic projection has also 
been proposed as a tool for pattern 
recognition{\footnote{We thank T. Ypsilantis (INFN, Bologna) for this information
and for having drawn our attention on the corresponding references.}}
 with a  RICH detector which will use a  27 Kton
water target and radiator in order to detect neutrino oscillation at Gran Sasso, using
a long base line  neutrino beam from CERN \cite{LBLRICH}. 
 
In working conditions, one is faced with another problem, specific to the DIRC~: generally,
the part of the Cerenkov cone populated by observed  photons represents a relatively small
azimuthal window compared with 360$^{\circ}$. In any representation of the data, this 
corresponds to a small  circle arc populated
with measured points. Taking into account the relative magnitude of the errors affecting 
these points and the circle radius, this generally prevents a direct use of standard circle fit 
algorithms.  Indeed, in such conditions, the fit quality is poor, even in absence of background,
as the radius is systematically underestimated. In order to circumvent this difficulty,
a further information has to be used and it happens that the charged track direction
is a good candidate for this purpose. This is basically the idea of Ref. \cite{BAILLON},
but in a completely different context. Indeed, one
can always assume that we have some knowledge about the charged track direction provided
by some tracking device located in front of the DIRC. However, this implies a deep change
in the circle fit algorithm structure.
The pattern recognition problem of the long base line RICH
of Ref. \cite{LBLRICH}, even if not trivial, is much simpler than for the DIRC as the circle arc
to reconstruct is generally quite large and the errors affecting photon directions are much smaller.

In this way, one is led to introduce as measurements to be fit the angular distances
between each of the various photon directions and the charged track direction. This 
introduces strong correlations which have to be
carefully accounted for when constructing the expression for the $\chi^2$ to be minimised.
These correlations are partly due to the measurement errors on the charged track direction
and partly due to multiple scattering effects undergone by the charged track inside the radiator~; 
this last kind of correlations can be important in identifying tracks of very low momentum 
(typically below 1 GeV/c) and they survive even if the circle arc populated by photons
is large enough that we need not worry about circle center information. 

Finally, having defined appropriate tools for pattern  recognition and particle identification,
it remains to define a procedure able to provide a set of photons really associated with the track.
As the number of constraints available for photon recognition is small, tools allowing
to control background contamination effects in this sample are needed. 

\vspace{0.7cm}
The paper  is organised as follows. In Section \ref{dircdet}, we briefly outline the DIRC structure
and properties relying on known literature \cite{DIRC0,DIRC}, in its aspects relevant to reconstruction
and particle identification. In Section \ref{stereo}, we recall the properties of the stereographic
projection and the connection with Cerenkov cone reconstruction. Section \ref{circlefit} is devoted
to describing the circle fit algorithm and recognition procedure when the circle arc populated by 
measured points is small~; we also
sketch here how to deal with photon contamination control and  background removal. In section \ref{simdirc1},
we describe the Monte Carlo we have coded in order to check the full procedure~; all effects affecting
the recognition procedure are included and varied from minimum (to check the basic model properties)
to a maximum (photons produced by several tracks in the same bar, with or without additional flat background). 
The procedure of photon selection and background removal is described in details in section \ref{photsel}, it is
more specific to the DIRC problem than the fit itself. This procedure
relies on using first a clean subsample of photons (unambiguous photons) to start an iterative 
recognition procedure. In section \ref{MCresults} we describe, the working of the procedure and show
that it behaves well, even under very large background conditions~; the effects of
correlations are specifically illustrated and it is shown that they cannot be neglected.
Of course, photon selection relies on cutting out photons candidates~; then, Section \ref{cutadjust} illustrates
how cut levels can be adjusted and tuned in such a way that the probability distributions are not
too much affected. We thus show that pull distributions and probability distributions allow
to perform the background removal with a controlled quality.
 
Finally, two appendices contain a full treatment of the multiple scattering effects and correlations. 
We describe here also our modelling, where all effects are taken into account at first order only.
Monte Carlo results show that the effects of this approximation are small in the region where a specific
particle identification is expected  to be reliable with a DIRC device, {\it i.e.} above 500 MeV/c.

\section{Outline of a DIRC Device and Properties}
\label{dircdet}

\indent \indent
The DIRC (Detection of Internally Reflected Cerenkov light) is a new type of detector which will be
used as the main particle identification system in the BaBar experiment \cite{BABAR} at PEP2 collider (SLAC).
A prototype of the DIRC \cite{DIRC} has been extensively tested in the CERN PS beam in 1995/1996,
demonstrating the validity of the DIRC concept.
It will not be described here in full details. We will rather briefly outline the general principles
involved in its conception and operation, more information can be found in recent literature on the
BaBar DIRC (see Refs.\cite{DIRC},  \cite{DIRC0} and \cite{DIRCNEW}). We mainly limit
ourselves  to the aspects of relevance for the procedure we develop and the Monte Carlo we will
use in order to test it.

\vspace{1.cm}

The originality of a DIRC device is that, contrarily to most other Cerenkov ring imaging detectors, it
makes use of the Cerenkov light generated in the radiator medium by trapping photons (through total internal
reflection) into the radiator itself and guiding them toward a set of photomultipliers (PMTs) for detection~;
this also allows the detector to be quite thin in the direction of the incoming tracks, because the Cerenkov cone
expands outside the main sensitive area  of the detector.

\begin{figure}[ht]    
\begin{center}
\epsfig{file=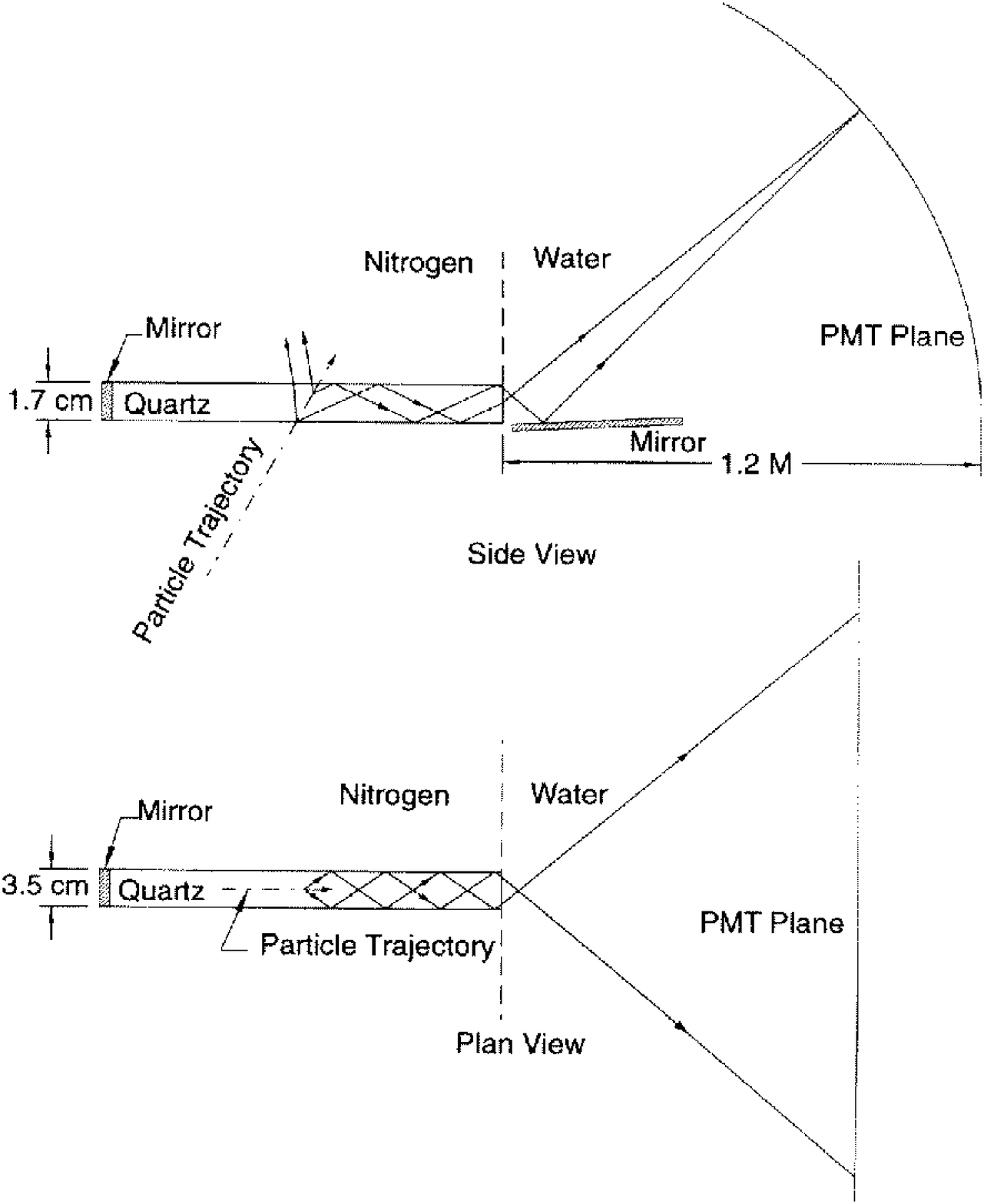,width=15cm}
\caption{A schematic drawing illustrating the DIRC principle~: photons produced in the quartz
bar radiator are transported to the bar end because of total internal reflection. Photon angles
are preserved till bar exit.}
\label{dircscheme}
\end{center}
\end{figure}

In the DIRC, the radiator is made of long rectilinear fused silica bars of modest rectangular section, a material
chosen mainly because of its high refractive index (1.474) \cite{DIRC,DIRCNEW} compared to air or nitrogen and its long 
absorption length in the UV region.
As sketched in  Fig. \ref{dircscheme}, the quartz/air interfaces of the bars act as perfect
mirrors for a wide range of Cerenkov photons incidence angles, and, for a sufficient optical and geometrical
quality, they are able to transport the photons to the bar exit, with unperturbed directions except for 
reflection symmetries  with respect to the bar faces.

The number of photons produced in the quartz bars depends on several parameters. Interestingly, it tends
to increase with the incidence angle (since the path length inside the radiator
medium grows), which is a behavior rather opposite to more traditional ring imaging detectors, where a large
part of the Cerenkov light is lost {\it because} of the internal reflection inside the radiator.

To avoid loosing too much light at the bar exit, the array of photomultiplier windows
and the light quartz bar exits are immersed in pure water, which refractive index (1.34) is 
close to the quartz one (1.474), so that photons crossing the quartz/water interface have a 
low probability of being reflected back into the bar. 

There is also a reflecting device parallel to the bottom bar surface put at the bar end  (in the water)
in order to redirect toward the PMTs the photons emerging from the bar with either downward or too upward
going directions~; the other bar end is equipped with a small mirror in optical contact with the quartz
for the same purpose. 

\begin{figure}[ht]    
\begin{center}
\epsfig{file=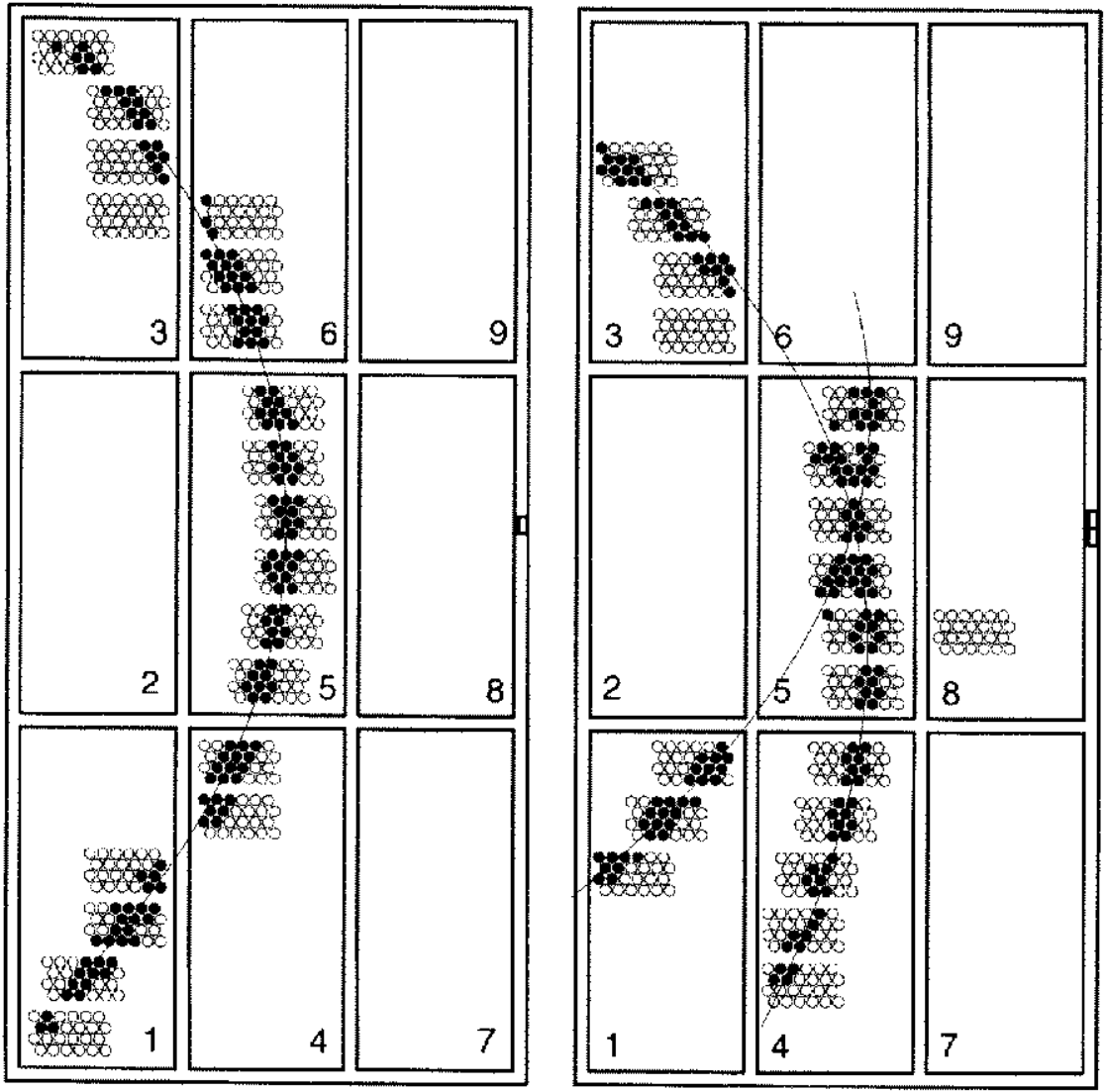,width=15cm}
\caption{Display of a single track event in the DIRC prototype showing the superposition of reflections of the
original Cerenkov cone.}
\label{dircimage}
\end{center}
\end{figure}

Once the Cerenkov image is detected on the PMTs array, two informations are available~: the
location of the hit PMT and the photon detection time. This image (see Fig. \ref{dircimage} for an example
from the DIRC prototype tested at CERN \cite{DIRC}) is in fact a superposition of different reflections of the
original Cerenkov cone by the bar reflecting surfaces. So, a given hit can belong to the Cerenkov cone
centered on the original track direction, or to any of its images with respect to the reflecting 
surfaces of the bar.
In the case of the prototype of Ref. \cite{DIRC}, 8 reflected photon directions (due to 3 possible symmetry
reflections planes) are to be taken into account. This ambiguity problem is specific to the DIRC~: given a
definite photon direction, there are as many track--photon associations as reflection planes and it may be
that several combinations are physically acceptable. Fortunately, it may also happen that only one solution is
acceptable~; this defines unambiguous solutions.
Fig. \ref{dircimage} also illustrates that the arc populated by the hits can be relatively
small (of the order of 60 degrees).

Using the spatial information, one is able to reconstruct uniquely the original Cerenkov angle of
a photon (which is emitted on the Cerenkov cone) if the ambiguities related with the various
{\it possible} reflections that could affect the photon during its propagation can be disentangled.
The timing information is mainly used for a preliminary recognition of background photons, generally out of 
time for a given track, as the information  provided by 
the individual photon detection time is very poor compared to the spatial information given by the PMTs.

\vspace{1.cm}
Thus, a possible strategy for pattern recognition in the DIRC is effectively to try discriminating spatially
between  photon ambiguities in order to determine the correct symmetry assignment, and then  use the set of non
ambiguous resulting photon Cerenkov angles (from the whole image) in order to compute the relevant quantities we are 
interested in~: the track Cerenkov angle and its error. This is achieved through a fitting procedure, and these
informations allow in a second step to refine the choice among the ambiguous photon solutions left
provisionally aside, and take part of them into account in a final fit.
Among the difficulties that are to be met, the smallness of the arc length populated by photons should be noticed
and has to be especially addressed.  

In the following sections, we detail the different parts of the algorithm implemented in order to achieve this goal.

\section{Pattern Recognition Using the Stereographic Projection } 
\label{stereo}

\indent \indent In a transparent medium of index $n$,
Cerenkov photons are emitted by a charged particle with 
 an angle $\theta_C$ with respect to the charged track
direction and  this angle is given by~: 

\begin{equation}
\cos{\theta_C}=\frac{1}{n \beta}
\label{basic1}
\end{equation}

\noindent where $\beta$ is the speed of the particle. As the photon
wavelength{\footnote{ We postpone to  Appendix A commenting 
on the influence of chromaticity fluctuations (dependence of $n$ on the photon 
wavelength $\lambda$) which affects each photon direction.
}} is generally not measured, this turns out practically to 
assume that the refractive index $n$ is known with a random error $\delta n$,
independently for each photon. 

Let us associate to each of the charged track and photon directions 
 a $unit$ vector, and draw all of them from
a common origin denoted $(0,0,0)$. All end points of these vectors lay
on a unit sphere, and all photon directions generate
a cone of half aperture $\theta_C$ around the charged track direction. 
From now on we shall always refer to photon
and charged track directions  only in this representation.

\subsection{The Stereographic Projection}

\indent \indent
The intersections of the photon directions
with a plane perpendicular to the charged track direction at unit distance 
from the origin define a set of points 
distributed  along a circle of radius $\tan{\theta_C}$, centered at the intersection
of the charged track direction with this plane. 
However, if the $actual$ charged track direction is not the one  chosen
in order to define the plane, the figure represented by the intersections of the
photon directions with this plane is no longer a circle but an ellipse, and departure 
from a circle may become large if the charged track direction is poorly known\cite{LHCB},
because of systematic errors in the measurement of the charged track direction, 
or misalignments effects. 

\begin{figure}[ht]  
\begin{center}
\epsfig{file=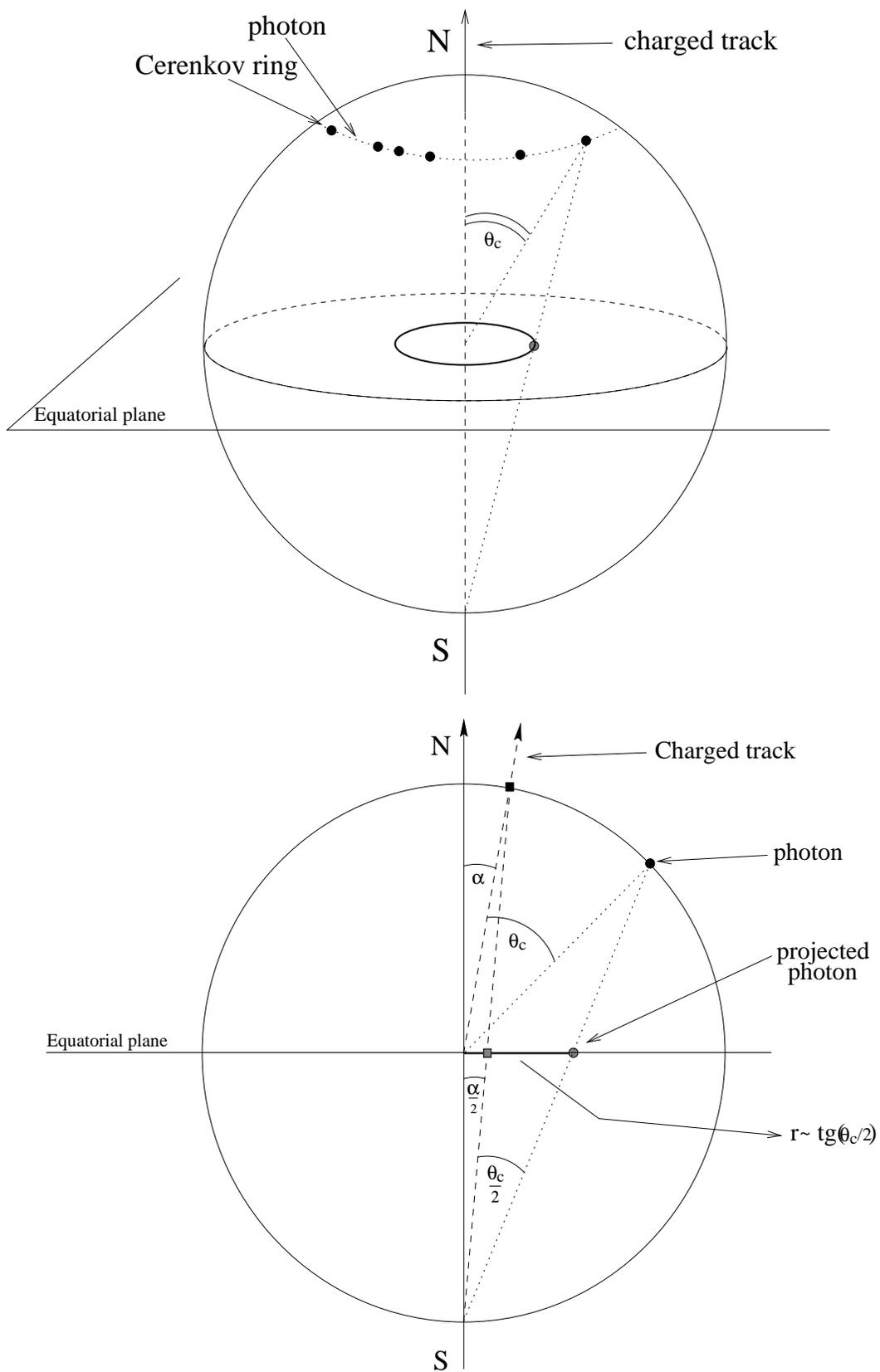,width=14cm}
\caption{Stereographic projection.}
\label{sphere}
\end{center}
\end{figure}

A way to circumvent this problem (or minimize it at least) is to use the
stereographic projection{\footnote{Basically, it is a standard conformal mapping
of the sphere onto a plane, {\it i.e.} angles on the sphere are conserved
in the projection.}} \cite{JULIA}. 
Let us briefly recall it. Let us choose on the
unit sphere defined above the pole axis along the measured charged track direction.
The stereographic projection of points on the sphere is the intersection of the 
lines joining the south pole $(0,0,-1)$ to these points with the equatorial plane.
In this transform, a circle drawn on the sphere ({\it i.e.} the intersection
of the Cerenkov cone with the sphere) is projected out as a circle. Then
the Cerenkov cone centered along the charged track direction becomes a circle
of radius $\tan{\theta_C/2}$, centered at the origin. This origin is simply 
the image of the charged track direction, {\it i.e.} the projection of the north
pole as seen from the south pole. This is sketched in Fig. \ref{sphere}.
Ref. \cite{LBLRICH} prefers performing the stereographic projection onto
a plane tangent to the sphere at the north pole, rather than onto the equatorial plane~;
correspondingly, the algebra is slightly modified with respect to what will be presented just below.

In practical applications, however, we only know approximately the direction of the
charged track, and therefore the pole axis as defined above coincides only 
approximately with the $actual$ charged track direction. In order to illustrate
what happens, let us assume that the $actual$ unknown charged track direction 
makes an angle $\alpha$, possibly large, with the pole axis ({\it i.e.} the
$reconstructed$ charged track direction). 
Then, by means of the stereographic projection, the images of the photon directions in the 
equatorial plane are still on a circle (see Fig. \ref{sphere}), 
the radius $R$ of which being~:

\begin{equation}
R =\displaystyle \tan{\frac{\theta_C}{2}} 
\frac{1+ \displaystyle \tan^2{\frac{\alpha}{2}}}
{ 1- \displaystyle \tan^2{\frac{\theta_C}{2}} \displaystyle \tan^2{\frac{\alpha}{2}}}
\label{basic2}
\end{equation}

\noindent and the circle center is shifted from the origin to a point located 
at distance $r_0$~:

\begin{equation}
r_0 =\displaystyle \tan{\frac{\alpha}{2}} 
\frac{1+ \displaystyle \tan^2{\frac{\theta_C}{2}}}
{ 1- \displaystyle \tan^2{\frac{\theta_C}{2}} \displaystyle \tan^2{\frac{\alpha}{2}}}
\label{basic3}
\end{equation}

\noindent close to the image of  the actual charged track direction, which is located at $\tan{\alpha/2}$.

It is clear from rels. (\ref{basic2}) and (\ref{basic3}) that the error on the circle center 
is of first order in $\alpha$, while the corresponding error on the  radius is only of second order.
Therefore, the robustness of the stereographic projection follows from the fact that the
analytical shape of the Cerenkov figure in the equatorial plane is always a circle, even
if the actual charged track direction is quite different from the measured one.
Moreover, the radius is affected at second order only by angular errors on the charged
track direction. If the angle $\alpha$ were large, it is clear from Rels. (\ref{basic2}) and (\ref{basic3})
that having determined by fit $R$  and $r_0$ allows anyway to reconstruct the correct Cerenkov angle.
Stated otherwise, the pole axis used in order to perform  the stereographic projection 
may be chosen independently of any assumption on the charged track direction. In general, we have~:

\begin{equation}
\tan{\theta_C}=\displaystyle \frac{2 R}{1-R^2+r_0^2}
\simeq \frac{2 R}{1-R^2}\left [ 1-\frac{r_0^2}{1-R^2} \right ]+{\cal{O}}(r_0^4)
\label{basicf}
\end{equation}
 
\noindent If $r_0$ is small enough, $R$ is negligibly affected{\footnote{If $\theta_C=500$ mr,
in order that $2 \arctan{R}$ provides an overestimate of at least 0.1 mr, the
error on the charged track direction should be, at least, $\alpha \simeq 30$ mr.
}} but can easily be corrected. 

The correspondence between the coordinates of a point on the unit sphere $(X,Y,Z)$, and 
those of its image $(x,y)$ into the  equatorial plane (through the stereographic projection) is
defined by~:

\begin{equation}
x=\frac{X}{1+Z}~~~~~, ~~~~~y=\frac{Y}{1+Z}~~~~,~~~~(X^2+Y^2+Z^2=1)
\label{basic4}
\end{equation}

This transform is never singular in our case as we have always $Z > 0$.

\subsection{Handling Measurements and Errors}
\label{errmeas}

\indent \indent
Let us assume that the above coordinates $X$, $Y$ and $Z$ (or any other quantity)
are measurable quantities with known covariance error  matrix, originating from random distributions
$\hat{X}$, $\hat{Y}$ and $\hat{Z}$~; assuming the measurement process
unbiased, we can write $\hat{X}=X_{true} +\delta X$, where $X_{true}$ is the true
(unknown) value of $X$ and $\delta X$ a random  variable of standard deviation $\sigma_X$ and of zero 
mean{\footnote{ 
We shall always use the notation $<f>$ for the expectation value of any random variable $f$.}}~:

\begin{equation}
\left \{
\begin{array}{lll}
<\delta X>=0 \\[0.5cm]
<[\delta X]^2>=\sigma_X^2
\end{array}
\right.
\label{basic5}
\end{equation}

\noindent Correspondingly, for any other measurable quantity, we  define
a centered error function carrying the  corresponding standard deviation
($\sigma_Y$ or $\sigma_Z$, for instance). As we assume the measurement process  unbiased,
we should have indeed $<\hat{X}>=X_{true}$. Using this language, true and measured values
are quantities which differ by first order terms ${\cal{O}}(\delta X)$.

The error functions  affecting the coordinates $x$ and $y$ (in the equatorial plane)
can be derived by differentiating Eqs. (\ref{basic4})~:

\begin{equation}
\left \{
\begin{array}{lll}
\delta x= \displaystyle \frac{1}{1+Z} \delta X - \frac{X}{(1+Z)^2} \delta Z \\[0.7cm]
\delta y=\displaystyle  \frac{1}{1+Z} \delta Y - \frac{Y}{(1+Z)^2} \delta Z \\
\end{array}
\right.
\label{basic6}
\end{equation}

When estimating errors using these expressions, $X$, $Y$ and $Z$ should be the corresponding
true values. As they are unknown, one classically uses instead the measured central values.
In terms of differentials, this lack of information affects second order terms (like $(\delta X)^2$).
So, at first order, it is legimate to use directly measured values while estimating coefficient functions.
It is clear that going beyond first order development in analytic expressions would raise a problem here,
as the additional terms would be competing with the (uncontrolled) terms introduced by using the measured values
instead of the true ones.

If the measurement $(X,Y,Z)$ is unbiased, the point $(x,y)$ is unbiased too at leading order
({\it i.e.} $<\delta x>=<\delta y>=0$). The covariance terms ($<[\delta x]^2>$, $<[\delta y]^2>$ and
$<\delta x \delta y>$) can be computed in terms of $X$, $Y$ and $Z$ and of their errors and
correlations~; when computing them, one has to take into account that $X^2+Y^2+Z^2=1$ and that
their error functions are not independent~: $X \delta X+ Y \delta Y+ Z \delta Z=0$.

\section{A Circle Fit Algorithm}
\label{circlefit}

\indent \indent 
As explained in section \ref{stereo}, using the stereographic projection, the 
directions of the Cerenkov photons emitted by a charged track are represented
by points in a plane laying on a circle. Up to second order terms, the circle center is nothing but 
the projection of the charged track direction onto the (equatorial) plane of the sphere.
Therefore, the reconstruction problem of the Cerenkov angle is replaced
by a problem of circle recognition in a plane.

It is a long standing problem to find the most suitable way to perform a circle
fit to a given set of points affected by measurement errors (see refs. 
\cite{CIRCLE1,CIRCLE2,CIRCLE3} for instance). The main problems addressed 
in (necessarily) approximate procedures are~:
 
\begin{itemize}
\item linearisation of circle parametrisation 
 
\item  non--gaussian character of the errors on the circle center coordinates and radius.
\end{itemize}

In addition to the above mentioned questions, we  address two more issues, connected
with the BaBar DIRC, namely~:

\begin{itemize}
\item  in any representation,
the measured points are not spread out onto the whole circle, but along a relatively
small arc (about 60$^{\circ}$ degrees). Taking into account the relative magnitude
of the error on the points and of the circle radius value, this happens to affect deeply
the circle fit quality, if no additional information on the circle center is accounted for. 

\item  there exist
correlations among photons as a consequence of the multiple scattering undergone 
by the emitting charged track. Accounting for further constraints (charged track direction measurement)
may introduce further correlations (see Sections 2 and 3 in Appendix B for instance).

\end{itemize}

\subsection{The $\chi^2$ for Fitting a Circle Arc}
 
\indent \indent
Let us assume that we have $n_{\gamma}$ measured points $(x_i,y_i)$ with
random errors $(\delta x_i,\delta y_i)$, not necessarily gaussian. 
As we restrict our study to a DIRC device where the points are actually photons,
we shall use indifferently the words photon and point.  
We do not state any assumption on these errors, except that the measurements are unbiased~:

\begin{equation}
<\delta x_i>=<\delta y_j>=0~~~,~~~ \forall i,j=1, \cdots ~n_{\gamma}
\label{fit1}
\end{equation}

Stated otherwise, the
expectation values  $<\delta x_i~ \delta x_j>$, $<\delta x_i~ \delta y_j>$ and
$<\delta y_i~ \delta y_j>$ which define errors and correlations are not constrained
and  no further assumption is needed on higher order moments.
In the approach we have followed,
the effects of multiple scattering are not affected to the  photons measurements but
rather to the track direction.

We assume  that we have, beside the circle points, also a $measurement$
 of the circle center coordinates  and its error~; we define our reference frame
in such a way that this measured center is located at  the origin.
The circle parameters we fit  are $a$ and $b$ (the center coordinates) and $R$ (its radius). 
This parametrisation  is examined in full details in Appendix B.

The $\chi^2$ to minimise in order to get the best circle fitting $n_{\gamma}$
points is defined by (see Appendix B 4)~:

\begin{equation}
\chi^2= \sum_{i,j} (d_i-R) (d_j-R) V^{-1}_{ij} + A^t \Sigma^{-1} A
\label{fit2}
\end{equation}
 
\noindent with~:

\begin{equation}
d_i =\sqrt{(x_i-a)^2+(y_i-b)^2}
\label{fit3}
\end{equation}

\noindent where $a$ and $b$ are the circle center coordinates  parameters 
(measured as  $(0,0)$, up to errors) to be fit.
We have denoted by $A$ the vector{\footnote{Actually, the
vector $(a,b)$ should be written $(a-a_{measured},b-b_{measured})$~; however we take
into account that the measurement has been conventionally set at $(0,0)$.}} $(a,b)$
of the  center coordinate and by $A^t$  its 
transposed. The error matrix $\Sigma$  depends on the reconstruction
errors of the track at the DIRC entrance ($\Sigma_0$), on the number of photons
which take part to the fit ($n_{\gamma}$) and on the multiple scattering undergone
by the charged track inside the radiator~; it is explicitly computed in Appendix A
and in Sections 2 and 3 of  Appendix B.

In usual approaches this second contribution to the $\chi^2$ 
is not considered \cite{CIRCLE1,CIRCLE2,CIRCLE3}~; however,  when the circle center is constrained
by an auxiliary measurement, it is legitimate to use it. Moreover,  it is harmless 
to remove it only if the circle length populated by
the measured points is large enough (typically greater than 180$^\circ$) and/or if
$\sqrt{<[\delta d_i]^2>}/d_i$ is small enough. In the case of the DIRC, none of these conditions
are practically met, and removing the constraint on the center represented by the second term in the RHS of
Rel. (\ref{fit2}) may simply lead to a complete failure of the circle fit, even in absence of fake 
photons. 

\vspace{1.cm}

The matrix $V$ which appears in Rel. (\ref{fit2})
is the error covariance matrix. It is also the matrix
of the error function expectation values~:

\begin{equation}
V_{ij}=<\delta d_i ~ \delta d_j>~~~,~~~ \forall i,j=1, \cdots ~n_{\gamma}
\label{fit4}
\end{equation}

\noindent The error function $\delta d_i$ affecting the measurement $i$  is given by~:

\begin{equation}
\delta d_i =\displaystyle \frac{x_i(\delta x_i+\delta a_i)+y_i(\delta y_i +\delta b_i)}{d_i}
\label{fit5}
\end{equation}

\noindent (see Appendix B4) up to higher order terms. In this expression, we use as central
values for the circle center coordinates the point $(0,0)$, while the error
functions on the circle center are estimated for each photon. It is the reason
why the error functions which appear in Rel. (\ref{fit5}) are $\delta a_i$ and $\delta b_i$,  
referring to the charged track direction when it emits photon $i$ and $\delta x_i$
and $\delta y_i$ are the measurement errors of the direction of photon $i$.
The form of the error functions for $\delta a_i$ and $\delta b_i$ is given
in Appendix B1, using preliminary results from Appendix A. 
The form of the error function $\delta d_i$ is explained in Appendix B4~; 
the elements of the matrix $V$ in Rel. (\ref{fit4})
are computed in Section B 5. One can see there how
correlations terms like $<\delta a_i \delta a_j>$, $<\delta a_i \delta b_j>$, \ldots
produce non zero correlation terms  in  $<\delta d_i \delta d_j>$. 
Consequently, one can interpret $a$ and $b$ in Rel.(\ref{fit3}) as the mean
values (to be fit) of the sets $\{a_i\}$ and $\{b_i\}$. Actually, the single
information -- except for errors -- we have on these sets are the measured values
at the bar entrance~; how this is accounted for is explained in details in Appendix B.

\subsection{Linearisation of the Circle Parameter Equation}

\indent \indent
 It remains to linearise  Eq. (\ref{fit2}). The procedure is quite
usual  \cite{CIRCLE1,CIRCLE2,CIRCLE3} and turns out to replace Eq. (\ref{fit2})
by~: 

\begin{equation}
\chi^2= \sum_{i,j} (d^2_i-R^2) (d^2_j-R^2) \displaystyle \frac{V^{-1}_{ij}}{4R_0^2} 
+ A^t \Sigma^{-1} A
\label{fit6}
\end{equation}
 
\noindent where $R_0$ is an estimate of the radius (a weighted mean of the
$d_i$ at start and, in the forthcoming steps of the iteration  the fit value of $R$ at the 
previous step of the iteration). The matrix $V$ depends on the circle center
coordinates. They are chosen here at start as $(0,0)$, and can also be updated
in the forthcoming steps of the iteration procedure.
The final step toward linearisation  is to define as 
fitting parameter $c=R^2-a^2-b^2$ instead of $R$, together with $a$ and $b$. 
Then Eq. (\ref{fit6}) becomes~:

\begin{equation}
\left \{
\begin{array}{lll}
\chi^2= \sum_{i,j} {\cal  C}_i {\cal  C}_j W^{-1}_{ij}
+ A^t \Sigma^{-1} A \\[0.7cm]
W_{ij}=4R_0^2 ~V_{ij}\\[0.5cm]
{\cal C}_i=x_i^2+y_i^2-2ax_i-2by_i-c\\[0.5cm]
\end{array}
\right.
\label{fit7}
\end{equation}

\noindent and this last expression for $\chi^2$ is quadratic in $a$, $b$ and $c$.
Eqs. (\ref{fit6}) and (\ref{fit7}) simply follow from the fact that near the minimum we have~:

$$ (d_i-R) = \displaystyle \frac{(d^2_i-R^2)}{(d_i+R)} \simeq \frac{(d^2_i-R^2)}{2R_0} $$

\noindent This relation fastly improves when iterating, and few iterations only are 
needed in order to be at less than $10^{-2}$  from $\chi^2_{min}$ (generally 2 steps
are enough for the accuracy just quoted).

The conditions defining the minimum are~:

\begin{equation}
\begin{array}{llll}
\displaystyle \frac{\partial \chi^2}{\partial a}=0~~~,
&\displaystyle \frac{\partial \chi^2}{\partial b}=0~~~,
&\displaystyle \frac{\partial \chi^2}{\partial c}=0
\end{array}
\label{fit8}
\end{equation}

\noindent and provide a linear system of equations for $a$, $b$ and $c$ which gives an optimum 
solution to the minimisation problem. On the other hand, denoting the variables $a$, $b$ and $c$
by $u_{\alpha}$, ($\alpha=1,2,3$), Eqs. (\ref{fit7}) can be written (summation over repeated indices
is understood)~:

\begin{equation}
\chi^2= T_{\alpha \beta}u_{\alpha}u_{\beta}+ Z_{\alpha}u_{\alpha}+K
\label{fit9}
\end{equation}

\noindent where the matrix $T$, the vector $Z$ and the scalar $K$ can easily be expressed
in terms of the (given) $W$ and $\Sigma$ matrices and of the photon coordinates $([x_i,y_i],
i=1,\cdots n_{\gamma})$ moments. In addition, the matrix $T^{-1}$ is the  error covariance matrix
for the fit parameters $(u_{\alpha}, \alpha=1,2,3)$ \cite{PDG}. This
covariance matrix gives the error contour at $\chi^2_{min}+1$, the 1$\sigma$ contour.

The number  of degrees of freedom associated with the $\chi^2$ in rel. (\ref{fit7}),
is $n_{\gamma}-1$ and then the fit probability is the value of the $\chi^2$ probability function
${\rm Pr}(\chi^2, n_{\gamma}-1)$.
One can also define a consistency check of the set of photons under consideration with 
the  set of track parameters~: the charged track direction providing the expected 
coordinates of the circle center $a_0=0$ and $b_0=0$ (the measured values) and the --five-- possible values
of the circle radius corresponding to the --five-- possible mass assignments for
the charged track, $R_k$ ($k=1, \cdots 5$). In this case, the  $\chi^2$
simplifies to~:

\begin{equation}
\left \{
\begin{array}{lll}
\chi^2= \sum_{i,j} {\cal  C}_i {\cal  C}_j W^{-1}_{ij}\\[0.7cm]
W_{ij}=4R_k^2 ~V_{ij}\\[0.5cm]
{\cal C}_i=x_i^2+y_i^2-c_0\\[0.5cm]
\end{array}
\right.
\label{fit10}
\end{equation}

\noindent where $c_0$ takes five possible values $c_0=R_k^2$, each corresponding
to one of the possible values of $R_k=\tan{\theta^k_C/2}$.
In this way, one can check the consistency of the  set of photons considered
with the  measured charged track parameters ($a_0=0$, $b_0=0$, $R_k$) for each of the five possible 
mass assignments. The $\chi^2$ just above
corresponds exactly to $n_{\gamma}$ degrees of freedom. One can then decide to choose the 
best assignment as being the one which corresponds to the lowest $\chi^2$, provided
it is above some significance threshold (a lower probability cut). We illustrate
in the following that the corresponding probability distributions have all expected
properties.

\subsection{Fit Likelihoods, Fake Photons and Contamination} 
 
\indent \indent Let us denote $\chi^2_{n_{\gamma}-1}$ and  $\chi^2_{n_{\gamma}}$, the
$\chi^2$ defined resp. by Rels. (\ref{fit7}) and (\ref{fit10}). Given a set of photons,
it is clear that, using standard formulae, the former $\chi^2$ allows to define a maximum 
likelihood including the center  measurement, while the latter leads to the likelihoods of the
full measured track with the full photon set considered, for each possible mass assignment.

In practical applications, however, it should be noted that defining the {\it true} 
(maximum) likelihood, implies that~: 

\begin{itemize}
\item the photons considered are indeed photons, 
\item the photons are actually connected with the track considered,
\item the photons errors are correctly estimated. 
\end{itemize}

This sets the problem of background photons. Indeed, in the DIRC,
actual photons have errors which are well approximated by gaussians and
can be computed, more or less accurately, with known information
(geometry, chromatic errors, \ldots). If an observed hit
is not an actual photon (noise) or if it is an actual photon but
emitted by another track, possibly from another quartz bar, its error distribution
(and its standard deviation)  are completely unknown~;
therefore, their
actual contribution to any $\chi^2$ cannot be estimated with any
controlled accuracy.

Stated otherwise, any procedure  aiming at providing a reasonable estimate
of the $\chi^2$ probability (or of any likelihood), should remove background photons
from the photon sample kept in the fit and the $\chi^2$ estimates. On  the other
hand, it is clear that, in presence of noise,  any estimate of the $\chi^2$
is altered, except if one would be able to remove background from the photon set at the 100\%
level, which is generally hopeless.
 
Fortunately, the level of actual contamination in any photon set left
by any cleaning up procedure can be checked statistically. Indeed, as can be seen
from Rels. (\ref{fit6}) --  (\ref{fit9}), the fit solution found for $R$ (denoted here $R_{fit}$) 
and its error $\sigma_R$, crucially depend on the photons used, their errors and their mutual
correlations, all informations which can be accessed with a good accuracy.
If the photon set is too much contaminated, the fit solution may depart
significantly from the expected solution.

The pull $(R_{fit}-R_{true})/\sigma_R$ should follow a centered gaussian law of unit standard deviation.
This property  provides a quantitative criterium to test the quality of background removal.
Indeed, the most likely effect of background
is to shift $ R_{fit}$ from its expected value~; this is reflected in the pull
distribution by a shift of its mean value from zero and an increase of its rms
with respect to 1. The magnitude of observable departures from the standard pull expectations 
clearly signals a more or less acceptable level of contamination, as soon as  a correct pull behavior has been
ascertained with noise--free samples. Therefore, checking the model with noise--free
samples is an unavoidable step in constructing the procedure.

When running with Monte Carlo data, all checks
can easily be performed~; when running with real data, several checks are still possible
using selected events samples like pions from $K^0_S$ decays or protons from $\Lambda$
decays, which can be  selected  using kinematics only, {\it i.e.} independently of the DIRC.
Assuming that background conditions are not especially dependent on the existence
of such particles in the events, the quoted pull can be plotted and the influence
of background contamination inferred.

\subsection{Pattern Recognition Using a Circle Fit Algorithm}

\indent \indent In practical applications we have
a collection of hits that are associated with Cerenkov photons emitted by a given track.
These hits are of three different kinds~:

\begin{itemize}
\item photons emitted by the track under consideration
\item photons emitted by other tracks than that under consideration
\item background hits associated with electronic noise of photomultipliers
or to unidentified tracks (which practically can be merged with noise). 
\end{itemize}
 
Hits of the first kind have generally well behaved error functions (not far from gaussians,
if not exactly gaussians), while the other two kinds of hits have practically unknown error
distributions and should be removed in order to give a statistical meaning to the circle fit.
Stated otherwise, the photon sample must be cleaned up. There is also, specific
 to the DIRC device, another category of fake photons (the ambiguity problem) which will be addressed
in Section \ref{photsel}.

Basically, the cleaning procedure of the photon sample relies on the fact that photons, actually
emitted by the track considered, have error distributions which can be well approximated by
gaussians. In this case, a suitable criterium in removing fake photons is to eliminate all hits
giving a contribution to the $\chi^2$ greater than some maximum value.

Therefore,  the recognition procedure turns out to perform a fit with a starting
sample of photons, in order to have an estimate of the circle parameters ($a$, $b$ and $R$),
then compute the $\chi^2$ distance of each photon to this circle, remove those which are too 
far, and restart the procedure with the surviving photons~; the procedure is repeated
up to convergence. At this point,
one can consider that the circle parameters are reliable and reexamine all possible
photons in order to keep those which are at acceptable $\chi^2$ distance from the expected
circle (typically less than about 9). In this way, one recovers $ambiguous$ photons
which were put aside in order to start the reconstruction procedure.
Using this new enlarged sample, one can then perform a final circle fit and get the
optimum circle parameter values.
Practically, in the case of the DIRC, there are some subtleties which allow to improve
background rejection and photon recovering~; they will be described with more details
in Section \ref{photsel}.

\section{Simulation of a simplified DIRC}
\label{simdirc1}

\indent \indent
A fast Monte Carlo program was written in order to test the reconstruction algorithm. 
We have coded this program in order to output all needed
information allowing to check the algorithm behavior in full details, which is 
not usually an easy task within a Monte Carlo simulating a complete experimental detector.
This program simulates only one quartz bar and a PMT array plane (3 cm diameter PMTs packed in a
rectangular lattice, located at approximately 120 cm from the bar exit). The angle of the PMT plane with respect
to the bar axis could be chosen. The bar itself was 5 meters long, which is approximately twice longer than the
DIRC prototype bar of Ref. \cite{DIRC}, but corresponds to the actual size of the bars
of the BaBar DIRC \cite{BABAR}. The distance between the bar exit face and the PMT plane has been chosen
following BaBar DIRC setup.

\vspace{1.cm}
The simulation included most part of effects that could hamper DIRC performances: 
\begin{itemize}
\item Errors on track direction and momentum, supposed to simulate the response imprecision of a tracker
in front of the DIRC bar. In current running conditions we have chosen $\sigma(p)/p=3.~10^{-3}$
and the generated angular errors have rms $\sigma(\theta)=\sigma(\phi)= 1$ mr.

\item Realistic detector geometry (geometrical uncertainties are important in the DIRC)
\item Chromaticity in the quartz radiator medium (dependence of medium index $n_Q$ -- and hence Cerenkov angle --
on photon wavelength)~; it corresponds to choosing $\delta n_Q/n_Q=6~ 10^{-3}$.

\item The  ratio $g$ of the water to quartz refractive indices was treated as independent of the photon wavelength.
This is close to the real situation, where $\delta g /g$ is typically 10$^{-3}$.

\item Track multiple scattering in the quartz bar.

\item Number of Cerenkov photons emitted along the track proportional to $\sin^2{\theta_C}$ and
to the charged particle path inside the radiator medium. The number of generated photons per
centimeter has been chosen \cite{DIRC,DIRCNEW} as $N_0=135$.

\item The bar end opposite to the  water tank bar exit window
is treated as a mirror. 

\item Full account of photon reflection and refraction properties inside the bar, providing a correct
bar acceptance simulation.

\item The active part of the PMT plane has been truncated to half a plane, in order to prevent getting
fully populated Cerenkov rings. 

\end{itemize}

In this simulation, the geometry and detector structure are incomplete (compared to the final BaBar
DIRC \cite{BABAR}): there is no reflecting wedge ending the quartz bar at the water tank entrance
(suppressing one reflection plane for photons) and geometrical imperfections of the bar
(expected to be very small anyway) are not simulated. The PMT array here is a
plane while in BaBar detector, it is umbrella shaped \cite{BABAR}. It is packed
as a rectangular lattice (while the BaBar packing is hexagonal, introducing  in
this way additional correlations among geometrical errors, however of limited
magnitude).  

In addition, no interaction of the track with the bar is taken into account (except for
multiple scattering), in particular no energy loss is implemented. The PMTs spectral response
function is also not used in the simulation, except for computing the mean water and quartz refractive
indices (and their dispersion). No photon absorption inside the bar is accounted for.
Finally, there is no magnetic field effect.

To summarize, there is no conceptually important difference between this simulated setup and that
of BaBar and, instead, all relevant features are accounted for.

\vspace{1.cm}

The main data sample used for this study is a set of 5000 single electron, muon, pion, kaon, and proton tracks having a
large bar incidence angle and momentum range ($20^{\circ} < \theta_{inc} < 70^{\circ}, 0.5 < P < 5$ GeV). This sample
could be used in several ways~: in normal mode, one event contained only a single track~; another mode of operation
allowed to superimpose several tracks in one event, for physics background studies ~; a last possibility
was to add a random and (spatially) uniform noise to the event tracks, for other noise studies. These background
conditions can thus be made quite severe compared to  normal conditions. The quality checks of the
whole procedure have been performed accurately by varying at will the magnitude of all errors, the background kind and
level, and the phase space window.

The momentum range where the algorithm has been fully tested goes down to 500 MeV, practically the kaon Cerenkov
threshold. At this momentum
the angular error due to the multiple scattering for a pion is about 10 mr (for a bar thickness of 1.7 cm)
this is quite comparable to the angular error due to the PMT window size (about  7 mr rms).
Going to lower momenta is possible~; however, in this case the angular error due to multiple scattering 
may become dominant (for a 200 MeV pion, it is about 30 mr, far above the PMT  geometrical error). 
In this case the procedure still works, however our first order estimates of the errors might become moderately
accurate and higher order terms might become necessary{\footnote{Actually, as below $\simeq 700$ MeV, the main PID
device in BaBar is the drift chamber by means of the dE/dx measurements \cite{DIRCNEW}, it does not seem useful to 
go to such complications. An overlap region of about 200 to 300 MeV, where an optimum  reconstruction can be performed 
with the DIRC and using dE/dx, allows already interesting cross--checks.}}.

\section{Photon Selection and Background Removal}
\label{photsel}

\subsection{Outline of the Procedure}

\indent \indent
The linearized $\chi^2$  fit we have described needs, as input, photons with
unambiguously defined direction with respect to the track momentum~: so, the 
main step of the photon selection procedure is to lower the number of 
ambiguities arising in the reconstructed photon direction due to the various
reflection symmetries of our problem (bar surfaces, mirrors)~; ambiguities should be selected
according to criteria which guarantee the symmetrized photon solutions correspond to 
possible or probable Cerenkov angles for the current track. Ambiguous photons cannot be
used directly by the fit and so they are ignored during the first steps of the 
procedure  (part of them will be recovered by a dedicated algorithm, see below).
This ambiguity removal is actually performed in several steps using different criteria
(detailed in section \ref{ambigid}).

The next step of the selection procedure is the removal of the possible
background photons contaminating the unambiguous photons population~; an iterative cut procedure
involving a median estimator is used for this purpose (section \ref{bkgrem}).

The algorithm requires a minimum number of 
unambiguous photons in order to go on (typically 3, but this number can be lowered to 2).
If there are enough unambiguous photons left, the parameters of interest (Cerenkov angle,
coordinates of the center of the Cerenkov circle in the equatorial plane) are fit to this
set of remaining unambiguous photons. This preliminary fit is made only in order to have a first 
approximation of these parameters and of their errors.

A second and analogous fit is applied to a photon population built by adding to the primary
set of unambiguous photons, photons which were originally flagged as ambiguous,
but which have parameters not too far from the parameters resulting of the primary fit, 
according to the $\chi^2$ distance criterium . These additional unambiguous photons should also
fulfill another condition which guarantees they are "unambiguous", at least to some extent~:
for each new photon, the two ambiguities closest to the central parameters should be themselves
sufficiently separated in the relevant $\chi^2$ distance (see section \ref{recovphot}).

This last fitting operation produces a new (and final) determination of the circle fit
parameters and errors.

\subsection{Ambiguity Identification and Removal}
\label{ambigid}

\subsubsection{Ambiguous -- Unambiguous Cerenkov Photons}
\label{ambigid1}

\indent \indent
In our problem, each hit recorded in the DIRC PMT array is associated with a reconstructed
track. This primary association is made using straightforward criteria~: for each hit, the 8 symmetrized
Cerenkov angles are selected by requiring that they lay into a physically meaningful interval (typically,
bounded by the Cerenkov angles corresponding to the extreme mass hypotheses).

This step is important, because it controls the total number of PMT hits/tracks that will be examined by the full
reconstruction algorithm, and hence influences significatively the time performance of the algorithm.
After this primary simple association step, one PMT hit/track pair usually still gives rise to several
Cerenkov angle solutions. These different solutions are called "ambiguities" hereafter~: 
assigning a unique Cerenkov angle to a photon for one track turns out to discriminate between these ambiguities.

\begin{figure}[ht]  
\begin{center}
\epsfig{file=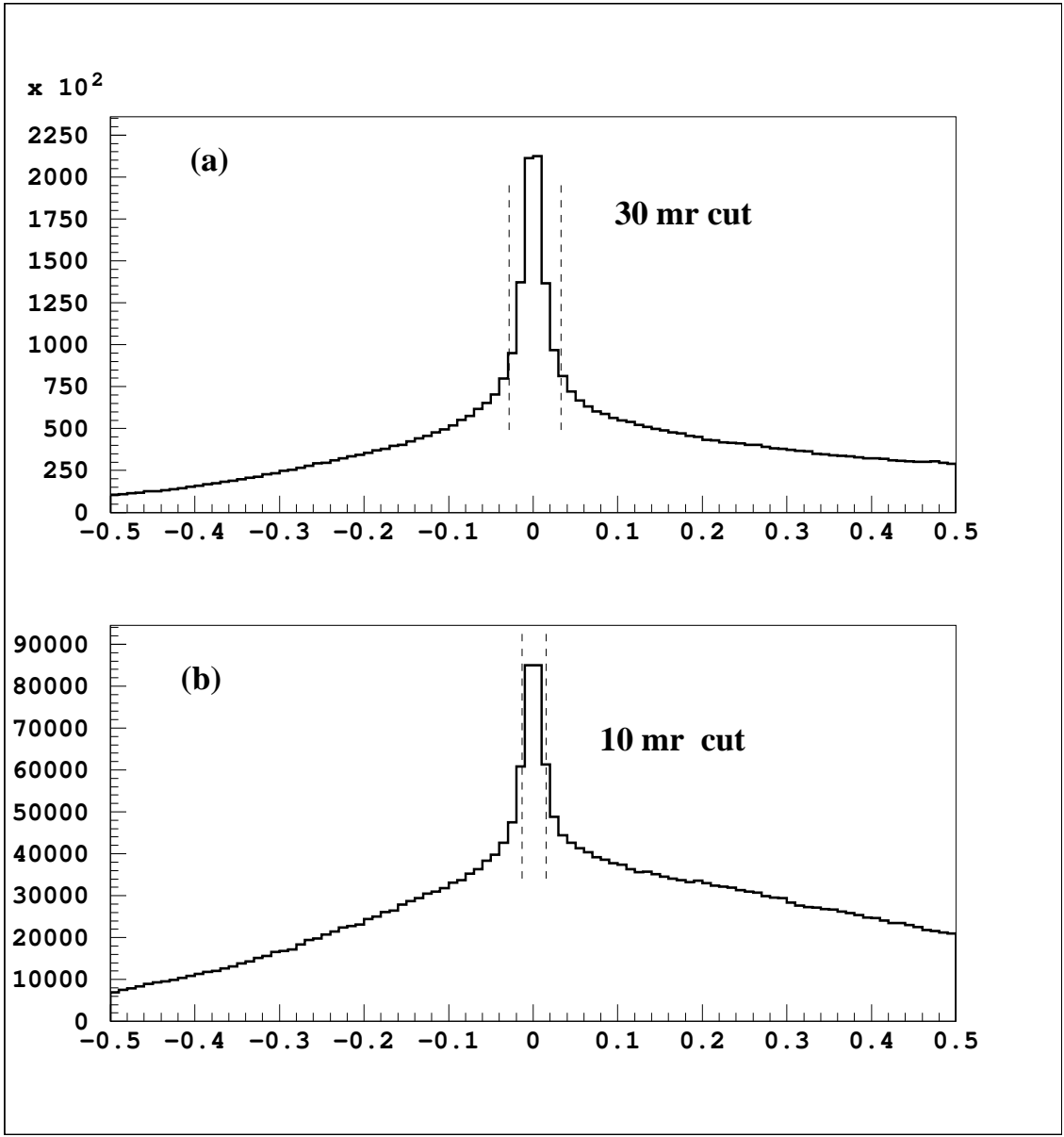,width=15cm}
\caption{Distribution of $\delta\theta$ (radian units) for the mixed track sample with one additional
reconstructed background track; in (a), $\delta\theta$ as measured from signal track; $\delta\theta$ in (b),
as measured from background track.}
\label{dthetacut1}
\end{center}
\end{figure}

Usually, this primary association step leaves an average number of approximately 2 ambiguities per hit,
so that a large part of the photons are still ambiguous.
To further select our sample of "solutions", we restrict even
more the allowed range of Cerenkov angles around the various mass hypotheses, through a cut in the
$|\delta\theta|=|\theta_{solution}-\theta_{hypothesis}|$ variable
computed for each solution Cerenkov angle. This cut is made typically at the level of 30 mr which corresponds
to a rather large angular range around each mass assignment~; Fig.
\ref{dthetacut1}a shows that this window can be naturally defined and checked on a track sample.
Solutions found in the allowed range around at least one of the five mass hypotheses are 
considered. As $\theta_{hypothesis}$ is computed from the track momentum and a mass assumption, it is
always in a valid range~; $\theta_{solution}$ however is computed from measured angles and mean refractive index,
then, even for the correct solution, it may be physically meaningless (i.e greater than the maximum Cerenkov angle).
Such values have nevertheless to be kept in order to prevent biasing distributions.

Since other studies \cite{BABAR} have shown that the main background in the DIRC are photons generated by
other tracks in the same event, another cut is made, on the same grounds, to cope with this "inter track" noise
in multitrack events~: each PMT hit/track solution surviving this 30 mr cut   
is tested versus the other {\it measured} tracks of the event to check if it can be associated with another
track according to the same $|\delta\theta|$ criterium
(at the more restrictive level of 10 mr in this case, see Figure \ref{dthetacut1}b). The second 
$|\delta\theta|$ cut discards only a limited number of photons but an  important number of background solutions.

In order to be classified unambiguous, each photon should give one and only one reflection solution
relative to the track considered, such that $|\delta\theta|<30$ mr and no solution
such that $|\delta\theta|<10$ mr, relative to any other track in the event.

The levels of the two $|\delta\theta|$ cuts (30 and 10 mr) have been adjusted such as to correspond
to a 3 $\sigma$ and a 1 $\sigma$ cut; this means they have been kept at a quite loose level,
leaving about 99\% of the signal visible by the algorithm. Obviously, these criteria can easily
be adjusted without a Monte Carlo, as illustrated by Fig. \ref{dthetacut1}.
 
\subsubsection{Perpendicular Tracks Recovering}
 
\indent \indent
The previous classification of photons as "ambiguous" or "unambiguous"  is not well
adapted to the case when the track direction is almost perpendicular or parallel to some surface of 
the quartz bar~: in this case, trivial ambiguities appear systematically because of the intrinsic
symmetries in the bar + track system, {\it i.e.} one allowed Cerenkov angle for this PMT hit/track pair
will always generate two valid solutions per trivial symmetry plane, since these reflections have in fact
(almost) the same Cerenkov angle.

To cure this problem, a narrow cut on the incidence angle of the track with respect to the bar 
is made which, for this kind of tracks, runs a procedure discarding systematically these trivial
additional solutions, keeping randomly one of them. The magnitude of this cut is mainly related
to the geometry of the quartz bar + PMTs system, and in particular it is chosen 
quite small compared to the angular size of one PMT as seen from the water side bar exit. In our case, the PMTs have
a diameter of 3 cm and are distant from approximately 120 cm from the bar exit~: the corresponding angular
aperture is around 25 mr (corresponding to about 7 mr rms)~; the cut is set at 7 mr. 

This cut is slightly biasing but, as the trivial ambiguity kept is random, this bias is surely
limited. Monte Carlo studies do not show any clear signal of bias related with this cut, moreover
the proportion of almost perpendicular tracks can be expected small at least for phase space reasons.

\vspace{1.cm}  
After having applied the two $|\delta\theta|$ cuts and the "perpendicular tracks" recovering cut,
photons which still allow for several Cerenkov angle solutions are declared "ambiguous", and photons
admitting one and only one solution are "unambiguous". The latter ones are used directly in the
rest of the procedure. The three cuts already described are clearly independent of the particle kind.

\subsection{Cut on the Number of Unambiguous Photons}

\indent \indent
After the preliminary step described above
(step "A"), a first cut on the number of unambiguous photons is made~: we require
the fit to run with at least 2 photons as input. In principle, one should require 3 photons to be sure
to keep a non singular $3\times3$  matrix in the minimisation procedure~; nevertheless, 
the constraint put on the circle center, allows to lower safely this limit to 2 input photons.
The mean ratio of the number of unambiguous photons to real photon hits (having survived the primary
association cuts) at this point of the reconstruction is around 55\%~; this ratio obviously depends
on the cuts one is using.

When the number of unambiguous photons for the current event is lower than 2, in order to avoid loosing 
systematically the track, the algorithm tries  repeating step A using a less restrictive 
$|\delta\theta|$ cut level which is lowered from 30 to 20 mr in several steps. This part of the algorithm
concerns only a few percents of the events, and no significant bias could be traced back to it.
The gain in efficiency of the algorithm after this "smoothed" $|\delta\theta|$ cut is around 5\% of
events with two tracks.

\subsection{Background Removal using a Median Cut }
\label{bkgrem}

\indent \indent
In the previous step, we have already performed the removal of some background effects.
Indeed, the ambiguous photons represent a kind of combinatorial background, which is identified and left
aside provisionally, awaiting improved informations on the Cerenkov circle from the set of unambiguous photons.
Moreover, the 10 mr cut on $|\delta\theta|$ drops out already most of the contamination
produced by photons associated with other reconstructed tracks, which can affect the unambiguous photon
set associated with the correct track,

\vspace{0.7cm}
\indent \indent
After the construction of the unambiguous photons sample, it is possible to improve the removal of background
photons, or, at least, to remove outlier photons.
Further background may originate from several sources~: 

\begin{itemize}

\item Photons created by {\it unreconstructed} tracks (low energy tracks produced in the detector materials,
backsplashes from a nearby calorimeter, \ldots)

\item Improperly accounted for photon solutions (i.e. wrong reflection hypotheses considered as 
correct unambiguous solutions).

\item Background originating from the accelerator, PMT noise, cosmic rays 
\ldots is expected  lower than the other types of background.
 
\end{itemize} 

The first type of background is event dependent and
may have  a complicated structure, whereas the second and third types
are rather flat in Cerenkov angle space (at least on a large range around the nominal Cerenkov angle).
However, we have found no significant difference in the reconstruction behaviour between them and thus, they 
are treated likewise.
The second kind of background has been found relatively easy to deal with, at the expense of loosing 
for the fit strongly ambiguous photons ({\it i.e.} solutions which are consistent with more than one image of the 
Cerenkov cone, even with  the improved accuracy allowed by the fit  estimation of the circle parameters).
These strongly ambiguous photons have not been used in the circle fit~; 
they can nevertheless be counted in order to improve the information on the total number of photons 
associated with the track.

\vspace{0.7cm}
In any case, it is possible to cut out noise outside a window centered at the median Cerenkov
angle (ideally, around the correct Cerenkov angle) of the set of unambiguous photons 
associated with a given track. The same window will be used afterwards as the area where the fit will be performed.

The median is used here instead of the mean as central location estimator as it is less sensitive to noise and
allows a better window setting (the "mean" location estimator is known to be non robust with respect to
outlier tails, see for example \cite{MEDIAN1,MEDIAN2}).

In conditions where the background is low (and/or mostly produced by other reconstructed tracks),
the cut to be performed around the median corresponds to a gaussian cut{\footnote{$\sigma$
being defined by a theoretical computation of the gaussian errors.}} of about 3$\sigma$~; 
in much harder background conditions this cut may have to be lowered to about 2$\sigma$, affecting the probability
distribution shape only in the low reconstruction probability region.

The median cut procedure is iterated on the unambiguous photon sample till convergence is reached
({\it i.e.} until no new photon is removed), which usually happens after 2 or 3 iterations.
The ratio of the number of photons, unambiguous at this point of the procedure,
to photons having survived the first association step cuts is approximately 50\%.

\vspace{0.7cm}

In case of even higher background (like for example when many unreconstructed tracks are present in the
event) the median procedure itself behaves poorly, since background photons may accumulate at several locations
in Cerenkov angle space, simulating signal for almost all mass hypothesis.
In this case, it is meaningless to rely on simple location estimators (like the median): in particular,
the preceding procedure has an identification performance which depends on the track momentum when this one
is close to the Cerenkov threshold of the various mass hypothesis. In such heavy background conditions, it is better
to perform a fit only in restricted Cerenkov angle areas, and not on the full Cerenkov angle range: in this way,
the response becomes more uniform. The choice beetween the two selection methods for the fit window 
depends on the noise level, and can be done automatically. 

\vspace{0.7cm}  

After the median step, the sample of surviving photons  is supposed to be free at least of the influence of
unwanted tails which could spoil the subsequent fit. This preliminary fit  gives a first
estimation of the Cerenkov angle and the circle center in equatorial plane
for our sample of filtered unambiguous photons.

A last "cleaning" cut is applied after this first fit in order to remove outlier photons which have  
individual Cerenkov angles
far from the fitted Cerenkov angle~; these photons could still exist and could degrade the results of
the rest of the procedure. Typically, a fixed 3 $\sigma$ cut is applied for this purpose.   

\begin{figure}[ht]
\begin{center}
\epsfig{file=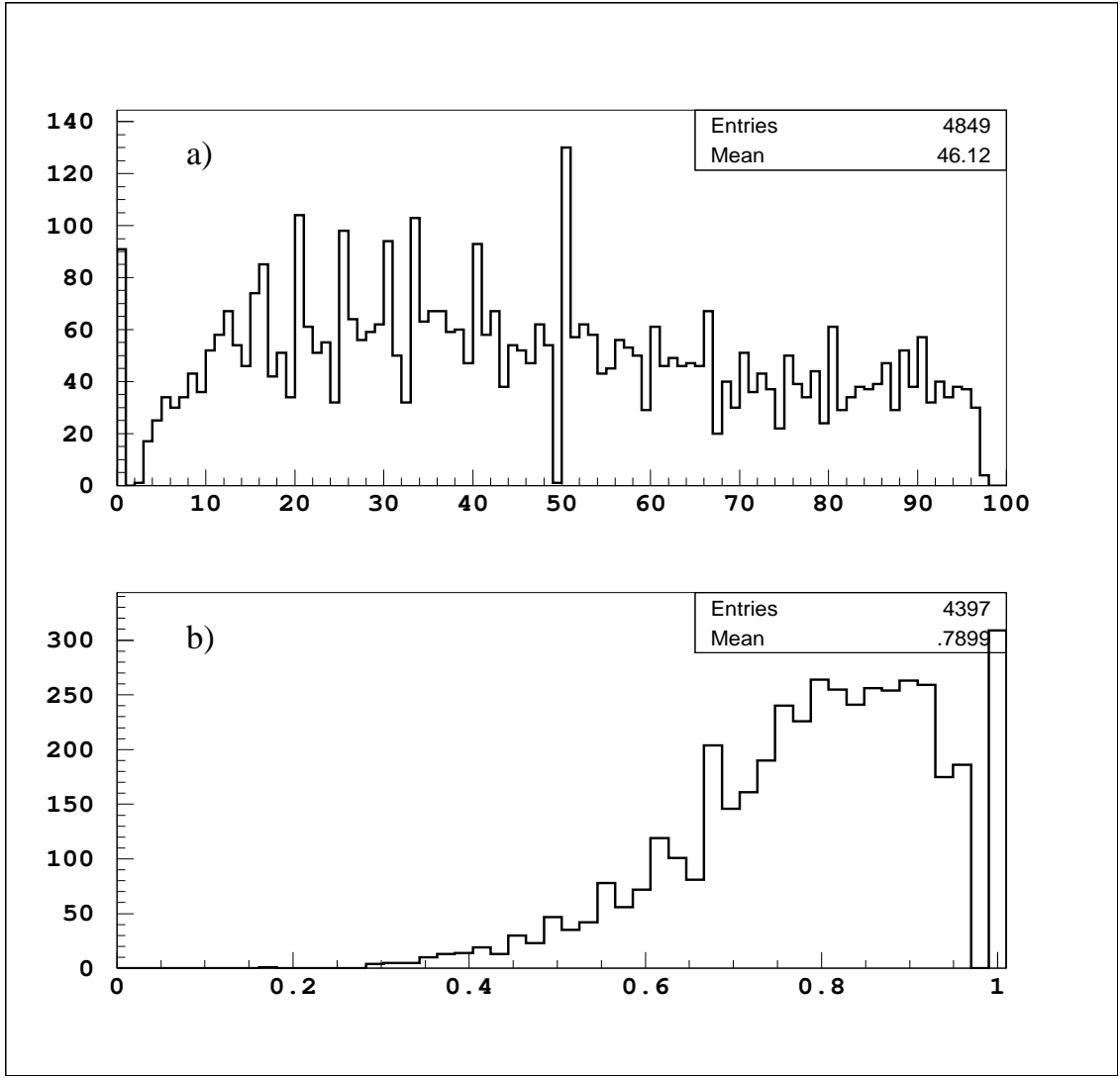,width=15cm}
\caption{In a), histogram of the proportion (in percents) of unambiguous photons relative
to the total number of detected photons (all are signal photons)
for each track before the photon recovering step. Non accepted events appear as a peak at zero. 
In b), histogram of the same quantity after the recovering procedure. 
In both cases no flat background and no additional tracks have been added.}
\label{recov_ratio}
\end{center}
\end{figure}

After this last cut, the unambiguous photons represent about 45\%
of the actual population of photons associated with the track. The corresponding spectrum is
shown in Fig. \ref{recov_ratio}a.

\subsection{Ambiguous Photons Recovering}
\label{recovphot}

\indent \indent
It is possible to get more photons in the final fit after recovering
part of the photons previously flagged as ambiguous, through a procedure using a $\chi^2$ criterium~: 
those among them which have a contribution to the $\chi^2$ (estimated at the reconstructed 
Cerenkov angle and circle center computed by the first fit) which
is not too high are declared unambiguous and included into the fitting sample,
with the additional condition that their two best solutions are not too close to each other
according to $\chi^2$ distance.

Practically, this means that the photon solutions are not farther away from the first
fit Cerenkov angle than typically 3 $\sigma$ and that the closest among the other solutions
is beyond, typically, 3.5 or 4 $\sigma$.

This "double $\chi^2$" criterium allows to input in the second fitting step a sample of photons of the order of 65\%
of the original sample (under large background conditions) or about 80\% (no additional background). The gain
represented by the recovering procedure thus amounts to 50\% to 100\% of the
photon number compared to the sample size before the recovering step.
The proportion of finally used photons wrt the detected photons associated with a track is plotted 
in Fig. \ref{recov_ratio}b for the sample with no additional background.

One may imagine to replace this double $\chi^2$ criterium by a single one~: keep the best solution provided
it is below $\simeq 3 \sigma$, whatever is the $\chi^2$ of the second ambiguity solution. In 
conditions where the background is small, this increases significantly the number of photons -- 
then, the accuracy on the reconstructed angle -- and the reconstruction quality. 
In very hard background conditions, however, one has to study carefully to which extent the
subsequent gain in photons does not degrade the reconstruction quality. It is not done here.

\section{Monte Carlo Results}
\label{MCresults}

\indent \indent
The results presented in this section have been obtained using the set of cuts defined in Section 
\ref{photsel}. The numerical values of these cuts have been tuned depending on the background 
conditions affecting each of the Monte Carlo samples.
We postpone to Section \ref{cutadjust} the discussion on cut handling and tuning. 
Here we examine the results obtained from analysing these samples, in order to draw conclusions
on the fit quality and the various aspects of background removal.

Another important aspect is the algorithm performance for what concerns the separation power between the
various mass hypotheses, more directly related to the use of the reconstruction for physics.
Many ways to estimate this performance can be devised~; here we will discuss only simple criteria.
Indeed, even if background effects are realistic in our Monte Carlo, and sometimes pessimistic,
actual performances can be precisely defined only with a detailed simulation of a given experiment or with real data.
Moreover, the "performance" requested depends strongly on the physics goals one is willing to achieve.

\subsection{One Track Event Display}

\begin{figure}[ht]
\begin{center}
\mbox{\epsfig{figure=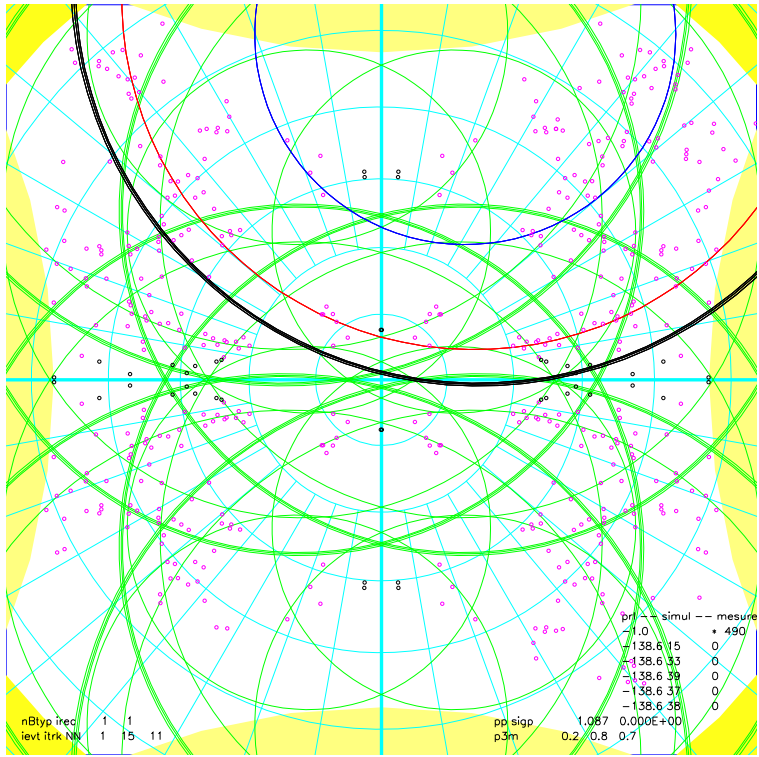,width=10.cm,height=10.cm}}
\par
\vspace*{0.5cm}
\mbox{\epsfig{figure=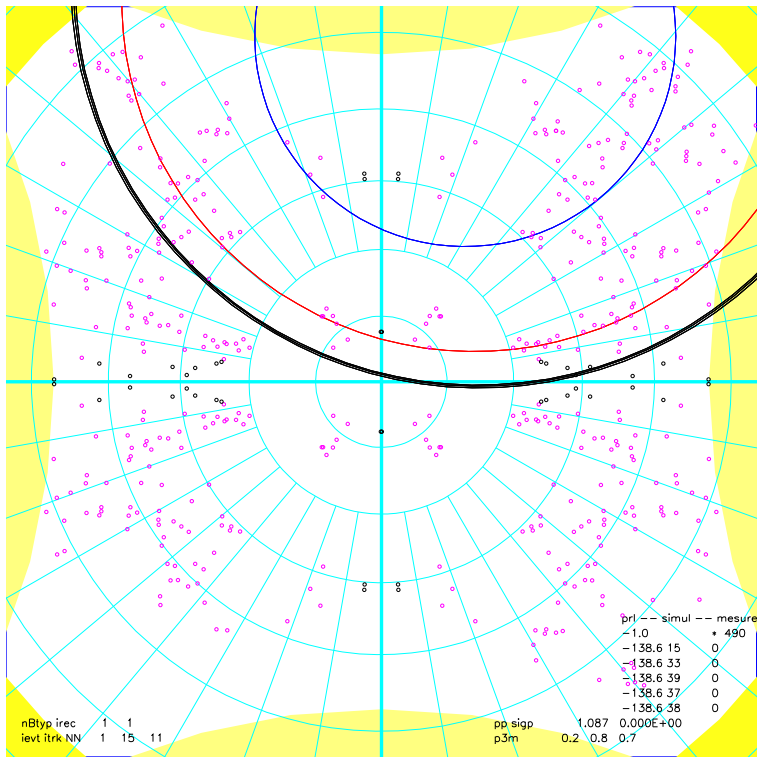,width=10.cm,height=10.cm}}
\caption{Display of DIRC Cerenkov (solution) photons for one kaon track produced together 
with a low flat background (see text). An equatorial projection wrt the {\it bar} axis is used here.
Top~: all photon solutions surviving the primary selection cuts are displayed with the various reflected circles~; 
Bottom~: same view, keeping only the circles associated with the original track direction.}
\label{event_display1}
\end{center}
\end{figure}

\begin{figure}[ht]
\begin{center}
\mbox{\epsfig{figure=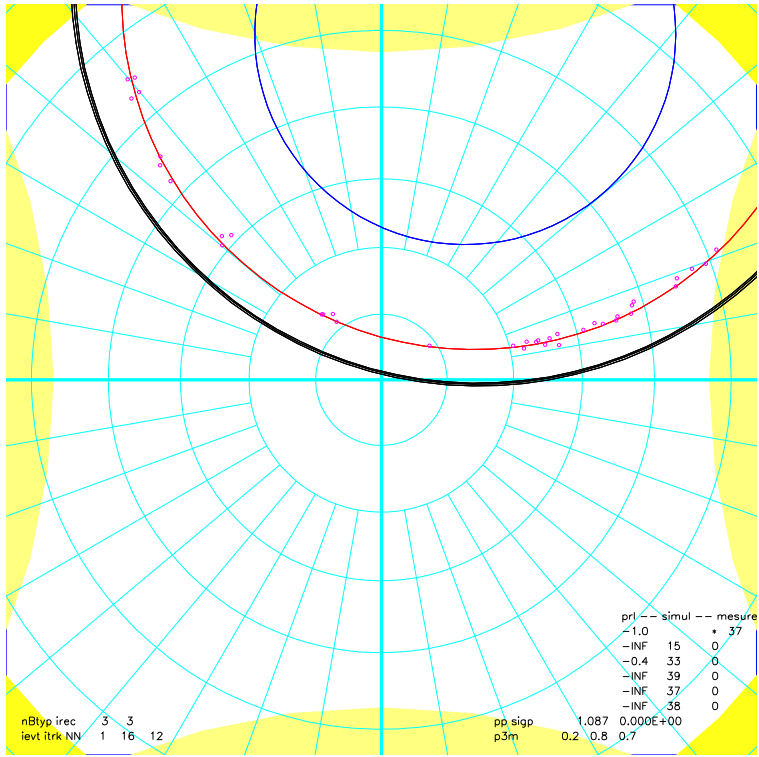,width=10.cm,height=10.cm}}
\par
\vspace*{0.5cm}
\mbox{\epsfig{figure=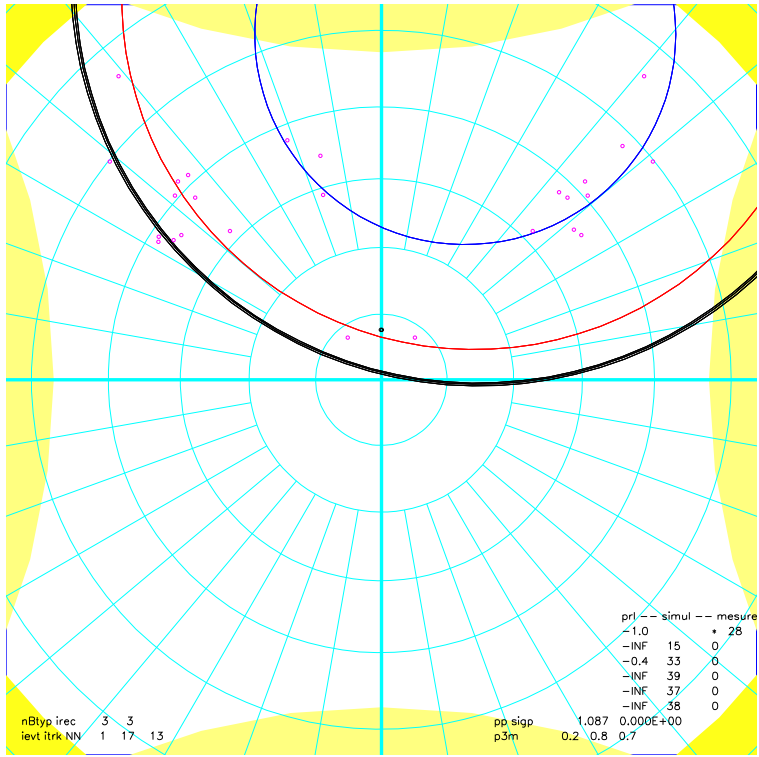,width=10.cm,height=10.cm}}
\caption{Display of DIRC Cerenkov (solution) photons for the event shown in Fig. \ref{event_display1}.
Top~: unambiguous photon solutions found by the algorithm~; all lay along the kaon circle.
Bottom~: all ambiguous photon solutions to be examined by the recovering procedure~:
they are spread along all possible circles.}
\label{event_display2}
\end{center}
\end{figure}

\indent \indent 
In order to substantiate the problem of pattern recognition and background removal when dealing with
the DIRC, it is not useless to display some events. For this purpose, we show in Figs. \ref{event_display1} 
and \ref{event_display2} an event with one track (actually a kaon of 1.087 GeV/c momentum) 
superimposed with a low flat background
(additional number of PMT hits  at the level of about 20\% of the signal PMT hits).
The stereographic projection has been performed with the polar axis along the bar axis and then,
the Cerenkov circles are not centered at the origin. 
Circles corresponding to the original track are drawn thick (the outer one looks even thicker as the
$e$, $\mu$ and $\pi$ assumptions gives circles nearly superimposed), their images with respect to all 
symmetry planes are the shaded circles shown in only  the upper Fig. \ref{event_display1}.
The lower Fig. \ref{event_display1} displays the same region with only the circles
associated with the original track direction under the various mass assumptions. The points
represented are the solutions surviving the primary association cut (see Section above)~;
the shaded area is the region of zero acceptance.

Fig. \ref{event_display1} illustrates the task of the recognition and expresses as clearly as possible 
the usefulness of the charged track information~; most of the background shown 
is produced by ambiguities. This means that this event is relatively clean.
The algorithm described above extracts the unambiguous photon subsample (see upper Fig. \ref{event_display2})~;
in the present case all unambiguous photons are located along the circle corresponding to
the kaon assumption. This is the effect of the median cut referred to above~; indeed,
by looking at Fig. \ref{event_display1}, one clearly sees that unambiguous photons
belonging to the proton circle (the innermost one) have been removed by the procedure.

The subset of ambiguous photons is shown in the lower Fig. \ref{event_display2}~;
the photon solutions which are to be examined by the recovering procedure belong to all mass assumptions.
The extracted ambiguous photons which will be added to the unambiguous photon sample  
belong only to the kaon assumption circle.

\subsection{ One Track Events with no Background}

\indent \indent
We first analyse single track events generated with no external background. 
This allows to study the algorithm performances and get quality checks under optimum
conditions. The track momentum range goes from 0.5 to 5 GeV/c and the sample contains equal numbers 
of $e$, $\mu$, $\pi$, $K$ and $p$.

No (flat) random noise is superimposed to the event--tracks and no additional track embedded
in these events, but this does not mean that they are background free. Indeed, the ``combinatorial''
background represented by wrongly assigned reflection assumptions to hits pollute the sets of
unambiguous photons. Therefore, we can check simultaneously the behaviour of the algorithm and the removal
of wrongly assigned photons. In this case, the median cut for background removal can be put
at a very loose value ($\simeq 4 \sigma$) compared to harder background conditions.

Fig. \ref{poinref}a shows that the reconstruction probability is close to be flat~: the mean value is
0.52 (expected 0.50), while its rms is 0.277 (expected 0.288), both parameters being close to
expectations for a flat distribution. The bias produced by slightly overcutting in order to
remove wrongly assigned (true) photons is thus negligible. The reconstruction quality is illustrated
in Fig. \ref{poinref}b, where the Cerenkov angle pull is represented~;  it is very close to a 
centered gaussian distribution of unit standard deviation. This means that errors are well understood
and correctly accounted for, including multiple scattering handling. This also means that
the identification is close to optimum{\footnote{Of course, as expected, the $\pi$ --$\mu$--$e$ separation
is poor in the track momentum range explored.}} (pions were identified at the level of 95\% and
kaon contamination was about 4\%). Finally,  Fig. \ref{poinref}c
shows the bias ($\theta_C^{true}-\theta_C^{fit}$)~; the mean bias found is about 0.3 mr, and the
spread is about 3 mr. The number of tracks sent to the procedure was 5000, the fraction
which was found with at least 2 unambiguous photons, and thus reconstructed{\footnote{
Among the 12\% events lost, about 2.2\% are protons below the Cerenkov threshold.}}, was 88\%. 
Therefore the single significant effect of the combinatorial background is to reduce by about 10\%
the number of events with at least 2 unambiguous photons. Nevertheless, if reconstruction quality
has to be considered, tracks with at most one unambiguous photon
(even when lowering the $|\delta\theta|$ cut to 20 mr) look somehow suspicious.

\begin{figure}[ht]
\begin{center}
\epsfig{file=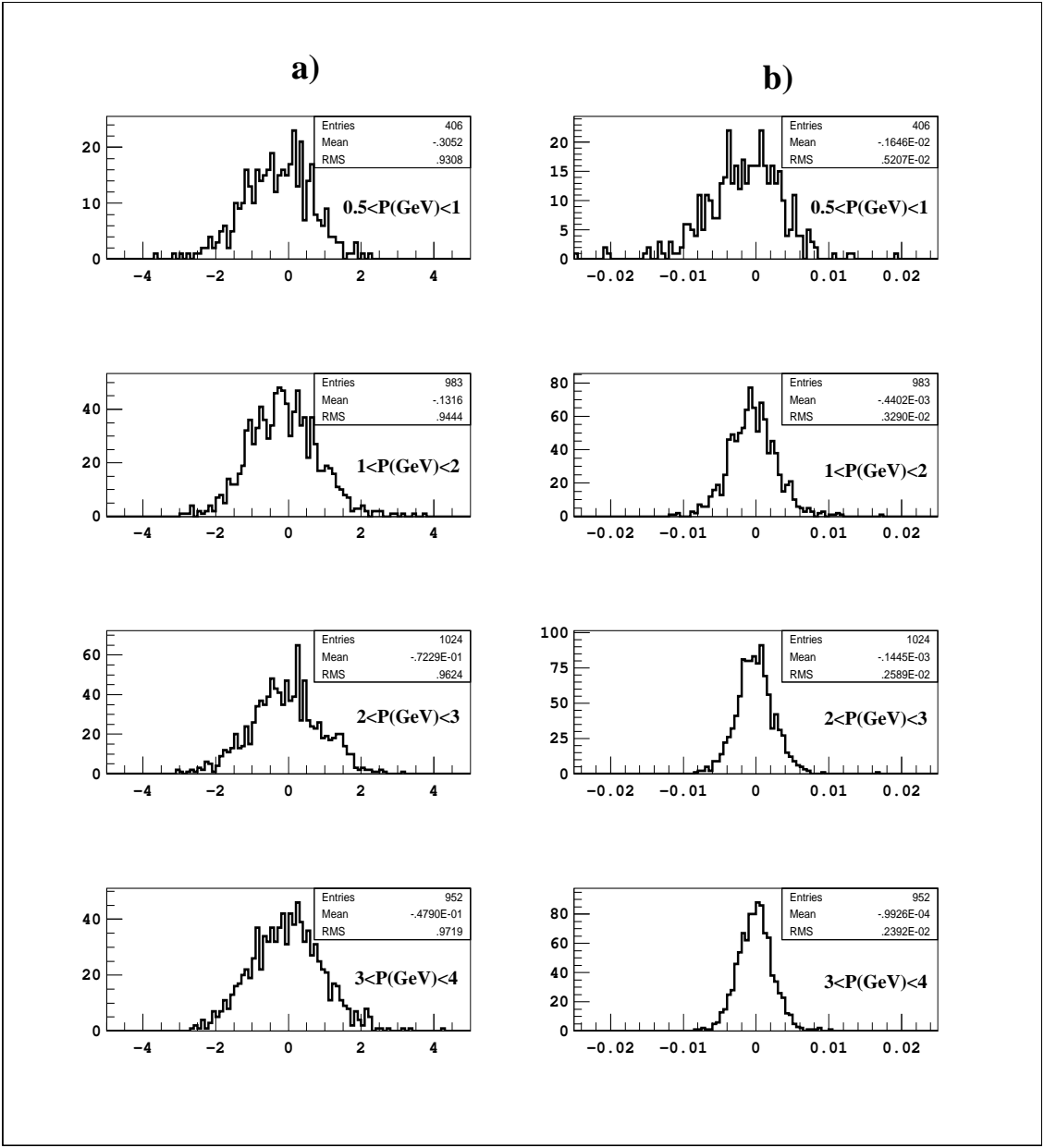,width=15cm}
\caption{Estimated error behavior. In a) the {\it estimated} pull dispersion (rms) 
of reconstructed Cerenkov angle around the expected Cerenkov angle in various track momentum bins~; 
b) rms dispersion of the
reconstructed Cerenkov angle (in radians) also in the same track momentum bins.
Notice in a) the trend to go towards a pull of zero mean and unit rms, when going
to higher and higher momenta~; this typically reflects the decreasing effect
of multiple scattering.}
\label{thcpulbias_vsp}
\end{center}
\end{figure}

\vspace{1.cm}
The reconstructed Cerenkov angle pulls  are plotted 
for a few bins of track momentum in Fig.
\ref{thcpulbias_vsp}a~: the rms of these pulls are quite stable as a function of the track momentum, 
and are close to 1. 
This stability implies that error estimation is correct, even in relatively small track momentum
 bins ~: the non gaussian behaviour of the Cerenkov angle  distribution, even if not smoothed by an averaging over the
track momentum, looks quite limited. There seems to be a systematic bias (at the 30\% level) at low 
track momentum which decreases with increasing momenta~; this should be attributed to
harder multiple scattering effects. Indeed, one should remark in  Fig. \ref{thcpulbias_vsp}a
that, going to higher and higher momenta, allows at the same time to reduce the histogram bias
to smaller and smaller values, while the rms become closer and closer to 1.

Absolute deviations from the simulated Cerenkov angle are also plotted in the same track momentum bins on
Figure \ref{thcpulbias_vsp}b, showing the dependence of the errors on the track momentum, as noticed before.
These plots also show that the general shape of the errors distributions in each momentum bin is correct,
{\it i.e.} not far from a true gaussian.

\begin{figure}[ht]
\begin{center}
\epsfig{file=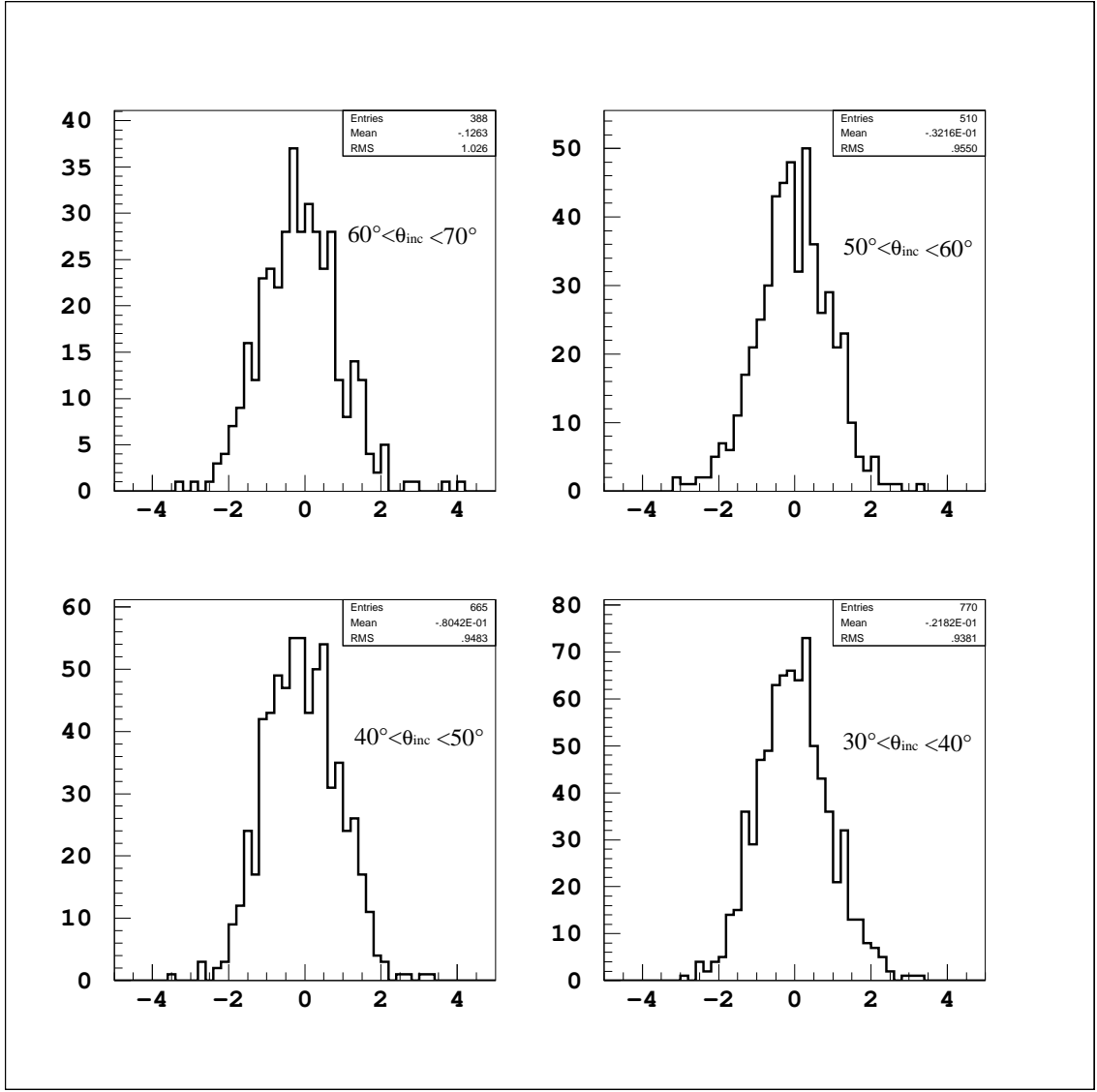,width=15cm}
\caption{Reconstructed Cerenkov angle pull in a few bins of
track incidence angle on the bar. The mean and rms are close to the expected values, 
except in the $60^{\circ} < \theta_{inc} < 70^{\circ}$ bin,
where the mean is slightly too large. }
\label{thetacpul_vsdip}
\end{center}
\end{figure}

There is an analogous situation when one examines the dependence of the errors and biases of the reconstructed
Cerenkov angle with respect to the track incidence angle on the quartz bar~: in this case,
Fig. \ref{thetacpul_vsdip} shows clearly the existence of a systematic bias for tracks hitting the bar
with a high incidence angle{\footnote{Angles are expressed with respect to
the perpendicular to the bar face from which the particle enters in the quartz.}}.
This seems to corresponds to the already noticed effect with low momentum tracks, {\it i.e.}
larger multiple scattering effect, but due to longer paths inside the quartz bar. Indeed, in this case,
tracks with high incidence angles may suffer stronger changes compared with low incidence angles.
The rms dispersion of the pulls looks here also quite stable when the track incidence angle varies,
as demonstrated in the same figure.  

\vspace{1.cm}

\begin{figure}[ht] 
\begin{center}
\epsfig{file=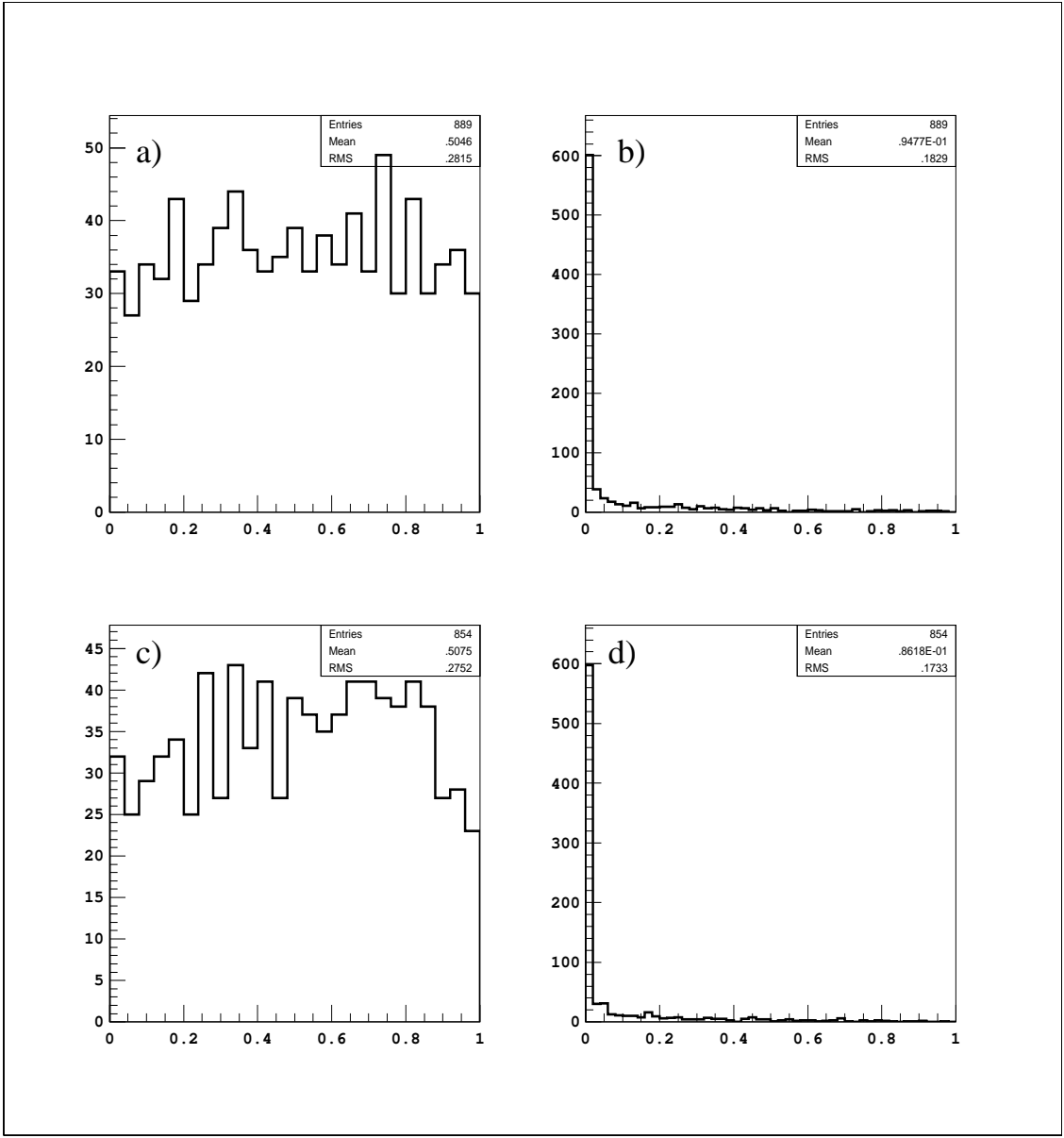,width=15cm}
\caption{$\chi^2$ probabilities of pion and kaon identification. In a), pion hypothesis probability for true pion tracks~;
in b), kaon hypothesis probability for true pion tracks~; in c), kaon hypothesis probability for true kaon tracks~;
in d), pion hypothesis probability for true kaon tracks.}
\label{prob_pik}
\end{center}
\end{figure}

Instead of fitting the circle parameters
on the photon sample extracted from data, one can decide to look for the $\chi^2$ probability
(with $n_{\gamma}$ degrees of freedom) by fixing completely the parameters to their values inferred
from the charged track direction (see Rel. \ref{fit10}). The result of this exercise is shown
in Fig. \ref{prob_pik} for true pions and kaons. The probability distribution happens to
be flat under the right mass assumptions, while it becomes sharply peaked towards 0, under
the wrong mass assumptions. Then, in addition to unaccurately reconstructed tracks, 
the very low probability bins may be enriched with tracks carrying wrongly assigned mass. It is usual
to take this into account by a probability threshold~; in the present case, a cut at about 2\%
looks enough.

\subsection{Effect of Correlations}

\indent \indent
The previous sample, affected only by the minimum (combinatorial) background, allows to study at various levels
the effects of correlations. To be more precise, the data sample is intrinsically affected by correlations~; a  
method to study their effects is to remove correlations terms in the analysis, {\it i.e.} in the algorithm. 
It was remarked in Appendix B5, that photons close in azimuth are strongly correlated. On the other
hand, we know that the circle arc populated by the Cerenkov photons can be small. Then, one can guess
that the net effect of correlations should be indeed strong.

\begin{figure}[ht] 
\begin{center}
\epsfig{file=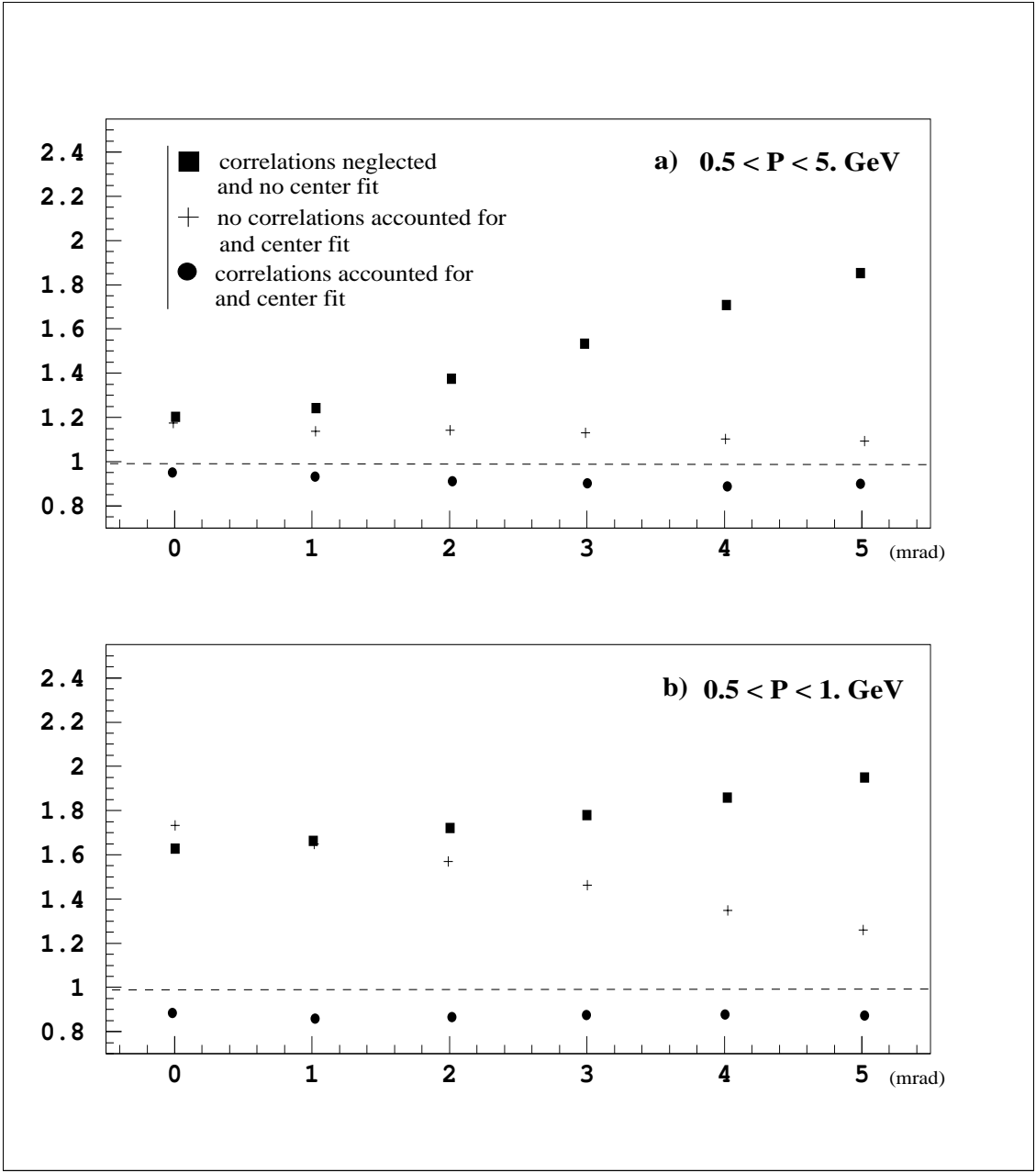,width=15cm,height=17cm}
\caption{An example of the Cerenkov circle fit and the inter photon correlations effect when the errors
on the charged track direction are changed.  
In a), the full track momentum range is used, whereas in b) only
low momentum tracks are considered. Along the horizontal axis, the dispersion of the track polar angles
in milliradian units ($\sigma_{\theta}=\sigma_{\phi}$)~; along the vertical axis, the RMS dispersion of
the pulls of the reconstructed Cerenkov angle. Square markers represent the pulls dispersion value
when the correlations are neglected and the center is not fitted, cross markers show the same quantity when the
center is fitted~; finally, round markers show the same quantity when correlations are taken into account and
the center is fitted. Fitting the center allows a partial account of correlations (see text).}
\label{correl_dth}
\end{center}
\end{figure}

A first way to compute the Cerenkov angle and its error, neglecting all correlations, is to set~:

\begin{equation}
\displaystyle R=\sigma_R^2 \sum_i \frac{d_i}{\sigma_i^2}~~~,~~~\frac{1}{\sigma_R^2}=\sum_i \frac{1}{\sigma_i^2}
\end{equation}  

\noindent and knowing $R_{true}$, one can compute the pull rms. In these expressions, $\sigma_i^2$ is the
squared error of $d_i$. This corresponds to fixing the charged track direction to its 
central value given by the tracking device. In such a computation one actually neglects  correlations
in estimating $\sigma_R$, since otherwise we would have used $ 1/\sigma_R^2=\sum_{i,j} V^{-1}_{ij}$
(see Rels. (\ref{fit4}) and (\ref{appb6}))
and not only the  trace of the inverse of  V, which becomes diagonal when correlations are neglected.
The corresponding pull is always well centered (reflecting the fact that the charged track direction measurement
is unbiased) while the pull rms varies dramatically with the spreads $\sigma_{\theta}$ and $\sigma_{\phi}$ as
illustrated by the upper curve in Figs. \ref{correl_dth}.

\vspace{0.7cm}
Another way to proceed is to neglect correlations in the expression for $\chi^2$ (Rel. (\ref{fit7}))
and solve at minimum. In this case, the result provides a fit value $R_{fit}$ practically not biased.
Its error $\sigma_R$ is computed from the solution at minimum $\chi^2$ and from the matrix $T^{-1}$ 
(see Rel. (\ref{fit9}))
which gives the errors and correlations for $a$, $b$, $c$, from which one can deduce $\sigma_R$ using~: 

\begin{equation} 
\displaystyle \sigma_R^2=\frac{1}{4R^2_{fit}} <[\delta c +2 a \delta a + 2 b \delta b]^2>
\end{equation}

\noindent and $R^2_{fit}=c+a^2+b^2$, {\it i.e.} one takes into account the fact that the actual center 
is not the origin, but is closer to its fit value. 
The corresponding rms pull is the middle curve in Figs. \ref{correl_dth}. 
It shows already a much better behaviour than  the previous method result which
neglected {\it all} correlations {\it stricto sensu}. This full account of the
fit center location mainly explains
the interesting behaviour of the pull rms at large values for $\sigma_{\theta}=\sigma_{\phi}$. 

Finally, the lowest curve in Figs. \ref{correl_dth} shows the solution taking into account all
correlations as explained in the sections above and in the appendices. In this case, the pull rms
remains always close to 1 and departures are never worse than about 10\% only{\footnote{Actually,
going to much larger values for $\sigma_{\theta}=\sigma_{\phi}$, the lowest curve remains
at about 0.9 while the middle curve crosses 1 and goes on decreasing.}}. 

Comparing the two Figs. \ref{correl_dth}, it is clear that the mean effect is much more
dramatic if correlations due to multiple scattering are harder (at low mean momentum). 
It is also clear from these figures that there is a small systematic effect (10 \% of the pull rms)
that our model does not account for. The reasons for this systematic effect are several~; first, 
errors and correlations due to
multiple scattering  are treated statistically{\footnote{We take into account only mean effects 
due to multiple scattering. Actually, the emission time sequence of the detected photons is not 
known and, moreover, the real multiple scattering effect undergone by the track 
before each photon is emitted is also unknown. Therefore, it seems hard to imagine
how to go beyond averaging.}}~; second, at low momentum and/or at large angular errors
on the incoming charged track direction, non--linear effects become visible. Figs. 
\ref{correl_dth} show however that these effects remain of limited influence. 

\begin{figure}[ht] 
\begin{center}
\epsfig{file=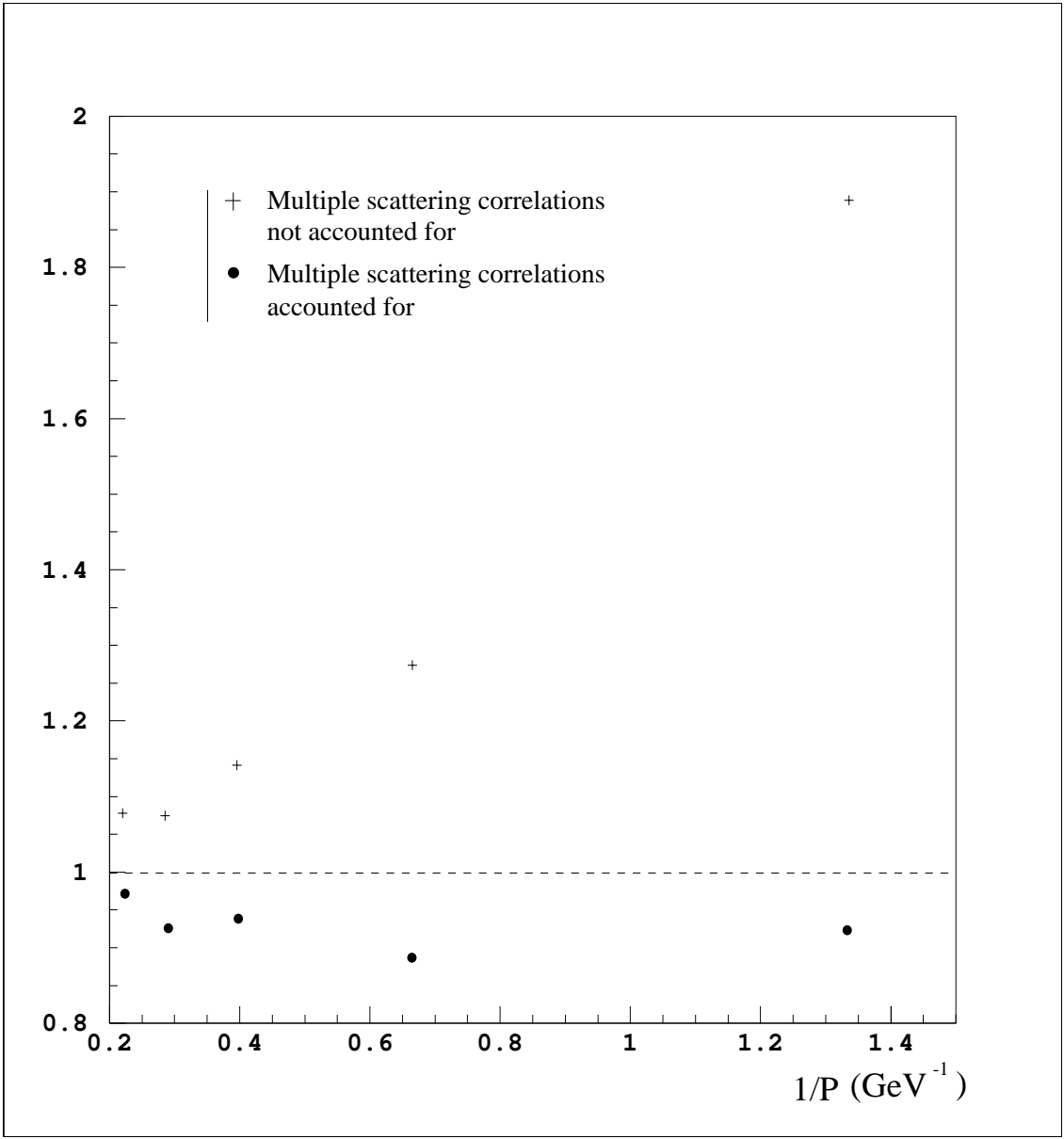,width=15cm}
\caption{Effect of inter photon correlations due to multiple scattering (the dispersion of track direction
measurement is set to zero). The reconstructed Cerenkov angle pull rms is plotted as a function of
the inverse track momentum~: cross markers represent the pulls dispersion when no correlations are accounted for~;
the round markers show the stability of the pulls when the correlations
are correctly taken into account.}
\label{correl_ms}
\end{center}
\end{figure}

One may wonder  why the second method which neglects correlations gives
a pull rms which improves with large angular errors on the incoming charged track 
direction, a behaviour quite different from the simple uncorrelated mean (first
of the methods above). This is actually due to the peculiar origin of this kind
of correlations compared to multiple scattering. In Fig. \ref{correl_ms}, 
the pull rms is plotted as a function  of $1/p$ ({\it i.e.} for 
increasing angular errors dues to multiple scattering)~; crosses correspond to the second 
method, dots to the third (standard) method. It is clear herefrom that neglecting  
correlations  gives a worse and worse result when multiple scattering 
errors increase.

The different behaviour of correlations due to multiple scattering only
(case A) and to errors on the incoming track direction only (case B)
can be understood to some extent. Let us assume that we are in
case B~; if we have had an exact knowledge of the actual charged track direction,
correlation effects would vanish when using it to subtract the center
position from the photon coordinates.
Using the fit center, in place of of the measured center (here the origin),
improves somehow the approximation made of the actual charged track direction~;
this is well reflected in Figs. \ref{correl_dth} by the difference
of slope between the curves with square and cross markers. 
Instead, in case A, there is clearly no longer one and only one actual charged track direction
associated with the photon set, and consequently, neglecting correlations should degrade the
result more and more as multiple scattering effects increase~; this is reflected by the behaviour
shown in Fig. \ref{correl_ms} by the (cross) curve. Real life stays in between cases A and B,
since actually we are in a mixed case.

\begin{figure}[ht]  
\begin{center}
\epsfig{file=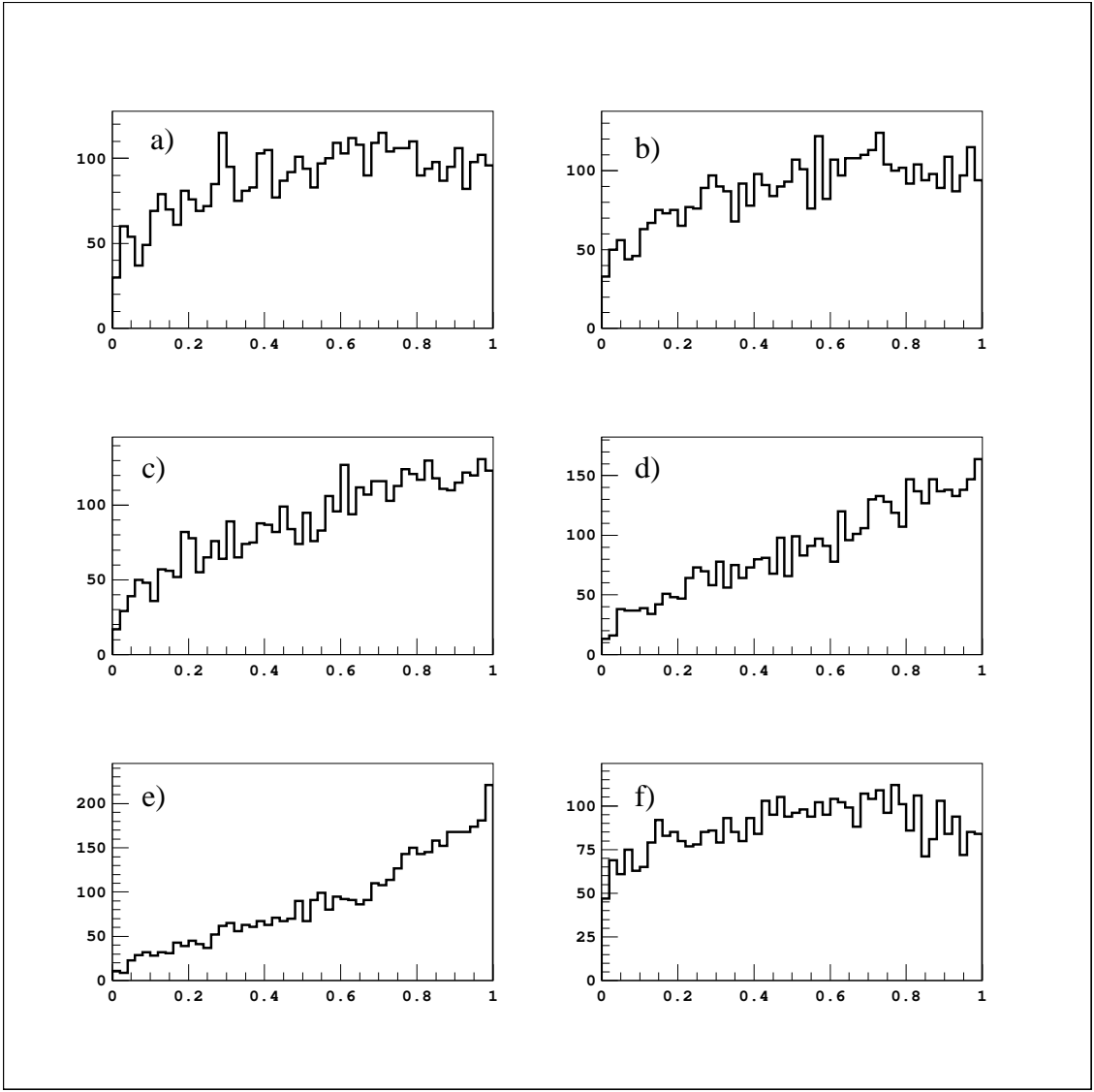,width=15cm}
\caption{Influence of the errors ($\delta\theta, \delta\phi$) on the incoming track direction on the
$\chi^2$ probability distribution when the inter photon correlations are not taken into account.
In a) $\delta\theta=\delta\phi=0$ mr; in b), $\delta\theta=\delta\phi=1$ mr;
in c), $\delta\theta=\delta\phi=3$ mr; in d), $\delta\theta=\delta\phi=4$ mr;
in e), $\delta\theta=\delta\phi=5$ mr. In f),  the same quantity for $\delta\theta=\delta\phi=5$ mr, when the
correlations are accounted for. The adjustable cuts used are the same in each case.}
\label{chi2prob_nocor}
\end{center}
\end{figure}

In any case, as the ``physical region'' for $\sigma_{\theta}$ and 
$\sigma_{\phi}$ is  expected to be
around 1 to 2 mr, it is clearly preferable to account for all correlations. 
Depending on the mean track momentum, the pull rms departure from 1 may be as large as $\simeq$ 50\%
instead of the $\simeq$ 10\% mainly due to non--linear effects.
This problem is reflected by  the $\chi^2$ probability distributions. Focusing on the mixed
track sample with $0.5$ GeV $\leq p \leq 5$ GeV and neglecting only the correlations in the $\chi^2$
(second of the methods just discussed), we get the distributions shown in Fig. \ref{chi2prob_nocor}.
Here are  displayed the probability distributions corresponding
to various values for $\sigma_{\theta}=\sigma_{\phi}$ from 0 --Fig. \ref{chi2prob_nocor}a --
to 5 mr --Fig. \ref{chi2prob_nocor}e -- compared with the case where all  correlations
are normally accounted for (Fig. \ref{chi2prob_nocor}f), corresponding to 5 mr. We see that, already for 
small angular errors on the charged track direction, the probability distribution is badly  distorted
compared to flatness, while accounting normally for correlations gives an acceptably flat 
probability distribution{\footnote { The difference in flatness between Fig. (\ref{poinref}a)
and Fig (\ref{chi2prob_nocor}f) is simply due to different median cuts~: In the former case
it was set at the very loose level of 4$\sigma$, while in the later it is 3$\sigma$.
One can furthermore compare these two figures and remark that a tighter median cut
mainly affects the low probability region by depopulating it somehow.}}, even at 5 mr.
One should also notice from Fig \ref{chi2prob_nocor}f that the systematic effects already noticed,
which survives our treatment of errors and correlations at $\sigma_{\theta}=\sigma_{\phi}=5$ mr 
is not hard enough that it would spoil the shape of the distribution probability.

One thus has to notice the dramatic effect of a bad estimate of the center coordinates 
on the rms pull and then on probability distributions. This effect on the pull
rms is actually due to a wrong estimate of the errors and correlations.
Indeed, as noticed in Section \ref{stereo} when discussing Rel. (\ref{basicf}), 
this effect affects much less the central estimate of the radius and hence, 
of the Cerenkov angle than the error estimation itself.

\subsection{Track--Events with various Background Conditions}

\indent \indent
Here we examine samples contaminated by various kinds of background~: flatly distributed noise on the
PMT detection plane, or merging of the track with one or sometimes two (unidentified or identified)
tracks which produce additional photons.
The chosen momentum range is still 0.5 to 5 GeV/c and the population contains the five possible particle
kinds in equal numbers.
This kind of background conditions can be considered the hardest as these additional tracks enter the same DIRC bar,
sometimes with directions very close to that of the track under identification.

\begin{figure}[ht]
\begin{center}
\epsfig{file=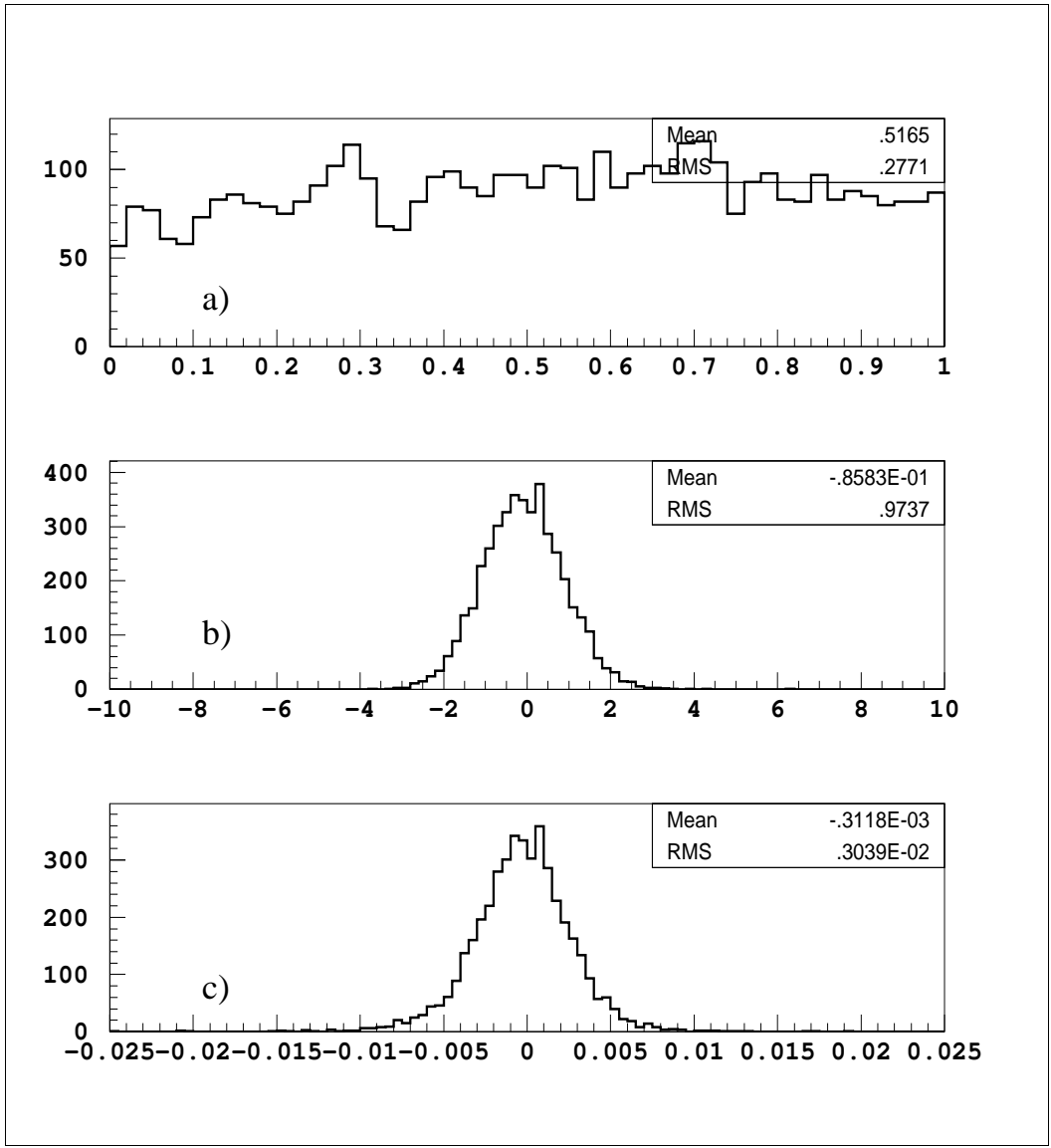,width=14cm}
\caption{Distributions obtained using  Monte Carlo data generated without additional noise. 
In a) the $\chi^2$ probability distribution~; in b), the reconstructed Cerenkov angle pulls~; 
in c), the absolute biases (radians). Notice in b) and c) the smallness of the tails.}
\label{poinref}
\end{center}
\end{figure}

\begin{figure}[ht]  
\begin{center}
\epsfig{file=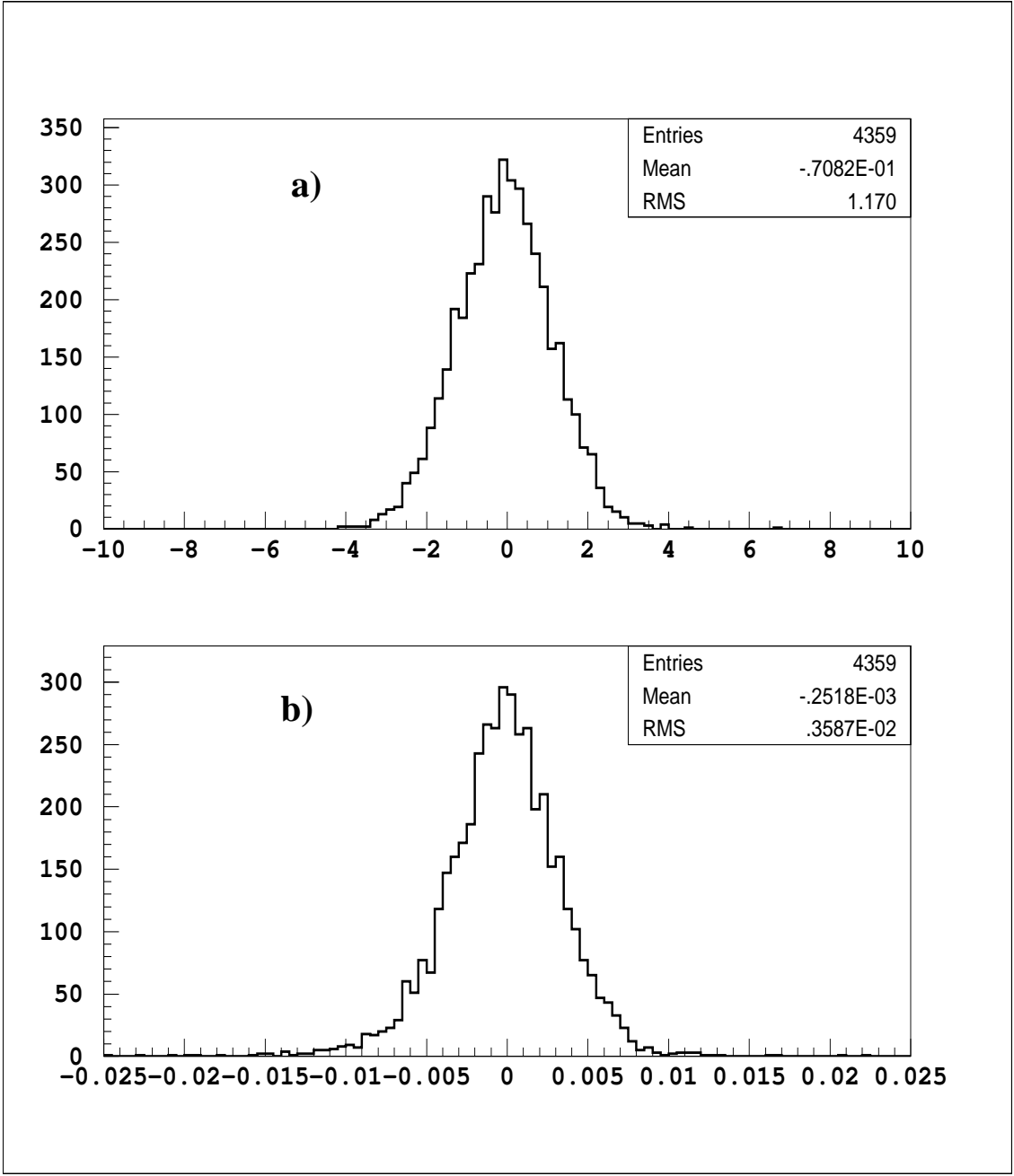,width=15cm}
\caption{Reconstructed Cerenkov angle pulls (in a) and biases (in b) for the mixed particle sample, one
identified additional track has been superimposed as background.}
\label{thetacres}
\end{center}
\end{figure}

The pull of the  Cerenkov angle for the sample of mixed particles  with one  identified additional
track is plotted on Fig. \ref{thetacres}a. This plot gives two important informations~: on the one hand,
the pull bias remains as small as when there is no background (see Fig. \ref{poinref}b)~;
on the other hand, the pull rms is close to one (typically 1.2), as one expects if the
error model is correct. A simple gaussian fit to the distribution in Fig. \ref{poinref}b gives 
 $\sigma= 0.95$ with a very good fit probability. All this shows that
the tails to the distributions remain limited and do not affect the fit quality.

Figure \ref{thetacres}b, a plot of the difference $\theta_C^{true}-\theta_C^{fit}$, allows to estimate the 
absolute dispersion of this quantity. Compared with the case with no background (see Fig. \ref{poinref}c),
one sees clearly that the bias is unchanged and limited (0.3 mr), whereas the dispersion is slightly
increased (3.6 mr instead of 3 mr).

\begin{figure}[ht] 
\begin{center}
\epsfig{file=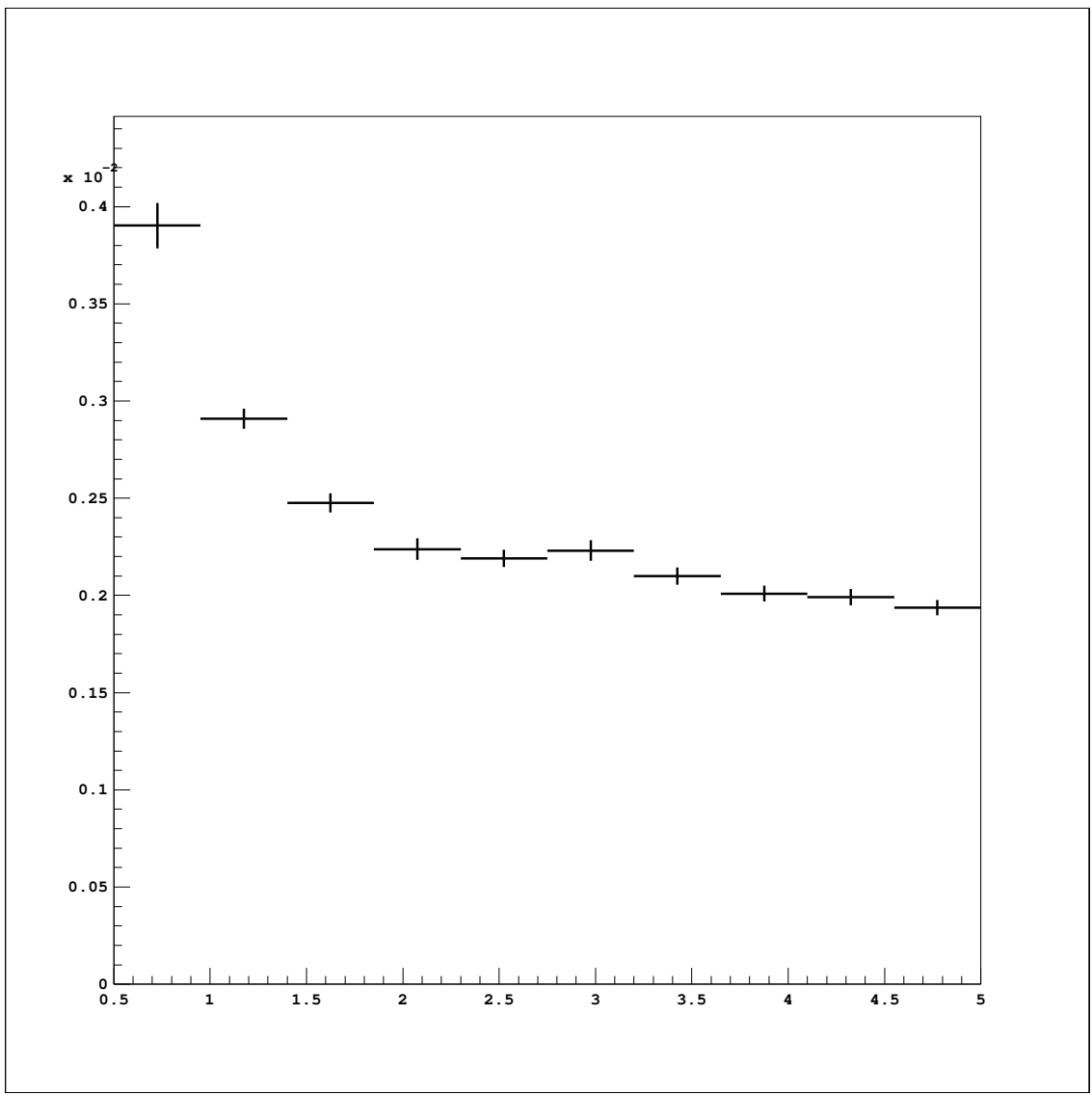,width=15cm}
\caption{Theoretical Cerenkov angle mean errors (in radian units) vs track momentum (GeV)~;
notice the rise at low momentum, mainly due to the increase of multiple scattering effects.}
\label{thetacerrtheory}
\end{center}
\end{figure}

This dispersion depends, among other things, of the track momentum and dip angle. With this respect,
Fig. \ref{thetacerrtheory} shows the mean theoretical error on the reconstructed Cerenkov angle as a 
function of this
momentum~; one can check that this error increases noticeably at low momentum (mainly because of
multiple scattering effects). Errors become smaller and reach a minimum plateau at high track momentum
because of the Cerenkov angle saturation, the greater number of Cerenkov photons (proportional
to $\sin^2{\theta_C}$), and smaller multiple scattering effects.

\begin{figure}[ht] 
\begin{center}
\epsfig{file=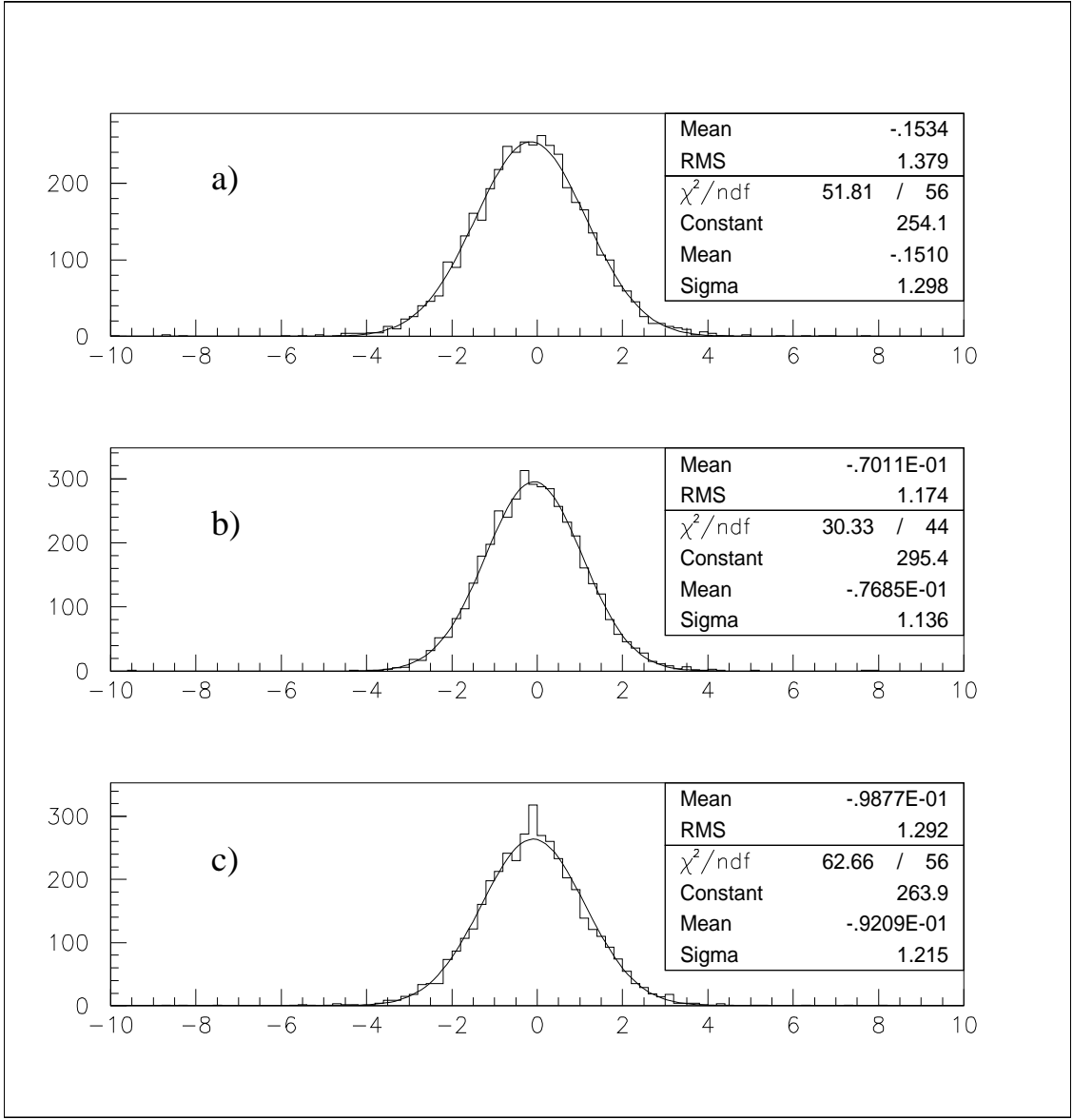,width=15cm}
\caption{Effect of different types of background on the pull distributions of the reconstructed Cerenkov angle.
In a), a flat random noise is superimposed to the signal track~; in b), a second track is mixed with the signal
track~; in c) two additional tracks are superimposed to the signal track. The effect of noise seems more important
in case a).}
\label{thetacpul_bkg}
\end{center}
\end{figure}

\vspace{1.cm}
The fit results obtained when no background was added to the track were good
(see Figs. \ref{poinref}, \ref{thcpulbias_vsp} and  \ref{thetacpul_vsdip}). 
They deteriorate only slightly if noise is added~: Fig. \ref{thetacpul_bkg} indicates that the
algorithm still behaves well under background conditions connected with the presence of other (measured) tracks
in the event. In Figs. \ref{thetacpul_bkg}b and \ref{thetacpul_bkg}c, the reconstructed Cerenkov angle pulls are displayed
when one or two other tracks are superimposed to the signal track (their existence is known to the algorithm,
which uses this knowledge to reject part of this background as ambiguities, as explained in Section
\ref{ambigid1})~; the dispersion and bias of the reconstructed angle are found close to the normal ones.

For the case of Figure \ref{thetacpul_bkg}a, where flat background has been superimposed to the signal
track, the algorithm seems to be more affected, since a sizeable bias of 15\% appears, the rms dispersion
increases but stays close to 1. This bias still is limited, even if noise conditions are quite severe compared to
what is expected with a real detector.

\begin{figure}[ht] 
\begin{center}
\epsfig{file=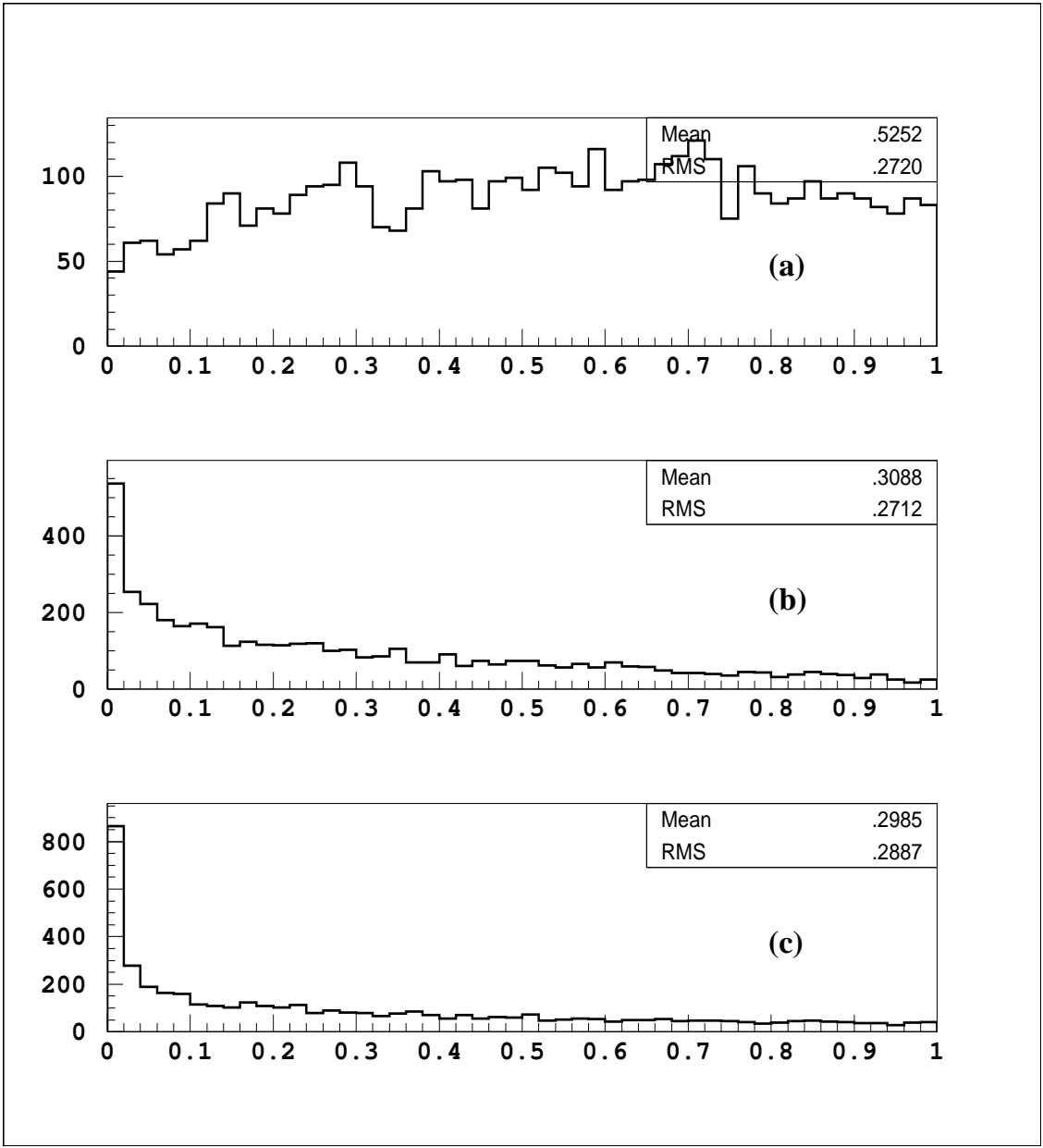,width=15cm}
\caption{$\chi^2$ probability distribution for the sample of mixed particles when using the same set of cuts~;
in (a), only signal photons are considered~; in (b), a flat random photon noise has been superimposed to
signal photons; in (c), photons from one additional track (considered as background) are superimposed to
the original track photons.}
\label{chi2prob1}
\end{center}
\end{figure}

Fig. \ref{chi2prob1} shows the $\chi^2$ probability distribution for $n_{\gamma}-1$ 
degrees of freedom for an equal mixture of single electron, muon, pion, kaon and proton tracks.
It illustrates the effects of the same cuts on different background conditions (the cuts used have been
calibrated on events with only a flat background).
The plots  are shown for several cases of actual background conditions: a) no
background, b) random flat background on the detection plane (at the level of 100\% of the signal photons),
c) 1 other track considered as background ({\it i.e.} no secondary $|\delta\theta|$ cut, no knowledge
of the existence of the second track, see section \ref{ambigid1}).

Fig.\ref{chi2prob1}a shows that the actual effect of the cuts on background free events is 
mainly concentrated in the low probability bins which appear slightly depopulated, while the
rest of the distribution looks acceptably flat. Figs. \ref{chi2prob1}b and \ref{chi2prob1}c
show that the main effect of background photons is to provide a large peak at small probability,
and thus the need for a threshold probability. These last two figures tends also to show that 
a large flat background is harder to account for than the background associated with photons
generated by an unmeasured track~: in the former case, the low probability peak extends up to
$\simeq$ 15\%, while in the later case the effect of this peak is negligible above $\simeq$ 5\%.
This kind of plots, which can be produced with any data set, are tools  allowing to 
tune the minimum probability above which the fit of the Cerenkov ring is considered.

\vspace{1.cm}

\begin{figure}[ht] 
\begin{center}
\epsfig{file=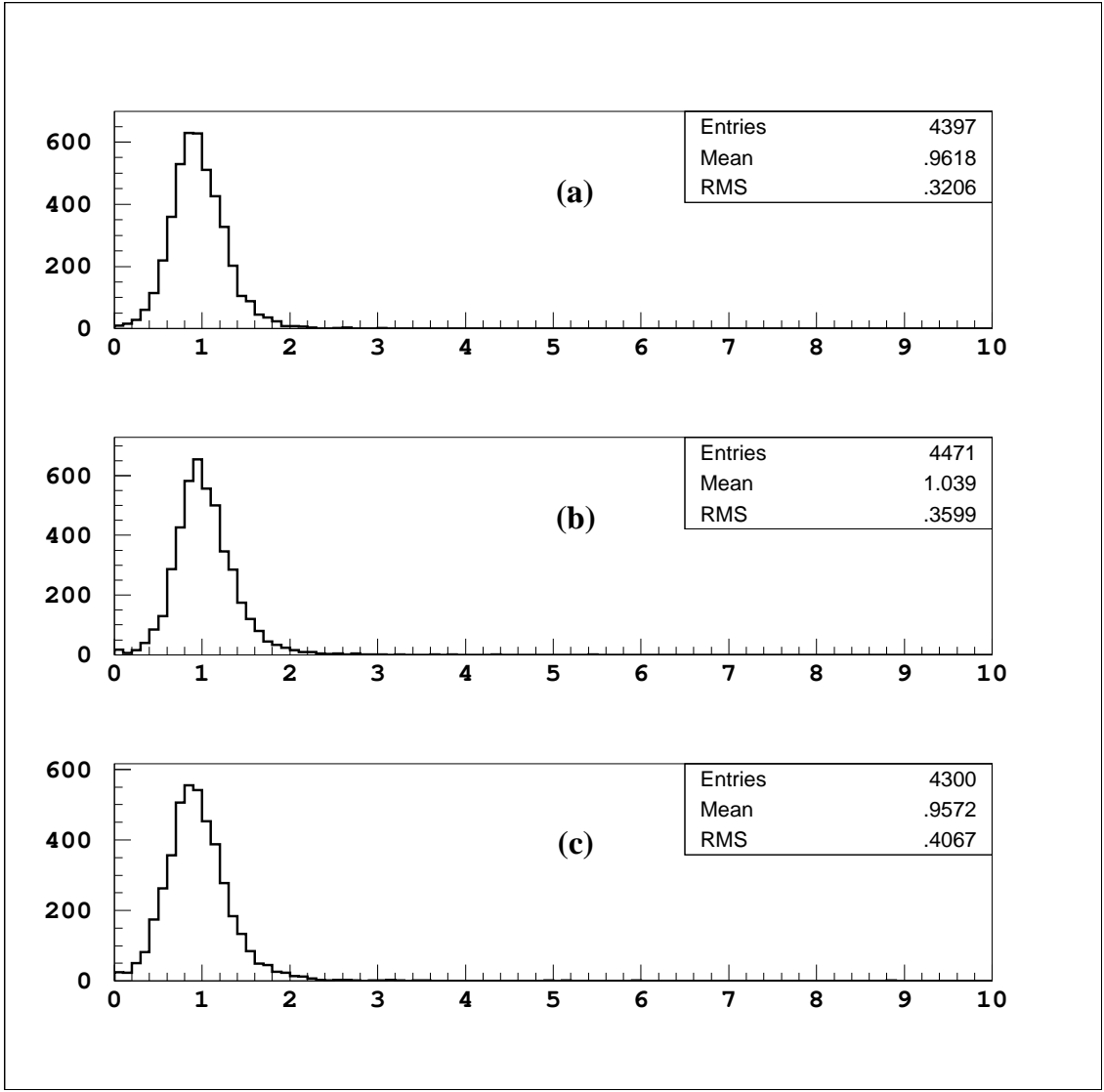,width=15cm}
\caption{Adaptation of cuts to background conditions, reflected in the $\chi^2$ per $n_{dof}$ distribution
for the sample of mixed particles when cuts have been correctly set~; in (a), only signal photons are
considered~; in (b), a flat random photon noise has been superimposed to signal photons~; in (c),
photons from one additional track (which existence is supposed to be known) are superimposed to
the original track photons. Accordingly, the probability distributions for each of these cases
are very close to be flat.}
\label{chi2pndf}
\end{center}
\end{figure}

As a global check,
Fig. \ref{chi2pndf} shows a plot of the $\chi^2$ per number of degrees of freedom obtained after running the
procedure on the same sample of single tracks for several noise environments~: a) without any background, b) with a
random flat background, c) with several tracks per event considered as background for the original track. One
can see that after the cut adjustment, the mean $\chi^2$ per ndof is close to 1. The rms of the $\chi^2$ per ndof
distribution behaves also as expected, for example in the case (a) the mean is $0.96$ and the rms of the
distribution ($0.32$ for a mean number of degrees of freedom of $21$) is effectively close to 
$\sqrt{\frac{2}{ndof}}$ ($0.31$). The number of reconstructed events varies between the three plots because
of different cut levels adapted to different noise conditions.

\subsection{Particle Identification}

\indent \indent The final goal of pattern recognition and circle reconstruction is particle identification.
In order to illustrate how the procedure behaves, we present results obtained for one track, assuming
another track has crossed the same (and single) bar. The angle between the two tracks is random
between 0$^{\circ}$ and 50$^{\circ}$.

\begin{figure}[ht] 
\begin{center}
\epsfig{file=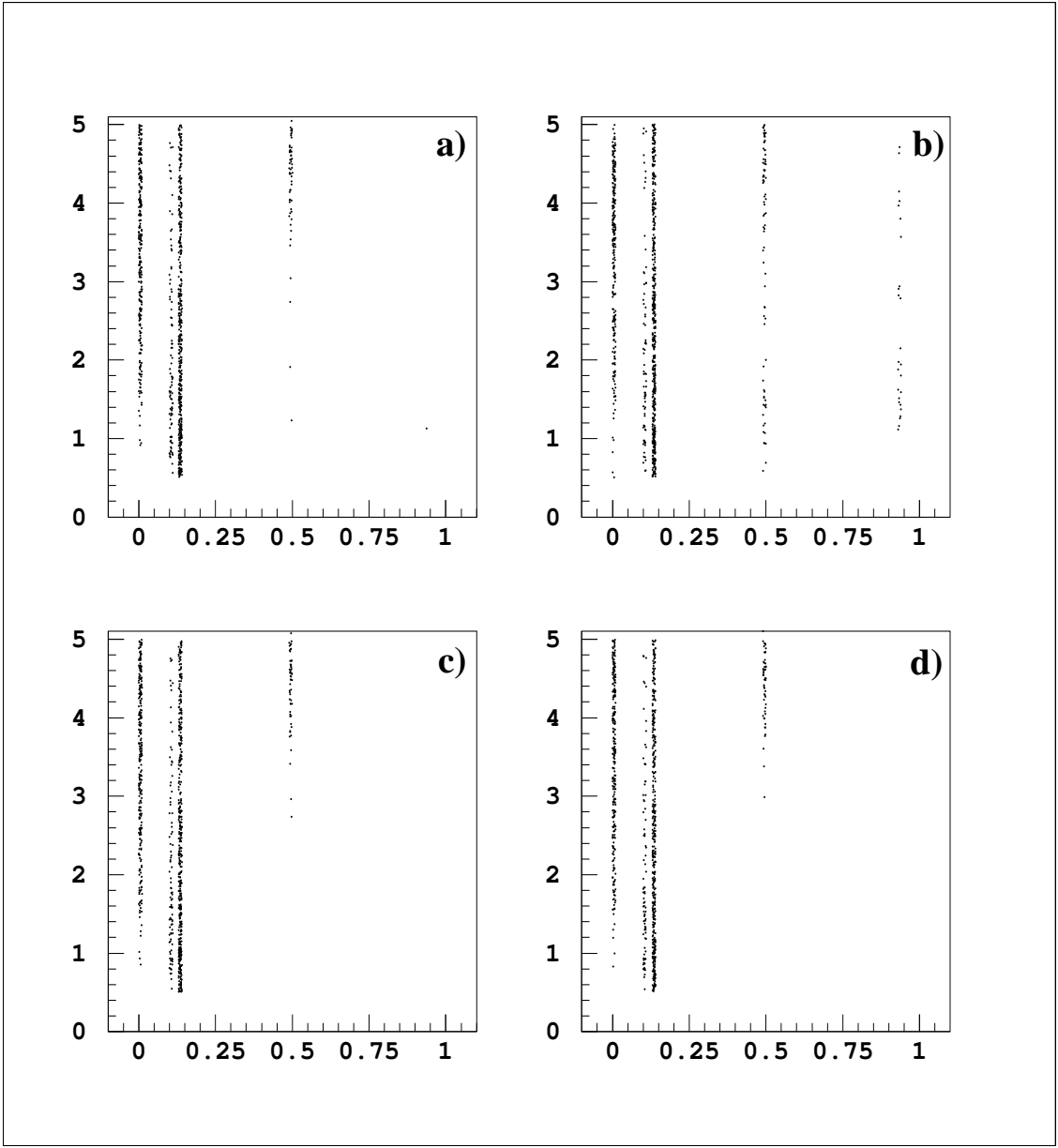,width=15cm}
\caption{Effect of the $|\delta\theta|$ secondary cut and $\chi^2$  probability cut.
The track momentum is plotted versus the identified mass~; units are GeV/c and GeV.
In (a), the scatter plot is made for generated pion tracks
with one additional {\it identified} background track; in (b), same plot when the additional track is considered as
unidentified (i.e: no $|\delta\theta|$ secondary cut). In (c), $|\delta\theta|$ secondary cut
applied (as in (a)), but requiring additionally a $\chi^2$ probability greater than 0.01. In (d),
same plot as in (c), but with a $\chi^2$ probability cut level of 0.03.}
\label{midvsp}
\end{center}
\end{figure}

In Fig. \ref{midvsp} we show the reconstruction of generated pions and  present (in a)) the
case when the accompanying track has been reconstructed and the secondary cut on 
$|\delta\theta|$ has been applied. In (b), we assume this track has not been measured
and then one cannot use this further cut. In both cases, no threshold probability has been
required and the identification is attributed to the largest probability (which can thus be quite small).
The correct identification (except for $e$--$\mu$--$\pi$ degeneracy) is good, above 90\% anyway.
One can further see that the secondary cut on $|\delta\theta|$ produces a large
improvement by reducing severely the misidentification of pions as kaons and protons.
Despite the $e$--$\mu$--$\pi$ degeneracy, below $\simeq$ 950 MeV pion
misidentification as electron looks negligible and, then, electron identification below this threshold
is possible with a good accuracy. Misidentification is low below $\simeq$ 3.5 GeV/c.

In Fig. \ref{midvsp}c, we have applied both the secondary cut on $|\delta\theta|$ 
and a threshold probability of 1\% which cleans up almost completely the momentum range below
$\simeq$ 3.5 GeV/c. It is easy to see that, strengthening this threshold to 3 \%, sharply reduces the
misidentification which becomes significant only above $\simeq$ 4 GeV/c.
 
All this illustrates that the largest difficulties are due to unmeasured tracks or unstructured noise.
As soon as a limited additional information is available (secondary track directions), the quality
of the identification sharply improves. Setting a threshold probability at a low level
appears naturally to be a suitable cleaning up tool, as a non--negligible part of the
misidentification, and hence of contamination, is produced by low probability reconstructions.

\subsection{Summary}

\indent \indent
As a matter of conclusion, even under severe background conditions, it is possible to remove 
background and select photons associated with a given charged track with a good efficiency
(still about 90\% of the tracks are fit) and with a limited contamination (about 95 \%
of particles identified as pions are indeed pions and the misidentification as kaons is
always low, sometimes very low).  This is achieved by starting with using
the subset of unambiguous photons~; indeed fitting with them the Cerenkov ring gives access
to refined values of the circle parameters which are used when reexamining the ambiguous photons. 
These improved parameter values allow a refined treatment of the photons left aside
as ambiguous.

The procedure described in Section \ref{photsel}
could possibly be improved, but basically it contains most of the needed ingredients.
Moreover, the fit algorithm presented in Section \ref{circlefit} looks well adapted
in order to fit circle arcs and to provide $\chi^2$ distances for fake photon removal.
This method allows to use all basic statistical tools (probabilities,
likelihoods \ldots).
Of course, we know that fake photons cannot be removed at the 100\% level, however, we have proved
that their residual contamination, in severe enough conditions, can be made low enough that the fit quality 
and its statistical meaning are not spoiled. 

\section{Cut Handling}
\label{cutadjust}

\indent \indent
The various cuts described in Section \ref{photsel} represent finally a set of 8 parameters~:
two cuts on  $|\delta\theta|$ allowing to define the starting sample of unambiguous photons,
the median cut level for background removal, 
the "perpendicular tracks" cut used to remove trivial ambiguities,
the two $\chi^2$ distance cuts in the ambiguous photons recovering step,
the minimum number of unambiguous photons, and the tails removal cut after the first fitting
step.

Some of these cuts can be adjusted in order to adapt the algorithm to different noise and background conditions,
some are nearly fixed by the detector characteristics ($|\delta\theta| < 30$ mr, perpendicular
track cut), or by algebra (the minimum number of unambiguous photons
is set at the smallest admissible value).
Also, one can verify that the last two cuts only influence marginally the reconstruction procedure performance. 

The 3 $\chi^2$ cuts have a well defined statistical meaning and the procedure outputs usual $\chi^2$ probabilities.
Simple criteria can be used in order to adjust these cuts, in such a way that they filter the background without 
spoiling too much the signal. Indeed, too stringent cut values may bias dramatically the fit quantities or
spoil the probability density function of the fit results. Too loose cuts may instead
result in very low reconstruction efficiency and poor fit quality. As, in the DIRC problem, errors are
close to gaussians,  $3\sigma$ in tuning these cuts is a magic value of well defined meaning.
When having to cut below this level one has to check for potentially harmful effects (biases or unacceptable
changes in the probability density function of the fit results). 

\subsection{Tuning of the Adjustable Cuts}

\indent \indent
In fact, only the first three parameters in the preceding list can be considered as adjustable by 
the user to adapt the algorithm to different noise and background conditions.

As illustrated above, the main tool in tuning the various cut levels is the probability
distribution of reconstructed rings. From the most common experience, one knows that if errors and
correlations are well understood, these distributions are flat between 0 and 1; one can always
assume this has been checked on clean samples.
Clean samples can indeed be constructed using a Monte Carlo. Tagged samples of identified
(by other means) particles, or low multiplicity samples, can always be extracted from data. 

Then, once the reliability of the error/correlation handling is ensured,
any departure from flatness has to be attributed to background. The set of cuts
defined above is devoted to the removal of the various kinds of background photons.
They are tuned by asking the particular value of the cut to optimize the flatness
of the probability distribution. Ideally, when all cuts are at optimum values,
the probability distribution has mean value 0.5 and rms 0.289.

In real life however, one knows that an actually flat distribution always  exhibits a peak at low
probability, reflecting the fact that there always exist configurations where events are
improperly reconstructed. This situation has been already met here (see Figs. \ref{chi2prob1}).
Therefore the above rules have to be slightly modified.
Flatness has to be requested above some threshold value $\alpha$, which can be relatively
large for the purpose of tuning ($\alpha=10\%$, 20\% or more are as good). One can always compute
the mean value and rms of the probability distribution above this threshold as a function of
the cut value and fit it (or compare) to respectively $(1-\alpha)/2$ and $(1-\alpha)/\sqrt{12}$.

\begin{figure}[ht]
\begin{center}
\epsfig{file=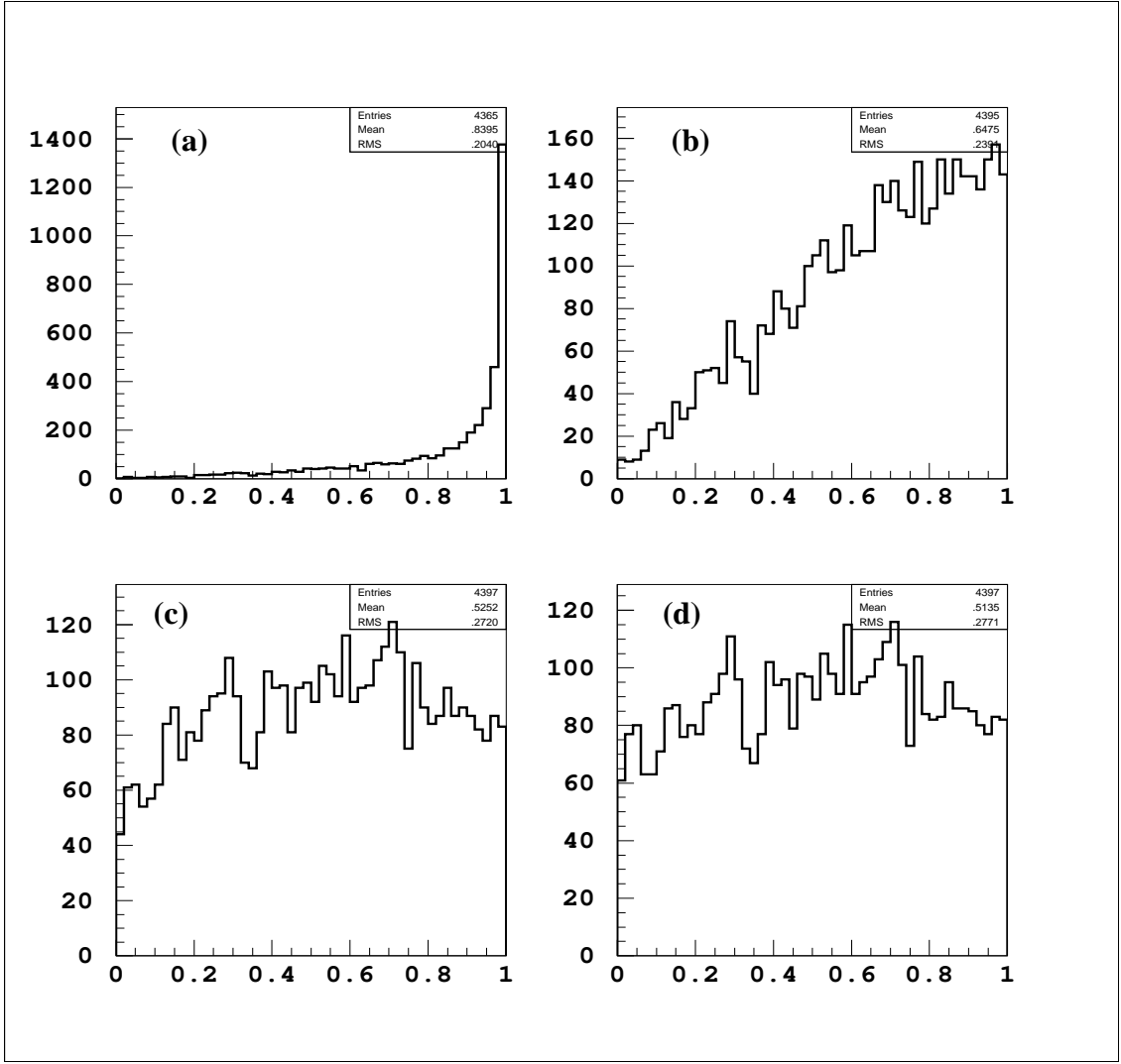,width=15cm}
\caption{Effect of median cut on the $\chi^2$ probability distribution for the sample of mixed particles with
no background of any type~; in (a), 1 $\sigma$ cut~; in (b), 2 $\sigma$~; in (c), 3 $\sigma$~; in (d), 4 $\sigma$.}
\label{chi2prob2}
\end{center}
\end{figure}

Fig. \ref{chi2prob2}, which deals with a sample of single tracks with no additional background,
illustrates how this tuning can be performed for the median cut. When fixing the cut level at 1$\sigma$,
we are clearly cutting too tight and then the probability distribution exhibits a huge peak at 1. Releasing
this cut allows to recover a non--biasing behavior at a cut level of about 3 to 4 $\sigma$ as one
can expect~: the mean probability is found close to 0.5 and its rms close to $1/\sqrt{12}$.
One can refine the tuning in an obvious manner. 

One may notice that the number of entries in the histograms is only marginally dependent on this cut
level~: the algorithm efficiency is not directly affected by it (but the final errors on the fitted
quantities are changing with this cut -- roughly like $\frac{1}{\sqrt{n_\gamma}}$ --  since the total
number of unambiguous photons entering the fit is cut dependent, even if the recovering procedure
finally limits its variation).

On the other hand, the low probability peak, as said above, reflects the existence of
improperly reconstructed rings. This is partly due to configurations where the errors
are not well estimated{\footnote{In the DIRC problem, the final errors are computed
by differentiating functions. They are geometry dependent and sometime
close to singular points of these functions.}} and partly due to the level of background
which survives the cuts. Usually, such a peak is made harmless by setting a 
threshold probability.

The two $|\delta\theta|$ cut play an important role in the magnitude of this peak.
The primary $|\delta\theta|$ cut, if it happens to be too loose, accepts
more easily background photons in the unambiguous photon subsamples. These background photons
will in turn degrade the fit quality and contribute to increase the peak at zero probability.
However, it cannot be set too tight (for instance $\leq 5$ mr) since in this case, the
number of unambiguous photons primarily associated with tracks may decrease too much
(below 2), and consequently the number of tracks lost will increase without
a significant improvement  at the zero probability peak.

\begin{figure}[ht]
\begin{center}
\epsfig{file=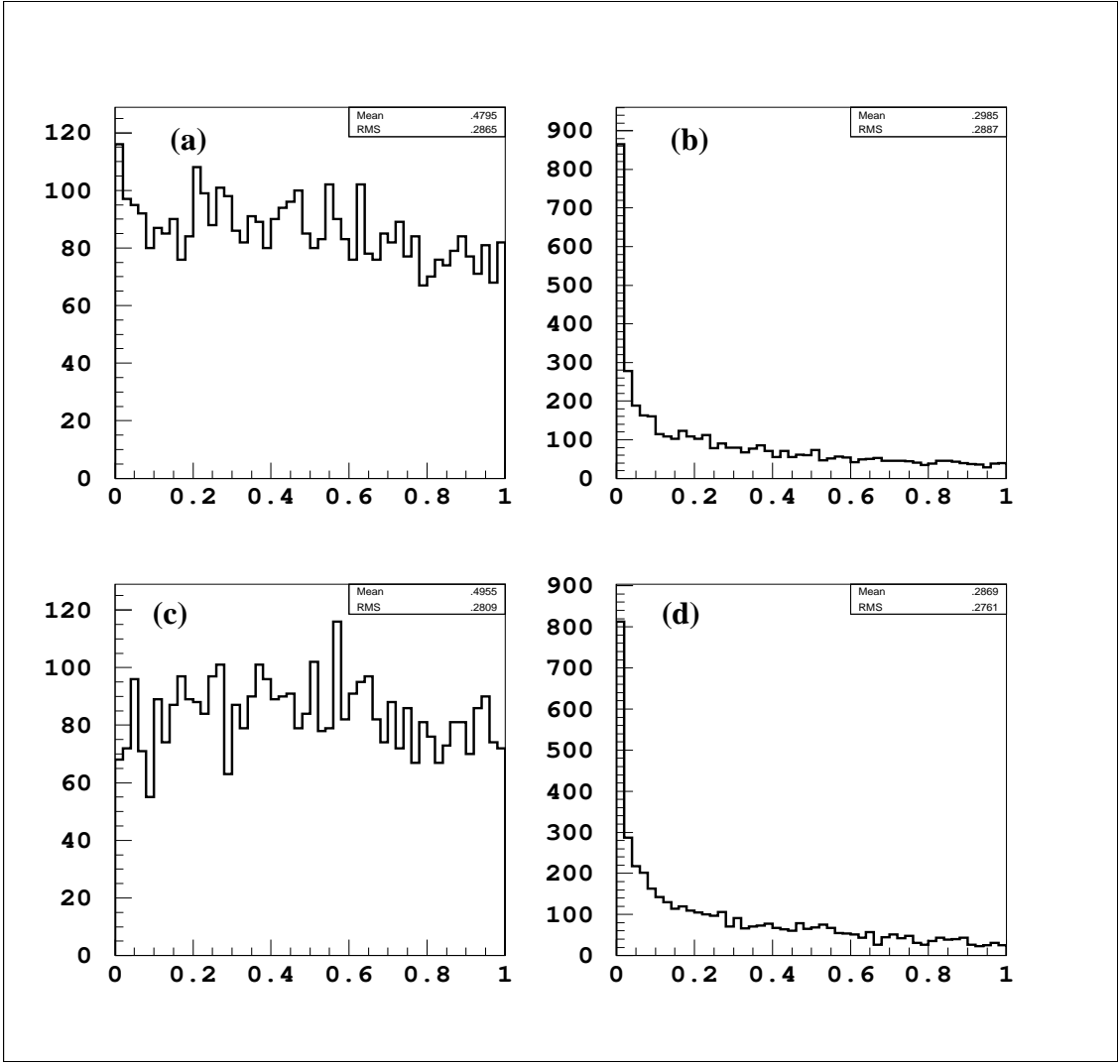,width=15cm}
\caption{Effect of the $|\delta\theta|$ secondary cut. The $\chi^2$ probability distribution for the
sample of mixed particles is shown with~: (a), 1 track background and $|\delta\theta|$ secondary cut~;
in (b), 1 track background and no $|\delta\theta|$ secondary cut~; in (c), 2 tracks background
and $|\delta\theta|$ secondary cut~; in (d), 2 tracks background and no $|\delta\theta|$
secondary cut. Note that the other cuts levels are different in cases a) and c).}
\label{dthetacut2}
\end{center}
\end{figure}

The secondary $|\delta\theta|$ cut acts likewise as can be seen
from Fig. \ref{dthetacut2}. Its strong power of rejecting the inter--track noise is 
illustrated by the vanishing low probability peak in quite hard background conditions. 

These examples illustrate how a given cut level can be tuned on any sample of tracks~; here again the procedure
applies to simulation and real data as well. The cut levels are tuned in such a way that the probability
distribution has its expected flat shape. This tuning allows to recover at the same time a 
mean probability of $\simeq 0.5$ and a spread of $\simeq 1/\sqrt{12}$ rms. After discarding (eventually) the low
probability bins, it is easy to compute the efficiency
of any further cuts in probability, because of the flatness of the remaining region.

\subsection{Effect of non tunable cuts}

\indent \indent
The non tunable cuts, briefly considered before, are the perpendicular tracks recovering cut and the
minimum number of unambiguous photons cut. 
The former is tunable only in the sense that it should be adjusted on data or on Monte Carlo simulation not to spoil
the algorithm acceptance (eventually creating biases if the cut is too wide), and not being rendered inefficient
by a too stringent width (in this case, only a small fraction of the almost perpendicular tracks is concerned
by the cut). Anyway, as stated before, this cut has only limited effects~: its activation or deactivation is hardly
noticeable on the studied data.

\begin{figure}[ht]
\begin{center}
\epsfig{file=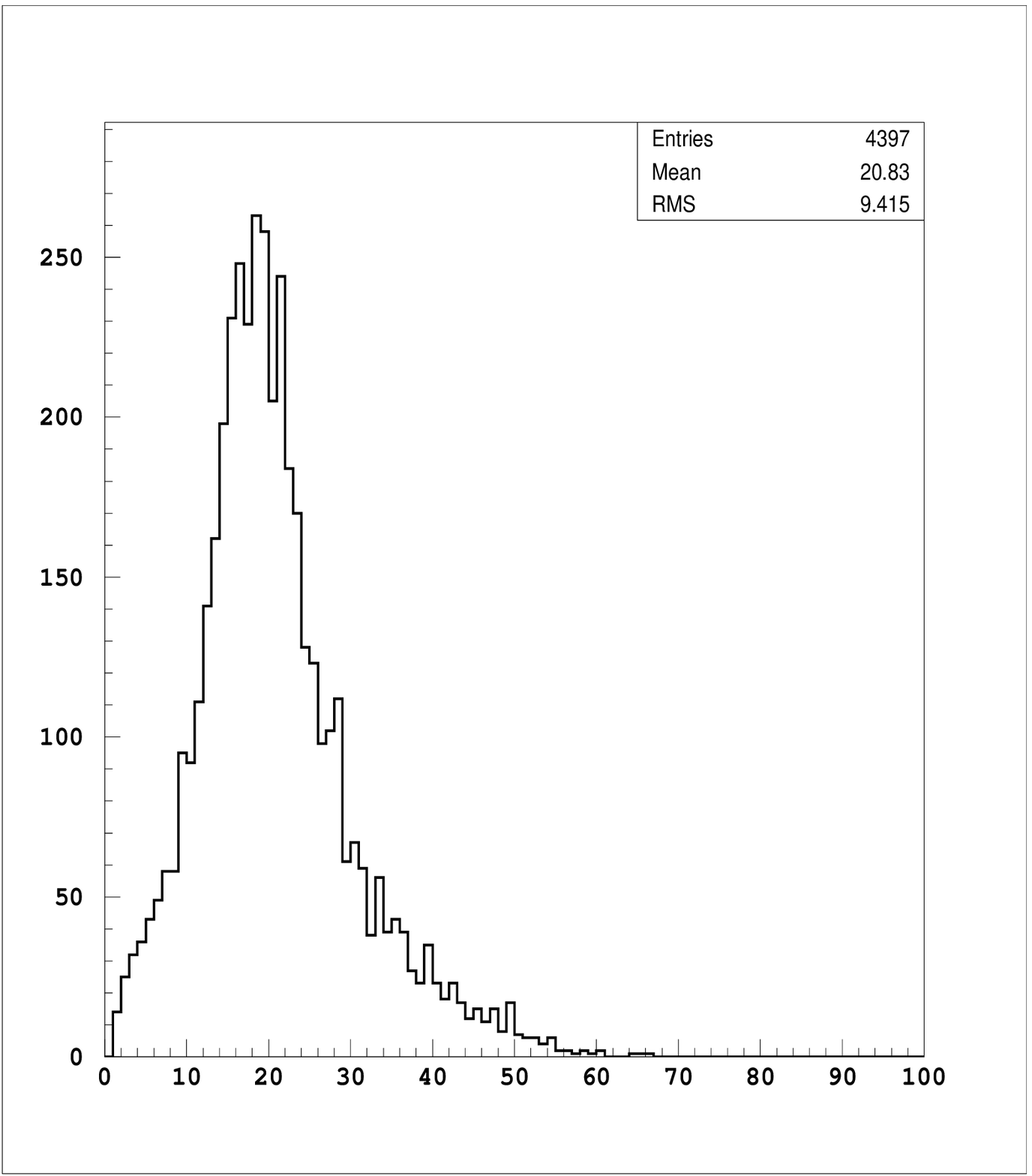,width=15cm}
\caption{The number of unambiguous photons per track obtained and the end of the algorithm for the mixed
particle sample (no background, no additional tracks).}
\label{nambgam}
\end{center}
\end{figure}

The second non adjustable cut has a rather strong influence on the global algorithm acceptance. Figure \ref{nambgam}
shows the distribution of the number of unambiguous photons at the end of the procedure~: it is clear that if the
cut level is increased starting from the normal value of 2, the number of non accepted tracks will be strongly
affected, since in this region the distribution is showing a strong dependence of the number of tracks upon
the number of unambiguous photons. Normally, there is no point in increasing the cut level above
2 or 3 unambiguous photons.

\vspace{1.cm}
To conclude this section, we can say that the choice of the cut level for both cuts is not really free, but
constrained by considerations depending strongly on the use of the algorithm and the type of data.

\section{Conclusions}

\indent \indent 
We have studied a procedure able to perform pattern recognition among photons in order to reconstruct
the Cerenkov angle associated with the charged track emitting them. 
The procedure, which has been developed for the case of the
DIRC, may apply {\it mutatis mutandis} to data from other ring imaging devices.
The basic requirement was that the procedure should provide a good approximation
of a $\chi^2$ value in order that a probability can be acceptably defined,
even in presence of a huge background mixed to the signal photons.
We have shown that such a procedure can indeed be constructed and proved
to work satisfactorily from the kaon Cerenkov threshold up to 4$\div$5 GeV/c.

We advocated the use of the stereographic projection which guarantees
that the figure to be fit is always a circle, whatever systematic
errors, misalignment problems, etc \ldots are. Moreover, we have shown that
such kind of errors affects negligibly the estimate of the central value for the radius
({\it} i.e. the Cerenkov angle).

The basic tool of the procedure is a fit corresponding to minimising
a $\chi^2$ expression, linearised as commonly
done. In the case of the DIRC,  this $\chi^2$ has
to be modified in order to take into account that the photon do not
populate the full circle, but rather a relatively small fraction of it.
The modification implemented turns out to take into account in the fit procedure
the existence of a measurement of the charged track direction, beside the photon
directions. In this way, systematic errors in the circle reconstruction 
can be avoided almost completely. We have also widely illustrated the role of
correlations and methods to estimate them. The fit procedure returns a value
of the circle radius (connected with the Cerenkov angle) and an improved
information on the charged track direction. This last information has been shown
to be crucial in cases where the charged track direction is poorly known, either intrinsically
(from the tracking device in front of the DIRC), or because of large multiple scattering effects.

We have shown, that using  this tool, it is possible to define an algorithm
able to solve ambiguities and remove efficiently background photons. It relies on
an iterative procedure, based on an intensive use of $\chi^2$ distances. 
It starts with unambiguous photons and recovers additional photons in a second pass.
Such a procedure depends on cuts, and it has been shown that one can check the effects of these cuts
and tune their levels on pull and probability distributions. This allows also to define
a calibration procedure which can be worked out with real data in a simple way
and optimized in real background conditions.

With this respect we have shown that the "combinatorial" background
generated by ambiguities can be easily overcome. We have also shown that
the background produced for a given track by photons associated with
other reconstructed tracks was easy to deal with.
The hardest background is  provided by non--reconstructed tracks
or flat background of various origins~; we have shown that it is possible to
deal reliably with them too, by using tighter cut levels.

As a final result, we have shown that the reconstruction and particle
identification are possible through a fit of the Cerenkov angle and particle
direction, with a remarkable efficiency (above 90 \%). We have also shown
that the fit probability is correctly estimated and has the expected flat distribution. 
This means that the cleaning up part of the procedure is able to discriminate efficiently between hits,
even when there is a large background together with the signal. The residual background contamination
level has been shown to be harmless under realistic conditions. Therefore one can use 
standard statistical tools in order to calibrate cuts and check the reconstruction quality.

\newpage

\renewcommand{\theequation}{\Alph {section} . \arabic {equation}}
\setcounter{section}{1}
\setcounter{equation}{0}
\section*{Appendix A : Errors and Correlations in a DIRC Device}
 
\subsection*{A1 Error Functions of the Charged Track Direction}

\indent \indent The 
charged track direction is affected by two different kinds of errors.
The first kind are the measurement errors on the track at the entrance into
the radiator (the quartz bar), the second is the error produced by its
multiple scattering inside the radiator, which makes that, for each emitted
photon, the Cerenkov angle $\theta_C$ is relative each time to a slightly
modified track direction. Let us denote the incoming track direction by 
$\vec{q}=(\sin{\theta} \cos{\phi}, \sin{\theta} \sin{\phi}, \cos{\theta})$~;
the condition $\vec{q}^2=1$ implies that $\vec{q} \cdot \delta\vec{q}=0$, and then
that the error vector $\delta\vec{q}$ is perpendicular to the track direction
$\vec{q}$. Let us write it~:

\begin{equation}
\delta\vec{q}=\delta_1\vec{q}+ \delta_2\vec{q}
\label{app1}
\end{equation}

\noindent where $\delta_1\vec{q}$ refers to the measurement errors and
$\delta_2\vec{q}$ to the multiple scattering. The error functions on the track
parameters $(\theta,\phi)$ being denoted $(\delta\theta,\delta\phi)$, we
have~:

\begin{equation}
\delta_1\vec{q}=\delta \theta ~\vec{v}+ \sin{\theta} \delta\phi ~\vec{w}
\label{app2}
\end{equation}

\noindent where $\vec{v}= \partial \vec{q}/ \partial \theta$ and 
$\vec{w}=[1/\sin{\theta} ] \partial\vec{q}/\partial \phi$ are unit vectors orthogonal to
each other and to $\vec{q}$. We may have $<\delta\theta\delta\phi> \ne 0$.
After a path of length $u$ inside the quartz, we also have~:

\begin{equation}
\delta_2\vec{q}= \displaystyle c \sqrt{\frac{u}{X_0}}
\left[\varepsilon_1(u)\vec{v} + \varepsilon_2(u)\vec{w}\right]
\label{app3}
\end{equation}

\noindent where \cite{PDG} $c=13.6~ 10^{-3}/\beta p[{\rm GeV}]$, $\beta$
is the particle speed and
$X_0$ is the quartz radiation length. The quantitites $\varepsilon_i(u)$
are gaussian random variables such that $<[\varepsilon_i(u)]^2>=1$ and
$<\varepsilon_1(u)\varepsilon_2(u)>=0$. Here and throughout the paper we neglect 
all departures from gaussian distributions \cite{PDG}.
From Rel. (\ref{app3}), we get~:

\begin{equation}
<[\delta_2\vec{q}]^2>= \displaystyle 2c^2 \frac{u}{X_0}
\label{app4}
\end{equation}

However, as we don't know where the photon has been emitted, the best estimate
of
\newline
 $<[\delta_2\vec{q}]^2>$  for this variance is its mean value
over the path followed, assigning an equal probability to each possible emission point{\footnote
{The notation here is obvious~: the inner $<\cdots>$ denotes the statistical mean
({\it i.e.} the expectation value)
already defined in the body of the text,  while the outer$< \cdots >_{xy\cdots}$ 
denotes the {\it additional} average performed over continuous variables $x,y,\cdots$.}}~:

\begin{equation}
<<[\delta_2\vec{q}]^2>>_u= \displaystyle \frac{1}{L}\int_0^L 2c^2 \frac{u}{X_0} 
du= c^2\frac{L}{X_0}
\label{app5}
\end{equation}
 
\noindent where $L$ is the total path length of the charged particle inside the
quartz.

 Let us assume that two photons are emitted after respectively paths
$u_1$ and $u_2$ inside the quartz. As the photon detection does not reveal where it has
been emitted, we can have with equal probabilities $u_1>u_2$ or $u_2>u_1$. Therefore, 
up to higher order corrections, we have~:

\begin{equation}
\left \{
\begin{array}{llll}
{\rm if ~~~ } u_2> u_1~~:
   \left \{
   \begin{array}{lll}  
	\delta_2\vec{q}(u_1)=& \displaystyle c \sqrt{\frac{u_1}{X_0}} 
	\left[ \varepsilon_1(u_1) \vec{v}+ \varepsilon_2(u_1)\vec{w} \right]\\[0.5cm]
	\delta_2\vec{q}(u_2)=&\delta_2\vec{q}(u_1) + \displaystyle c \sqrt{\frac{u_2-u_1}{X_0}} 
	\left[ \varepsilon_3(u_2) \vec{v}+ \varepsilon_4(u_2)\vec{w} \right]\\[0.2cm]
   \end{array}
   \right.
\\[1.0cm]
{\rm if ~~~ } u_1> u_2~~:
   \left \{
   \begin{array}{lll}  
	\delta_2\vec{q}(u_2)=& \displaystyle c \sqrt{\frac{u_2}{X_0}} 
	\left[ \varepsilon_1(u_2) \vec{v}+ \varepsilon_2(u_2)\vec{w} \right]\\[0.5cm]
	\delta_2\vec{q}(u_1)=&\delta_2\vec{q}(u_2) + \displaystyle c \sqrt{\frac{u_1-u_2}{X_0}} 
	\left[ \varepsilon_3(u_1) \vec{v}+ \varepsilon_4(u_1)\vec{w} \right]\\[0.2cm]
   \end{array}
   \right.
\end{array}
\right.
\label{app6}
\end{equation}

\noindent where the various $\varepsilon$'s carry unit variance and are statistically 
independent when they carry different indices and/or different arguments 
({\it i.e.} $< \varepsilon_i(u_j) \varepsilon_k(u_l)>=0$ only if $i \ne k$ and/or
$j \ne l$). 

One can check that $\delta_2\vec{q}(u_2)$ and $\delta_2\vec{q}(u_1)$ have the same variance
given by Rel. (\ref{app4}$\!$) (or by (\ref{app5}$\!$)) and it is shared equally
between their $\vec{v}$ and $\vec{w}$ components (remind that $\vec{v} \cdot \vec{w}=0$).
Moreover, we can now compute the expectation value~: 
\begin{equation} 
<\delta_2\vec{q}(u_1) \cdot \delta_2\vec{q}(u_2)>=<[\delta_2\vec{q}(u_1)]^2> \Theta(u_2-u_1)+
<[\delta_2\vec{q}(u_2)]^2> \Theta(u_1-u_2)
\label{app7}
\end{equation}

\noindent where $\Theta$ is the standard step function and the expectations values
on the RHS are given by Rel. (\ref{app4}) with the appropriate argument. It is
useful to have an estimate of this correlation coefficient~; it is 
obtained by averaging Rel. (\ref{app7}) upon $u_1$ and $u_2$.
This is easily achieved~:
\begin{equation} 
<<\delta_2\vec{q}(u_1) \cdot \delta_2\vec{q}(u_2)>>_{u_1u_2}=\displaystyle \frac{1}{L^2} 
\int_0^L \int_0^L du_1 du_2 <\delta_2\vec{q}(u_1) \cdot \delta_2\vec{q}(u_2)>=\frac{2}{3} c^2 \frac{L}{X_0}
\label{app8}
\end{equation}

 Therefore, the average correlation amounts to 2/3 of the average variance.  The covariance fraction 
 carried by each component are~:
 
\begin{equation} 
\left \{
\begin{array}{ll}
<<[\delta_2\vec{q}(u_1)\cdot \vec{v}][\delta_2\vec{q}(u_2)\cdot \vec{v}]>>_{u_1u_2}&=
\displaystyle \frac{1}{3} c^2 \frac{L}{X_0} \\[0.5cm]
<<[\delta_2\vec{q}(u_1)\cdot \vec{w}][\delta_2\vec{q}(u_2)\cdot \vec{w}]>>_{u_1u_2}&=
\displaystyle \frac{1}{3} c^2  \frac{L}{X_0} 
\end{array}
\right . 
\label{app9}
\end{equation}

\noindent while all other covariance mean values are zero.

\subsection*{A2 Finite Size Sample Corrections to  Multiple Scattering Errors}

\indent \indent
It is clear that the length of the path followed before the track emits any photon is inaccessible, 
nor the ordered time (or path) sequence of the detected photons. This implies that one has
to work with averaged quantities. In the previous subsection,  averaging is defined by Rels. (\ref{app5}) 
and  (\ref{app8}), for respectively the variance and covariance terms. Averaging by integrals
assumes the emission of an infinite number of photons along the charged track path
inside the radiator.

Here we present another method, based on a finite number of radiated photons~; it allows to
find the $1/n$ corrections to the above method, while checking it conceptually.

Let us assume  $n$ detected photons are emitted along the path of length $L$. 
Photon acceptance is mainly connected with their azimuth on the Cerenkov cone~; therefore
we can assume for simplicity,
that these photons are emitted after equal paths of length $L/n$. Let us also assume 
we work with each coordinate of the circle center in the equatorial plane (final results for variances have
to be multiplied by 2 for comparison with the preceding subsection).
We denote by $x_i$ the coordinate of the charged track direction (actually its fluctuation around $x_0$, 
the $true$ track coordinate{\footnote{Obviously, the true coordinate at the DIRC entrance 
is approximated by the mean value provided by the tracking device in front of the DIRC, but does
not coincide with it.}} at the radiator entrance) 
when it emits the $i^{th}$ photon in its ordered time sequence. 
Then we have~:

\begin{equation}
\left \{
\begin{array}{ll}
x_1=&x_0+ \epsilon_1 \\[0.5cm]
x_2=&x_0+ \epsilon_1 + \epsilon_2 \\[0.5cm]
x_3=&x_0+ \epsilon_1 + \epsilon_2+ \epsilon_3 \\[0.5cm]
\cdots & \cdots
\end{array}
\right.
\label{app10}
\end{equation}

\noindent where the functions $\epsilon_k$ are centered independent random variables~:
$<\epsilon_k >=0$ and $<\epsilon_k \epsilon_l> = \sigma^2 \delta_{kl}$ ($\sigma^2=c^2 L/(nX_0)$, 
$c$ being already defined). Let us also define $A=n \sigma^2$, a quantity independent of $n$,
which coincides with the standard $\theta_{rms}^{plane}$ of the Review of Particle Properties \cite{PDG}.
For simplicity, we choose from now on $x_0=0$. Trivially, writing  
$<\epsilon_k >=0$ for all $k$, does not mean it is true for any given track, but
that this is fulfilled by the mean values computed from a large sample of $tracks$. Indeed, 
$one$ set of $x_i$ corresponds to $one$ track and then to $one$ sampling of the $\epsilon_k $'s.

The ordered time sequence is unknown~; nevertheless, we can define the multiple
scattering variance of the sample by~:

\begin{equation}
V_m= \frac{1}{n}\sum_k x_k^2
\label{app11}
\end{equation}

Using Rels. (\ref{app10}), it is easy to find the expectation value of $V_m$~:

\begin{equation}
<V_m>= \frac{n+1}{n} \frac{A}{2}
\label{app12}
\end{equation}

\noindent which tends to half the value in Rel. (\ref{app5}), when $n \rightarrow \infty$,
as expected. Here we have to split up the result into two parts (correlated and uncorrelated).
The correlated part is the variance of the center of gravity~:

\begin{equation}
G= \frac{1}{n}\sum_k x_k
\label{app13}
\end{equation}

It is easy to compute it and get~:

\begin{equation}
<G^2>= \frac{(n+1)(2n+1)}{2n^2} \frac{A}{3}
\label{app14}
\end{equation}

\noindent
which tends to 2/3 of $<V_m>$ when $n \rightarrow \infty$, in agreement with Rel. (\ref{app8}).
The mean uncorrelated part (reduced variance of the sample) is the mean value of~:

\begin{equation}
W_m= \frac{1}{n}\sum_k (x_k-G)^2
\label{app15}
\end{equation}

\noindent which is~:  

\begin{equation}
<W_m> = \frac{n^2-1}{n^2}\frac{A}{6}
\label{app16}
\end{equation}

\noindent and tends to 1/3 of the variance when $n \rightarrow \infty$.
Therefore, the integral averaging presented in the previous subsection gives
indeed the large $n$ behaviour of the mean multiple scattering effects. The subleading
terms are of the order $1/n$ and small~; they can easily be read off the results in this subsection.
They show that the sharing uncorrelated--correlated parts of the variance departs from 1/3~:~2/3 by terms
of the order $1/(3n)$. They may become important only for number of photons of the order $n \leq 4 \div 5$.

\subsection*{A3 Handling of Other Errors}

\indent \indent There are qualitatively  two kinds of errors we have to deal with
in the DIRC problem. The first kind are mostly geometric errors due  
to the reconstruction of the photon direction~:  finite size of the photomultipliers,
finite size of the bar entrance window. There is no difficulty of principle in their 
reconstruction and propagation from the water tank and  photomultipliers back into 
the quartz bar and we will not comment on this any longer.

The second kind of errors (chromaticity) is due to the dependence of the
refraction indices (water and quartz as far as the DIRC is concerned)
upon the photon wavelength. This is due to the fact that photons
emitted by Cerenkov effect do not have a definite wavelength, which 
rather runs over a relatively wide spectrum. 

When refracting back the photon direction from
water (index $n_W(\lambda)$) to quartz (index $n_Q(\lambda)$), it is appropriate
to consider as basic random variable the quartz index~; then, the ratio $g$ of
these indices can be expressed as a function of $n_Q$ .
In the BaBar setup, this ratio can be considered constant over
the range of wavelengths to which the photomultipliers are sensitive. 

The quartz index is surely an appropriate variable because, thanks to Rel. (\ref{basic1}),
it affects directly the expected Cerenkov angle $\theta_C$ and its error  for each
photon. This is an important source of errors (it turns out to be equivalent
to treating the index as a random variable with a standard deviation corresponding to
$\delta n/ n \simeq 0.6 \%$). This error can be either attributed to the charged track
direction or to each photon direction separately, but, in the former case, one has to
take into account the fact that these photon errors are uncorrelated (this gives rise to corrections
in the error magnitude once applied globally to the track direction {\bf )}.  In
the course of the fit procedure, the $\chi^2$ which allows to reconstruct the circle radius
takes naturally this effect into account.
Moreover, when fixing the Cerenkov angles to the values 
expected from the charged track momentum, assuming the possible mass assignments
(the consistency check referred to in the body of the text),
it allows to take into account automatically the spread due to the 
photon wavelength spectrum.

\newpage
\renewcommand{\theequation}{\Alph {section} . \arabic {equation}}
\setcounter{section}{2}
\setcounter{equation}{0}
\section*{Appendix B : Basics for a Circle Arc Reconstruction Algorithm}

\indent \indent As noted in the main text, when fitting a circle having at hand points (subject
to measurement errors) spread out onto a relatively small arc (typically 60$^{\circ}$),
and if the error on each point is non-negligible compared to the radius of the circle{\footnote{
With this respect a ratio of 2\%, typical for the BaBaR DIRC, is large when points are on a circle 
arc of about 60$^{\circ}$. For  a track fit in a drift chamber, 60$^{\circ}$ is relatively large as
the ratio of error to radius is here of about 10$^{-3}$.}}, the fit quality becomes poor. 

Indeed, one can attempt a circle fit using standard algorithms \cite{CIRCLE1,CIRCLE2,CIRCLE3}.
However, fitting the 3 circle parameters under the typical conditions sketched above, using 
{\it only} the points on the circle leads to unexpected bad results. The center coordinate
in the direction associated with the two outmost ``experimental'' points
is relatively good~; however, the center coordinate in the direction toward the arc
is significantly and systematically displaced towards the arc
(compared with its expected value) and consequently 
the fit radius is systematically  underestimated, sometimes badly.

One way to circumvent this problem is to introduce  additional information about the circle center 
(the charged track direction as measured by another device than the DIRC, for instance a drift chamber).
In this case, the fit algorithm works much better (as illustrated in this paper) in the sense that the 
pull is found with the expected unit standard deviation and a negligible residual methodological 
bias{\footnote{Practically, the residual methodological bias is of the order 10$^{-4}$ of the
radius ({\it i.e.} about 0.1 to 0.2 mr for the Cerenkov angle) and then it is
overwhelmed by the bias originating from fake photons, which cannot be completely removed 
by any realistic procedure.}}.

\subsection*{B1 A Naive Estimate of the Error Matrix of the Charged Track}

\indent \indent The vector  $\delta \vec{q}$ defining the error on the charged track direction 
$\vec{q}$ is  given by Rels. (\ref{app1}),  (\ref{app2}) and (\ref{app3}). It is perpendicular to the 
charged track direction, and it 
projects out onto the equatorial plane (through the stereographic projection) as $\delta \vec{q}/2$.
Denoting $a$ and $b$ the coordinates of the circle center in the equatorial plane
in the directions parallel resp. to $\vec{v}$ and $\vec{w}$ (both orthogonal to $\vec{q}$), 
we can define the error functions on the circle center by~:

\begin{equation}
\left \{
\begin{array}{lll}
\delta a = \displaystyle \frac{1}{2} \delta \theta + 
\frac{1}{2} c \sqrt{\frac{u}{X_0}} \varepsilon_1(u)\\[0.5cm]
\delta b=\displaystyle \frac{1}{2} \sin{\theta} \delta \phi +
 \frac{1}{2} c \sqrt{\frac{u}{X_0}} \varepsilon_2(u)
\end{array}
\right .
\label{appb1}
\end{equation}
\noindent where $u$ is the path length followed by the charged track inside the radiator.
No average value over the path length has to be performed in this naive approach. Doing this way,
the error matrix $\Sigma$ for the charged track can be computed by taking the expectation value of
 appropriate second order terms. One can then choose, as a rule of thumb, to approximate the
error functions above by considering the standard deviations at $u=L/2$, {\it i.e.} at half the full
path of the charged track inside the radiator~; in this case, we have~:

\begin{equation}
\rm{rough ~estimates~:~~}
\left \{
\begin{array}{lll}
<[\delta a]^2>= \displaystyle \frac{1}{4} \left[ <[\delta \theta]^2>
+\frac{1}{2}c^2\frac{L}{X_0} \right]\\[0.5cm]
<[\delta b]^2>= \displaystyle \frac{1}{4} \left[ \sin^2{\theta}<[\delta \phi]^2>
+\frac{1}{2}c^2\frac{L}{X_0} \right]\\[0.5cm]
<\delta a \delta b>= \displaystyle \frac{1}{4}  \sin{\theta} <\delta \theta \delta \phi>
\end{array}
\right.
\label{appb2}
\end{equation} 

\noindent where the quantities $<[\delta \theta]^2>$, $<[\delta \phi]^2>$ and
$<\delta \theta \delta \phi>$ are the elements of the error matrix $\Sigma_0$ 
provided by the reconstruction procedure 
from the tracking device in front of the DIRC bar. The question now is to get a motivated 
 estimate of the full error matrix $\Sigma$ of the charged track direction 
by taking into account theoretical {\it a priori} information on multiple scattering.

\subsection* {B2 First Approach to the Charged Track Error Problem}

\indent \indent Let us consider only the $a$ coordinate of the circle center,
or restrict our problem to a one dimensionnal aspect. Solving the 2-dimensional 
case ($a,b$) will follow straightforwardly.

Actually, what is of relevance for our problem, is the mean value and error 
on the directions of charged tracks associated with the emitted {\it and} detected photons. 
There is some difference with the errors at half path inside the radiator, as
will be shown below.
 
For a charged track entering the radiator medium, the tracking device provides
a measurement of the direction with its error matrix ($\Sigma_0$ referred to above).
Let us associate with this measurement the origin in the equatorial plane where
all directions are projected out. 

 When the charged track emits photon $i$, it has a given (even if unknown
exactly) direction~; this direction varies from photon to photon
simply because of the multiple scattering the charged track undergoes.
The intersection of this direction  with the unit sphere has for image in the equatorial
plane the point of coordinate $a_i$ along the direction $\vec{v}$ (see Section A1)~;
what is of relevance  for the Cerenkov angle estimate is clearly
the mean value and standard deviation of the set $\{a_i~;~i=1, \cdots ~n\}$.

On the other hand, even in perfect cases (no multiple scattering)  the actual direction 
of a given track is obviously not the mean value (defined here as the central value given 
by the drift chamber reconstruction). Fitting the Cerenkov cone allows possibly to improve this 
last estimate using a (large) number of additional informations (photons), going thus
closer to its {\it actual} value. The effect of multiple scattering is that
the actual center seen from each photon is different and randomly distributed~;
then we cannot access $one$ actual circle center, but only define a mean value
of the set of actual centers.

Moreover, if photons can allow for improving the average estimate of the charged 
track direction, errors can be estimated from the (underlying) 
parent random distribution.

Therefore, we can write~:

\begin{equation}
a_i=a_0+\delta a_i~~~, ~~~i=1, \cdots ~n
\label{appb3}
\end{equation}

\noindent where $a_0$ denotes here the expectation value of the 
center of gravity of the $\{a_i\}$ set{\footnote {
Actually, this turns to define the $a_i$'s by relations
like Rels. (\ref{app10}), with an additional offset $a_0$ in place of $x_0$. 
}} and
where the random error functions $\delta a_i$ given by Rel. (\ref{appb1}) 
with different paths $u_i$, are unbiased ($<\delta a_i>=0$).
In connection with what said just above, it should be stressed 
that  the expectation value $a_0$ for all measured photons associated
with $one$ track is not necessarily zero, but the set of all $\{a_0\}$  (each associated
with a given track) is surely distributed around zero, with known deviations.  
Then, what is of relevance in our problem is
the value of $a_0$  and its standard deviation on a track by track basis. 

In order to get the mean value and the error of the direction set $\{a_i\}$,  one can
minimize the function (we define a vector $g$ of components $g_i=1, ~\forall i=1, \cdots n $
in order to match indices)~:

\begin{equation}
F(a)=(ag_i-a_i) V^{-1}_{ij}(ag_j- a_j)~~~~,~~{\rm{with~}} ~ ~V_{ij}=<\delta a_i \delta a_j>
\label{appb4}
\end{equation} 

\noindent where summation over repeated photon indices is understood. The zero of $dF(a)/da$
gives this minimum (using also Expression (\ref{appb3}))~:

\begin{equation}
\displaystyle a= \frac{g_i V^{-1}_{ij} a_j}{g_i g_jV^{-1}_{ij}}
=a_0+\frac{g_i V^{-1}_{ij} \delta a_j}{g_i g_jV^{-1}_{ij}}
\label{appb5}
\end{equation}

This expression gives the usual result for the estimate $a$  from a set
of measurements $\{a_i\}$ in the least squares approach. Its expectation
value{\footnote{As all $a_i$ have been fixed at zero (the measured central value) for the track considered,
this corresponds to take as sampling $\delta a_i=-a_0 g_i$.}} 
$<a>=a_0$ is unbiased  and its error function $ \delta a$, which
can be read off Rel. (\ref{appb5}), allows to compute its standard deviation 
$\sigma_a$ ($\sigma_a^2=<[\delta a]^2>$)~:

\begin{equation}
\displaystyle \frac{1}{\sigma_a^2}=g_i g_jV^{-1}_{ij}
\label{appb6}
\end{equation} 

When there is no correlation ($ V_{ij}\simeq \delta_{ij}$), this expression gives 
the usual result ($\sigma_a^{-2}=\sum_i \sigma_{a_i}^{-2}$).

In general $\delta a_i$ and $\delta a_j$ are given by expressions like in Rels. (\ref{appb1})
with two different path lengths $u_i$ and $u_j$, both unknown.  Correspondingly
the quantity $V_{ij}=<\delta a_i(u_i) \delta a_j(u_j)>$ can only be estimated
by performing the average as presented in Section  A1 of Appendix A. This gives~:

\begin{equation}
V_{ij}=<<\delta a_i(u_i) \delta a_j(u_j)>>_{u_iu_j}=\left [ B E + \frac{1}{24}A I \right]_{ij}
\label{appb7}
\end{equation} 

\noindent where $E$ is a rank 1 matrix 
such that each $E_{ij}=1$~; here $B=[<[\delta \theta]^2>+1/3A]/4$ and $A=c^2L/X_0$.
In order to compute $\sigma_a^2$ using Rel. (\ref{appb6}), we need to invert $V$
just defined. Indeed, using $E^2=nE$, it is easy to prove that~:

\begin{equation}
V= \lambda I + \mu E \Longleftrightarrow V^{-1}= \frac{1}{\lambda}[I- \frac{\mu}{n\mu+\lambda} E]
\label{appb8}
\end{equation}

\noindent 
where $n=$ dim $I=$ dim $E$ is also the number of points (photons).
Using this formula with the matrix in Rel. (\ref{appb7}), we easily find~:

\begin{equation}
\sigma_a^2= \frac{1}{4}[<[\delta \theta]^2> +\frac{1}{3}A+\frac{1}{6n}A] 
\label{appb9} 
\end{equation}

This relation could be expected beforehand. It shows that the uncorrelated part of the error
affecting each $a_i$  (2/3 of the variance associated with multiple scattering 
as shown by Rel. (\ref{app8}),  plus the variance provided by the device in front of the 
DIRC bar) is transfered to $a$ without changes, while the correlated part  
(1/3 of the multiple scattering contribution to the full variance) scales with $n$.
If we had neglected the terms generated by the multiple scattering effects
in $V_{ij}$ (for $i \ne j$), we would have got instead $\sigma_a^2=[<[\delta \theta]^2>
+A/(2n)]/4$ which can be considerably smaller at low track momentum.

Actually, the finite size sample corrections (see Section A2) may be accounted for. 
Concerning the uncorrelated part, this would amount to $1/n^2$ corrections and can be 
neglected~; the correlated part gives however a $1/n$ contribution which  corrects
the term $A/6n$ in Rel. (\ref{appb9}) by a factor of 4. Therefore, 
an improved expression for $\sigma_a^2$ taking into account all $1/n$ corrections is~:

\begin{equation}
\sigma_a^2 \displaystyle
\left|_{n} \right.= \frac{1}{4}[<[\delta \theta]^2> +\frac{1}{3}A+\frac{2}{3n}A] 
\label{appb9a} 
\end{equation}

It is also interesting to compare Rels. (\ref{appb9}) and (\ref{appb9a}) with 
the first Rel. (\ref{appb2}).
Indeed, one clearly sees that the  variance for $a$ is smaller than the variance
at mid path inside the quartz as soon as the number of photons is greater than 1
(large $n$ limit) or 4 (finite $n$ corrected). Then, in all practical applications,
Rels. (\ref{appb2})  do not reflect the sharing of the variance
between correlated and uncorrelated parts and leads to an overestimate of the
multiple scattering contribution to the center errors.

\subsection*{B3 The Full Charged Track Error Problem}

\indent \indent We have just treated (as a one--dimensional problem) the 
determination of the error on the the $a$ coordinate of the circle center,
taking into account the number $n$ of emitted photons. We have seen that
the uncorrelated part of its variance decreases as $1/n$ while the
correlated part is unaffected. However, our actual problem is two--dimensional
and because of correlations terms like $<\delta a_i \delta b_j>$, it is not equivalent
to the conjunction of two one--dimensional problems. 

In order to complete the treatment, let us display the following identity
for the inverse $V^{-1}$ of a symmetric matrix $V$
of rank and dimension $n$~:

\begin{equation}
V^{-1}=
\left (
\begin{array}{lll}
S_1 & \widetilde{C}\\[0.5cm]
C & S_2
\end{array}
\right )^{-1}
=
\left (
\begin{array}{lll}
(S_1-\widetilde{C}S_2^{-1}C)^{-1}  &- (S_1-\widetilde{C}S_2^{-1}C)^{-1}\widetilde{C}S_2^{-1}\\[0.5cm]
-S_2^{-1}C (S_1-\widetilde{C}S_2^{-1}C)^{-1} & S_2^{-1}+S_2^{-1}C (S_1-\widetilde{C}S_2^{-1}C)^{-1}
\widetilde{C}S_2^{-1}
\end{array}
\right )
\label{appb10}
\end{equation}

\noindent where the submatrices $S_1$ and $S_2$ are square matrices of dimensions
resp. $k~(<n)$ and $n-k$, while $C$ and $\widetilde{C}$ can be square or rectangular.
This identity allows to invert easily a large and structurally complicated matrix, when 
submatrices have peculiar forms, easy to invert (like  Rel. (\ref{appb8}), for instance).

Let us apply this relation to our problem. We have, by construction,
$S_{1~ij}=<\delta a_i \delta a_j>$, $S_{2~ij}=<\delta b_i \delta b_j>$ and
$C_{ij}=<\delta a_i \delta b_j>$, where the error functions are given by expressions
like Rel. (\ref{appb1}) with appropriate path lengths. It is easy to compute 
these (sub--)matrices~:

\begin{equation}
\left \{
\begin{array}{ll}
\displaystyle S_1=B_{\theta} ~E + \frac{1}{24}A~I &
~~~,~~~\displaystyle B_{\theta}=\frac{1}{4} [<[\delta \theta]^2> + \frac{1}{3} A] \\[0.5cm]
\displaystyle S_2=B_{\phi} ~E + \frac{1}{24}A~I &
~~~,~~~\displaystyle B_{\phi}=\frac{1}{4} [\sin^2{\theta}<[\delta \phi]^2> + \frac{1}{3} A ]\\[0.5cm]
\displaystyle C=B_{\theta \phi} ~E &
~~~,~~~\displaystyle B_{\theta \phi}=\frac{1}{4} \sin{\theta}<[\delta \theta \delta \phi]> 
\end{array}
\right.
\label{appb11}
\end{equation}

\noindent where the rank 1 matrix $E$ has been already defined  and still 
$A=c^2 L/X_0$. We clearly have $C=\widetilde{C}$ and all submatrices here are $n \times n$.
Rel. (\ref{appb10}) is useful in our case because these three submatrices
have each a special form. 

We can now define two sets of relations analogous to (\ref{appb3}) for the  $a_i$ and $b_i$
introducing this way $a_0$ and $b_0$,
and  the vector of dimension $2n$~: $(\cdots, ag_i-a_i,\cdots,bg_j-b_j,\cdots)$.  Then we can
define a function $\chi_c^2=F(a,b)$ in a way analogous to Rel. (\ref{appb3}),
using this vector and the matrix of Rel. (\ref{appb10}). Doing as previously,
it is easy to find that the solution which minimizes  $F(a,b)$ is a couple
of random variables $(a,b)$ of expectation values $(a_0,b_0)$ (to be fit) with an inverse of covariance matrix
defined by the sum of the elements of each of the four submatrices in Rel. (\ref{appb10}).
These sums can be easily computed knowing the submatrices $S_1$, $S_2$ and $C$ (Rels. (\ref{appb11})),
by means of Rels. (\ref{appb10}) and  (\ref{appb8}).

After tedious algebra, it follows from there that the center fixing term can be written~:

\begin{equation}
\chi^2_C=\left ( a_0,b_0 \right )
\left (
\begin{array}{ll}
\displaystyle \frac{1}{4}[<[\delta \theta]^2> + \frac{1}{3} A + \frac{1}{6n} A] &
\displaystyle ~~~~~~~\frac{1}{4} \sin{\theta} <\delta \theta \delta \phi> \\[0.5cm]
\displaystyle ~~~~~~~\frac{1}{4} \sin{\theta} <\delta \theta \delta \phi> &
\displaystyle \frac{1}{4}[\sin^2{\theta} <[\delta \phi]^2> + \frac{1}{3} A + \frac{1}{6n} A] \\[0.5cm] 
\end{array}
\right )^{-1}
\left (
\begin{array}{l}
a_0 \\[0.5cm]
b_0
\end{array}
\right )
\label{appb12}
\end{equation}

In writing this expression, we have used $a_0$ and $b_0$ instead of
$a_0-a_{measured}$ and $b_0-b_{measured}$, taking into account that the corresponding 
measured values are zero by definition.
This relation defines the error covariance matrix of the charged track direction 
associated with $n$  detected photons~; it differs from the matrix $\Sigma_0$
by its taking into account the multiple scattering undergone by the charged track inside the
radiator. It can simply be written $\Sigma=\Sigma_0+[A/3+A/(6n)]I$.

One can apply the finite size sample corrections found in Section A2,
as we did in the previous subsection. This 
amounts to change the expression for $\Sigma$ to $\Sigma=\Sigma_0+[A/3+2A/(3n)]I$.

Finally, it should be noted that the $a_0b_0$ covariance
term is not affected
by effects due to multiple scattering~; this could have been inferred from Rels. (\ref{appb1})
(and explains the covariance term in Rels. (\ref{appb2}),  as the expectation value
$<\varepsilon_1(u) \varepsilon_2(v)>$ is zero for any values of $u$ and $v$). 
From now on, we name for clarity the parameters to be fit $a$ and $b$
instead of $a_0$ and $b_0$. 

It should be noted, however, that multiple scattering effects imply that
there are as many centers (charged track directions) as photons. Therefore, 
any fit procedure can only provide a determination of the
$mean$ center coordinates for  each track considered, as an approximation
of the actual center value at the DIRC bar entrance.

\subsection*{B4 The Minimization Function and the Circle Parameters}

\indent \indent Given a set of points known each with some error, and assuming they
should be on a circle arc, the problem we state here is to define a function
$F(a,b,R)$, the minimum of which providing the circle parameters. An usual approach
is actually to choose as function $F$, a $\chi^2$. Denoting by $R$ the circle radius
and by $(a,b)$ the center coordinates, the function is~:

\begin{equation}
\chi^2_n=\sum_{i,j=1,n}(d_i-Rg_i) V^{-1}_{ij}(d_j-Rg_j) 
\label{appb13}
\end{equation}

\noindent where $d_i=\sqrt{(x_i-a)^2+(y_i-b)^2}$ is the distance of each point $(x_i,y_i)$
to the fit center, and  $g_i=1$ defines a constant vector $g$ introduced only in order to
have a correct matching of repeated indices. A priori, the covariance matrix 
is defined by $ V_{ij}=<\delta d_i \delta d_j>$.  Usually, the error functions
$\delta d_i$ are obtained by differentiating the expression for $d_i$~:

\begin{equation}
\delta d_i=\frac{(x_i-a)\delta x_i+ (y_i-b)\delta y_i}{d_i}
\label{appb14}
\end{equation}

\noindent where $\delta x_i$ and $\delta y_i$ are the errors functions affecting
the photon measurements $x_i$ and $y_i$.

In our case, the points are spread onto a small arc
and/or the relative size of the errors compared with the circle radius is large~;
then, one has to introduce additional information for reasons already quoted at several
places in the body of this paper. The most
obvious additional information which is available in our case refers to the circle center. 

We have as {\it a priori} information the measurement provided by the tracking device located in front
of the DIRC bar entrance~; this is summarized by a central value and a covariance error matrix (referred to
anywhere above as $\Sigma_0$). Obviously, this defines a distribution
(normal, assuming we are lucky) but not the actual location of the center 
indeed associated with the track under consideration.

The question is now~: how to introduce the approximate knowledge
$(0,0)$ of the given charged track direction and keep anyway its actual center coordinates
$(a,b)$ to be fit? In the previous case (no charged track information), the measured 
quantities could be written~: 

\begin{equation}
x_i=x^0_i+a+\delta x_i ~~,~~ y_i=y^0_i+b+\delta y_i
\label{appc15a}
\end{equation}

\noindent
where $x^0_i=R \cos{\varphi_i}$ and $y^0_i=R \sin{\varphi_i}$. Here 
$R$ is the true radius, $\varphi_i$ is the true azimuth on the circle and
$(a,b)$ the true center. Then the true value for the $x_i$ and $y_i$
are respectively $x^0_i+a$ and $y^0_i+b$. Fixing the center at the ``measured''
value $(a_i,b_i)$ (actually $(0,0)$), turns out to rewrite these equations~:

\begin{equation}
x_i=x^0_i+a_i+\delta x_i ~~,~~ y_i=y^0_i+b_i+\delta y_i
\label{appc15b}
\end{equation}

However, for a given well defined track, we can write~:

\begin{equation}
a_i=a+\delta a_i~~~,~~~~ b_i=b+\delta b_i ~~~~,~~~ \forall i=1, \cdots ~n
\label{appb15}
\end{equation}
 
\noindent where $\delta a_i$ and $\delta b_i$ are the error functions
which take into account the errors at the entrance of the DIRC bar $and$ the multiple 
scattering undergone by the charged track up to the point where it
emits photon $i$.  In this way $(a,b)$  is the $actual$ center  
when it  exists (no multiple scattering), otherwise it can be formally
defined as the mean value of the quantity corresponding to
$G$ in Rel. (\ref{app13}), which is then non--zero on a track by track basis. 
Then Eqs. (\ref{appc15b}) can be rewritten~:

\begin{equation}
x_i=x^0_i+a+ \delta a_i+\delta x_i ~~,~~ y_i=y^0_i+b+ \delta b_i+\delta y_i
\label{appc15c}
\end{equation}
 
The difference between the case when the center is left free and when it 
is constrained is transfered to the error functions which become $\delta a_i+\delta x_i$ and 
$\delta b_i+\delta y_i$ instead of respectively $\delta x_i$ and $\delta y_i$.
Conceptually, the difference comes from what is submitted to fit in both cases.
In the former case (free center), the measured quantitites submitted to fit are
the measured points $(x_i,y_i)$, while in the latter case (constrained center),
the quantitites
submitted to fit are actually $(x_i-a_{measurement},y_i-b_{measurement})$.
In the former case, the center $(a,b)$ is fully fit, in the latter case
one fits the departure of the actual center from the measured point $(0,0)$.

Then,  Rel. (\ref{appb14}) is still valid but should be rewritten~:

\begin{equation}
\delta d_i = \frac{(x_i-a)(\delta x_i+\delta a_i)+(y_i-b)(\delta y_i+\delta b_i)}{d_i}
\label{appb16}
\end{equation}

\noindent in order to keep $\delta x_i$ and $\delta y_i$ their original meaning
(errors due to the measurement of the photon direction without reference to the charged
track). 
If, moreover, we choose the origin in the plane in order that it coincides
with the image of the track direction provided by the tracking device, Rel. (\ref{appb16})
can be approximated by~: 

\begin{equation}
\delta d_i = \frac{x_i(\delta x_i+\delta a_i)+y_i(\delta y_i+\delta b_i)}{d'_i}
\label{appb17}
\end{equation}

\noindent where $d'_i$ in the denominator is $d'_i=\sqrt{x^2_i+y^2_i}$.
$\delta d_i$ in  Rels. (\ref{appb16}) and  (\ref{appb14}) differ only
at first order and only by the differentials for $\delta a_i$ and $\delta b_i$.

Therefore, the quantity $\chi^2_n$ (Rel. (\ref{appb13})) can be used with
$V_{ij}=<\delta d_i \delta d_j>$, where the errors functions are given by
Rel. (\ref{appb17}). As they depend on the path length followed up to the
emission of each photon, this expression has to be approximated by its mean value
$V_{ij}=<<\delta d_i \delta d_j>>_{u_iu_j}$ easily computable
using all information given above.

In this way, we are in position to define $\chi^2_n$ as a function of 
$(a,b)$ and $R$. It remains to account for forcing the coordinates 
$(a,b)$ to remain in the neighborhood of the measured point $(0,0)$~;
this is achieved{\footnote{One may ask oneself whether Rel. (\ref{appb18})
actually exhausts the problem. Indeed, one may be tempted to introduce
in the $\chi^2$ to be minimized, terms coupling $a$ and $b$ with the
$d_i-Rg_i$~; this is not studied here. Anyway, such additional terms
would surely degrade the algorithm speed. From the results already at hand, 
one may conclude that their effect is small for track momenta above 500 MeV/c.
}} 
by defining the following $\chi^2$~:

\begin{equation}
\chi^2 = \chi^2_n + \chi^2_C
\label{appb18}
\end{equation}

\noindent using Rels. (\ref{appb13}) and (\ref{appb12}), where 
the influence of multiple scattering is already taken
into account, including correlations. While $\chi^2_n$
is $n-3$ degrees of freedom, the $\chi^2$ just defined
is $n-1$ degrees of freedom.

One could ask oneself about the influence of the additional
term $\chi^2_C$ when minimising, if the circle arc happens to be large enough
that such a term is actually useless. We have checked numerically
this case using trivial simulations, where the populated arc length 
and the number of "measured" points could be varied at will. We have 
found that for large circle arcs (about 180$^{\circ}$ or more)
the additional term $\chi^2_C$ did not prevent to get the same solution as
when it is removed, with completely negligible fluctuations.

Another possibility could be considered. This turns to give up 
fitting the center, accepting the measured value $(0,0)$ as optimum.
Using the notation $d'_i \equiv d_i(a=0,b=0)$, this turns
to estimate the radius by~:

\begin{equation}
R=\displaystyle \sigma_R^2 \sum_{ij} g_i V^{-1}_{ij} d'_j ~~~, 
~~~\frac{1}{\sigma_R^2}=\sum_{ij} g_i V^{-1}_{ij} g_j
\label{challenge}
\end{equation}

This gives results close in quality to minimising Eq. (\ref{appb18}),
if the fixed center $(a,b)=(0,0)$ is correctly measured. If however, there
is any bias in this estimate, it may become much worser. Indeed, in case
when the measured center is slightly biased, minimising Eq. (\ref{appb18})
is safe as one recovers the correct center location, even starting from a 
wrong center~; instead, using Eq. (\ref{challenge}), the estimate of 
the radius suffers piling up effects which summarize in sensible effects. 
For instance, assuming 2.8 mr systematic error on the charged track direction
(known with statistical accuracy 2 mr rms),
the radius pull gets  a rms of 1.12 instead of 0.98,
under the same conditions, for the procedure we recommand.
This difference of behaviour degrades in presence of background 
(1.4 compared to 1.2) or if the statistical accuracy on the charged 
track direction worsens. 

\subsection*{B5 Correlations among Photons in the Equatorial Plane}

\indent \indent 

The radius $\tan{\theta_C/2}$ of the circle is approximated by each of the photon
distance to the common center ($d'_i =\sqrt{x_i^2+y_i^2}$~). Each such estimate
is affected by an error function which is given by Rel. (\ref{appb17}),
 with the measured center set at the origin in the equatorial plane.
From expressions given above, we can then deduce~:

\begin{equation}
\left \{
\begin{array}{llll}
<[\delta d_i]^2>&=&\displaystyle \frac{1}{d^{'2}_i}
\left[
x_i^2[<[\delta x_i]^2>  +\frac{1}{4}<[\delta \theta]^2>]+
y_i^2[<[\delta y_i]^2>  +\frac{1}{4}\sin^2{\theta}<[\delta \phi]^2>]+ \right.\\[0.5cm]
~&~&\left. \displaystyle 2 x_iy_i[<\delta x_i\delta y_i>+\frac{1}{4}\sin{\theta}<\delta \theta \delta \phi>]\right]
+ \displaystyle \frac{1}{8}c^2 \frac{L}{X_0}
\\[0.5cm]
<\delta d_i \delta d_j>&=&\displaystyle \frac{1}{4d'_id'_j}
\left[
x_ix_j<[\delta \theta]^2>+y_iy_j\sin^2{\theta}<[\delta \phi]^2>+ \right.\\[0.5cm]
~&~& \left.  \displaystyle (x_iy_j+x_jy_i) \sin{\theta} <\delta \theta\delta \phi>\right] 
+  \displaystyle \frac{(x_ix_j+y_iy_j)}{d'_id'_j}\frac{1}{12}c^2 \frac{L}{X_0}
\end{array}
\right.
\label{appb19}
\end{equation}

\noindent where we have assumed that the measurements ($x$ and $y$) for photons $i$ and $j$
are statistically independent for ease of reading.

These relations give the expression for the elements of the error covariance
matrix which enter the fit procedure described in the body of the  text and just above.
The variance for each estimate of the circle radius ($d'_i$) depends on the photon errors, 
the track direction errors and the multiple scattering it undergoes~;
the covariance term exhibits an interesting feature~: up to the fact that the metric along $\vec{v}$
and $\vec{w}$ are different, one sees a surprising correlation pattern. Qualitatively,
correlations are the strongest (and positive) for photons close to each other in azimuth,
correlations are the strongest (and negative) for pairs of photons opposite in azimuth, while
there is no correlation for photon pairs having azimuthal distance of $\pi/2$.

\subsection*{B6 Multiple Scattering Effects in the General Case}

\indent \indent When the arc to be fit is large enough (possibly $2 \pi$ radians),
fixing the circle center becomes irrelevant. In this case, a question remains
about the influence of multiple scattering on the error definition and the fit 
procedure.

For each photon, $d_i=\sqrt{(x_i-a_i)^2+(y_i-b_i)^2}$ remains the basic quantity which 
enters the fit procedure. Whatever is the way to express the
problem, we  have as free parameter the charged track direction
and as ``data'' the angular distance of this direction with photon directions.
As noted above, when multiple scattering is active, the direction of the charged
track varies from photon to photon with theoretically known statistical fluctuations.
Therefore the estimate of the radius provided by each photon inherits the
fluctuations of the charged track. Stated otherwise, the error due
to multiple scattering can either be treated separately (as we did)
or included in the error  function of  the measurement ($x_i,y_i$),
together with the other contributions (geometrical errors, chromaticity error).
This means that Rel. (\ref{appb17}) is still relevant. Therefore,
when there is no center  fixing term, 
it is equivalent to consider that the error
function on ($x_i,y_i$) is ($\delta x_i+\delta a_i,\delta y_i+\delta b_i$) ,
where the second term of each component is reduced to only the multiple 
scattering contribution. 

In this case, Rels. (\ref{appb19}) become~:

\begin{equation}
\left \{
\begin{array}{llll}
<[\delta d_i]^2>&=&\displaystyle \frac{1}{d^{'2}_i}
\left[ \displaystyle
x_i^2<[\delta x_i]^2> + y_i^2<[\delta y_i]^2> + 2 x_iy_i<\delta x_i\delta y_i>\right]
+ \displaystyle \frac{1}{8}c^2 \frac{L}{X_0}
\\[0.5cm]
<\delta d_i \delta d_j>&=&\displaystyle \frac{(x_iy_j+x_jy_i)}{d'_id'_j}
\frac{1}{12}c^2 \frac{L}{X_0}
\end{array}
\right.
\label{appb20}
\end{equation}

This  shows that correlations among photons always exist, only due however
to the properties of multiple scattering. Therefore, for low momentum tracks,
when the multiple scattering is dominant, correlations among photons can never
be ignored in any reconstruction procedure for devices like the DIRC.
This is clearly independent of the representation chosen for the data (here
the stereographic projection)~; it only relies on the fact that the measured
quantities are angles between photons and
a single charged track direction which changes in a correlated way from one photon to the other. 

\newpage
\section*{Acknowledgements}
\indent \indent We thanks our colleagues of the BaBar DIRC group for the
interest they manifested in the successive steps of this work. We also
warmly acknowledge J. Chauveau for important remarks and comments.

\newpage

\newpage

\newpage

\newpage

\newpage

\newpage
\clearpage

\newpage

\newpage

\newpage

\newpage
\clearpage

\begin{thebibliography}{9}
\bibitem{YPSILANTIS} T. Ypsilantis and J. S\'eguinot,  Nucl. Inst. and Meth.  {\bf A343} (1994) 30.
\bibitem{SEGUINOT} J. S\'eguinot {\it et al.},  Nucl. Inst. and Meth.  {\bf A350} (1994) 430.
\bibitem{BAILLON} P. Baillon, Nucl. Inst. and Meth. {\bf A238} (1985) 341. 
\bibitem{DIRC0} P.Coyle et al., Nucl. Inst. and Meth.  {\bf A343} (1994) 292.
\bibitem{DIRC} R.Aleksan et al., Nucl. Inst. and Meth.  {\bf A397} (1997) 261. 
\bibitem{BABAR} Babar Technical Design Report, SLAC---R--95--457, March 1995.
\bibitem{LBLRICH} D. Hatzifotiadou {\it et al.}, CERN--LAA/PI-94--17, Geneva 1994~;
T. Ypsilantis, J. S\'eguinot and A Zichichi, CERN--LAA/96-13, Geneva 1996.
\bibitem{DIRCNEW} I. Adam {\it et al.}, SLAC--PUB--7706, Stanford,  Nov. 1997;
preprint hep--ex/9712001, Dec 1 1997.
Invited talk presented at the 1997 IEEE Nuclear Science Symposium and Medical
Imaging Conference, Albuquerque, New Mexico, Nov 9--15 1997.
J. Schwiening {\it et al.} SLAC--PUB--7706, Dec. 1997;
preprint hep--ex/9712018, Dec 12 1997. Invited talk presented at the 5th International
Workshop on B--Physics at Hadron Machines (Beauty '97), Santa Monica, California,
October 13--17, 1997.
\bibitem{LHCB} LHCb Technical Proposal, CERN LHCC 98--4, LHC/P4, Geneva, Feb. 1998.
\bibitem{JULIA} G. Julia, El\'ements de G\'eometrie Infinit\'esimale,
Gauthier--Villars, Paris 1936.
\bibitem{CIRCLE1} J.F. Crawford, Nucl. Inst. and Meth. {\bf 211} (1983) 223.
\bibitem{CIRCLE2} N.I. Chernov and G..A. Osokov, Comp.. Phys. Comm. {\bf 33} (1984) 329.
\bibitem{CIRCLE3} V. Karim\"aki, Nucl. Inst. and Meth.  {\bf A305} (1991) 187.
\bibitem{MEDIAN1} P.L. Rosin, Pattern Recognition Letters {\bf 14} (1993) 661
\bibitem{MEDIAN2} P.J. Huber, Robust statistics (Wiley, 1981)
\bibitem{PDG} Review of Particle Properties, R.M. Barnett {\it et al.}, Phys. Rev. {\bf D54}
(1996)1.
\end{thebibliography}
\end{document}